\documentclass[%
 reprint, 
 amsmath,amssymb,
 aps, physrev,
]{revtex4-2}

\usepackage{graphicx}
\usepackage{amsmath}
\usepackage[titletoc]{appendix}
\usepackage{mathtools}
\usepackage{amsthm}
\usepackage{graphicx}
\usepackage{amssymb}
\usepackage{orcidlink}
\usepackage{bm}
\usepackage{hyperref}
\hypersetup{
  colorlinks   = true, 
  urlcolor     = blue, 
  linkcolor    = blue, 
  citecolor   = blue 
}

\begin{document}

\title{\textbf{Gravitational wave polarisations in nonminimally coupled gravity} 
}%

\author{Miguel Barroso Varela \orcidlink{0009-0006-9844-7661}}\email{Contact author: up201907272@edu.fc.up.pt}
\author{Orfeu Bertolami \orcidlink{0000-0002-7672-0560}}%
 \altaffiliation[Also at ]{Centro de Física das Universidades do Minho e do Porto, Rua do Campo Alegre s/n, 4169-007 Porto, Portugal}
\affiliation{Departamento de Física e Astronomia, Faculdade de Ciências, Universidade do Porto, Rua do Campo Alegre s/n, 4169-007 Porto, Portugal} 

\date{\today}

\begin{abstract}
The properties of metric perturbations are determined in the context of an expanding Universe governed by a modified theory of gravity with a non-minimal coupling between curvature and matter. We analyse the dynamics of the 6 components of a general helicity decomposition of the metric and stress-energy perturbations, consisting of scalar, vector and tensor sectors. The tensor polarisations are shown to still propagate luminally, in agreement with recent data from gravitational interferometry experiments, while their magnitude decays with an additional factor sourced by the nonminimal coupling. We show that the production of these modes is associated with a modified quadrupole formula at leading order. The vector perturbations still exhibit no radiative behaviour, although their temporal evolution is found to be modified, with spatial dependence remaining unaffected. We establish that the scalar perturbations can no longer be treated as identical. We investigate the scalar sector by writing the modified model as an equivalent two-field scalar-tensor theory and find the same scalar degrees of freedom as in previous literature. The different sectors are paired with the corresponding polarisation modes, which can be observationally measured by their effects on the relative motion of test particles, thus providing the possibility of testing the modified theory and constraining its parameters. 
\end{abstract}

\maketitle

\section{Introduction}
Since their prediction in the early 1900s, gravitational waves (GWs) were seen as a topic of theoretical interest with no hope of any observational detection, mostly due to their significantly small effect on most objects. This all changed with the first detection of GWs by the LIGO experiment in 2015 \cite{GWDetection1st}. Since their experimental confirmation, GWs are now taken as one of the leading opportunities for probing distant events with gravitational effects far beyond any of those that currently take place in our Solar System \cite{GWProbing,GWProbing2}. This has led to a surge in interest in their theoretical properties in General Relativity (GR) and modified theories, as current and future experiments may provide the means to measure these and thus shed light on a putative more encompassing theory of gravity \cite{GWProbing2}. \par
As established in their original prediction, GWs in GR are expected to be present in the form of two massless tensor polarisations that propagate luminally through the vacuum. However, modified gravity theories often lead to the presence of additional polarisations, as well as different propagation properties for the tensor modes \cite{MassiveMasslessGWsHigherOrderGravity}. For instance, higher-order effective field theories of gravity have been shown to predict deviations from luminal propagation of the tensor modes in the vicinity of Schwarzschild \cite{GWSchwarzschild} and Reissner-Nördstrom \cite{GWExtremal} black holes, along with additional deviations in the cosmological context \cite{SpeedOfGravity}. Many of these properties have been constrained observationally in Earth-based experiments, which have provided us with an upper bound on the mass of the graviton \cite{GravitonMassBounds}, along with lower and upper bounds on the speed of these waves \cite{GWSpeedBound}. These measurements can be used to tightly constrain parameters of modified theories or to even rule them out, which is a complex task in other contexts due to the sensible nature of gravitational effects at local scales \cite{LIGOLISAPolarisations,LIGOLISAPolarisations2}. Although measuring relatively small modifications to GW properties around exotic backgrounds such as black hole spacetimes seems unlikely in the near future, the cosmological large-scale dynamics may provide promising opportunities to analyse modifications to GR in the proximity of our planet \cite{CosmographyGWs}. \par

Besides analysing alterations to the already predicted tensor modes, the presence of additional polarisations of GWs in the context of modified gravity has been extensively researched in past literature \cite{GWModes,LIGOLISAPolarisations,LIGOLISAPolarisations2,GWPolarisationSpeeds,MassiveMasslessGWsHigherOrderGravity}. This follows from the modification of the dynamics of the scalar and vector modes one considers when decomposing the metric perturbations into their different helicity sectors \cite{GWDecompositionBook,GWPolarisations}. The investigation of the properties of these additional modes can be achieved by an explicit perturbative approach or by applying the Newman-Penrose tetrad formalism \cite{NPFormalism}, as seen, for example, in Refs. \cite{NMCGravWaves,ProbingFR_GWs}. In a closer parallel to the work presented here, it has been shown that minimally coupled $f(R)$ theories predict the existence of scalar gravitational waves \cite{FR_GWs,FRGWs,FRGWs2}, which served as motivation for deeper research into this topic.    \par

In this work, we consider the properties of metric perturbations in a cosmological expanding background governed by a modified theory of gravity with non-minimal coupling (NMC) of matter and curvature \cite{ExtraForce}. This advances the work conducted in Ref. \cite{NMCGravWaves}, where only constant curvature backgrounds were considered as an initial probe for the properties of these waves in the modified regime. This kind of NMC theory has also been extensively researched in the context of mimicking dark matter profiles \cite{NMCDarkMatter,NMCDarkMatter2}, analysing the modified theory with solar system constraints \cite{NMCSolarSystem,NMCSolarSystem2,NMCSolarSystem3}, sourcing cosmological inflation in the early Universe \cite{NMCInflation,NMCInflation2,NMCInflation3} and the creation of large-scale structure \cite{NMCCosmologicalPerturbations}. More recently, it has been shown that we may solve the open problem of the ``Hubble tension" \cite{HubbleTensionReview} by considering the late-time effects of the NMC on the model-dependent evolution of cosmic microwave background data used for indirect measurements of the Hubble parameter, while simultaneously providing an observationally compatible mechanism to source the accelerated expansion of the Universe \cite{NMCHubbleTension}. This further motivates the analysis of the propagation of NMC gravitational waves in the context of an expanding Universe, as it allows for testing of the mathematical consistency of the theory at these scales, as well as providing the means for additional observational tests of the modifications to GR \cite{CosmographyGWs,LIGOLISAPolarisations}. \par

For simplicity of the analysis carried out here, whenever considering an expanding Universe we use comoving time $\eta$, which we relate to cosmic time as $dt=a(t)d\eta$, and write the background FLRW metric in Cartesian coordinates as
\begin{equation}\label{BackgroundMetric}
    ds^2=a^2(\eta)(-d\eta^2+dx^2+dy^2+dz^2)
\end{equation}
or more concisely as $\Bar{g}_{\mu\nu}=a^2(\eta)\eta_{\mu\nu}$. We define the Hubble parameter by $H=\Dot{a}/a=a'/a^2$ as usual, with dots representing derivatives with respect to cosmic time $t$ and primes representing derivatives with respect to comoving time $\eta$. We shall mostly use the ``comoving Hubble parameter" $\mathcal{H}=a'/a=aH$ as this is useful for our analysis, although one can find instances in the literature where $H$ and $\dot H$ are used for easier interpretation in the context of the expanding Universe. Additionally, we define the full metric as 
\begin{equation}\label{MetricDefinition}
    g_{\mu\nu}=\Bar{g}_{\mu\nu}+a^2(\eta)h_{\mu\nu}=a^2(\eta)(\eta_{\mu\nu}+h_{\mu\nu}),
\end{equation}
which will simplify the form of the equations and the determination of the propagation speed of the perturbations in case they exhibit radiative behaviour. As done in this equation, we will write background quantities with a bar as $\bar \rho$ whenever the distinction is necessary, while often omitting this when a quantity is already multiplied by another that is explicitly of perturbative first-order. We use the $(-,+,+,+)$ signature and set $c=8\pi G=1$ for simplicity. \par

We have organised this paper as follows. We present the nonminimally coupled model, the relevant field equations and their linearly perturbed form in Section \ref{NMCSection}. The helicity decomposition of the metric and stress-energy tensors, along with their associated polarisations and observable effects on the relative motion of test particles, are described in Section \ref{DecompositionSection}. This is followed by the analyses of the dynamics of the perturbations in GR and in the nonminimally coupled theory, which are included in Sections \ref{GRDynamicsSection} and \ref{NMCDynamicsSection} respectively. We present an alternative method to determine the polarisation spectrum based on an equivalent scalar-tensor theory formulation of the NMC model in Section \ref{ScalarTensorSection}.
We conclude the work in Section \ref{ConclusionsSection}, where we discuss the obtained results along with possible extensions of our considerations.

\section{Nonminimally coupled model}\label{NMCSection}
\subsection{Action and field equations}
The nonminimally coupled $f(R)$ model can be written in action form as \cite{ExtraForce}
\begin{equation}
    S=\int dx^4 \sqrt{-g} \left[\frac{1}{2}f_1(R)+[1+f_2(R)]\mathcal{L}_m \right],
\end{equation}
where $f_{1,2}(R)$ are functions of the scalar curvature $R$, $g$ is the metric determinant and $\mathcal{L}_m$ is the Lagrangian density for matter fields \cite{ExtraForce}. The dynamics of General Relativity can be recovered by setting $f_1=R$ and $f_2=0$. The inclusion of a cosmological constant can be considered by choosing $f_1=R-2 \Lambda$. By varying the action with respect to the metric $g_{\mu\nu}$ we obtain the field equations \cite{ExtraForce}
\begin{equation}\label{FieldEquations}
\begin{aligned}
    (F_1+2 F_2\mathcal{L}_m)G_{\mu\nu}=&(1+f_2)T_{\mu\nu}+\Delta_{\mu\nu}(F_1+2 F_2 \mathcal{L}_m)\\
    &+\frac{1}{2}g_{\mu\nu}(f_1-F_1 R - 2 F_2 R \mathcal{L}_m),
\end{aligned}   
\end{equation}
where we have defined $\Delta_{\mu\nu}\equiv\nabla_\mu\nabla_\nu-g_{\mu\nu}\Box$ and $F_i\equiv df_i/dR$. As can be seen from the presented equations, the choice of the Lagrangian density is non-trivial, as it directly affects the resulting dynamics via the nonminimal coupling, while only appearing in the form of the related stress-energy tensor in General Relativity \cite{LagrangianChoice,LagrangianForm}. With this in mind, throughout this work, we follow the arguments discussed in Refs. \cite{LagrangianForm,LagrangianChoice2} and take the Lagrangian density to be $\mathcal{L}_m=-\rho$, with $\rho$ denoting the energy density.  \par
Applying the Bianchi identities $\nabla_\mu G^{\mu\nu}=0$ to Eq. (\ref{FieldEquations}) leads to the non-conservation law \cite{ExtraForce}
\begin{equation}\label{NonConservationEq}
    \nabla_\mu T^{\mu\nu}=\frac{F_2}{1+f_2}\left(g^{\mu\nu}\mathcal{L}_m-T^{\mu\nu}\right)\nabla_\mu R,
\end{equation}
which reduces to the usual stress-energy tensor conservation equation when we set $f_2=0$. This modification of the conservation equation follows directly from the non-minimal coupling of matter and curvature, as seen by its direct dependence on $f_2$ and its independence from the minimally coupled part of the theory described by $f_1$. In this work, as we focus on the effects of the NMC model independently of the minimally coupled $f(R)$ model, we set $f_1=R$ and single out the remaining effects with $f_2\neq0$. \par
Although the form of $f_2$ will be kept abstract throughout the discussion presented in this work, a useful choice is to consider a general power-law expansion in $R$. Indeed, writing this as 
\begin{equation}
    1+f_2(R)=\sum_{n=-\infty}^{\infty}\left(\frac{R}{R_n}\right)^n,
\end{equation}
where $R_n$ can be thought of as setting the scale for which the effects of each term in the series become considerable, we can capture the behaviour of the NMC theory at all orders of the curvature, with positive exponent terms dominating in the early Universe and negative exponent terms coming into play at late-times \cite{NMCHubbleTension}. This means that we expect the evolution of the perturbations considered in this work to be highly epoch-dependent and for the modifications from GR to be strongest at significantly early and late times of the Universe's expansion. 

\subsection{Background cosmology in NMC model}

As the behaviour of perturbations will be highly dependent on the background dynamics, it is important to review how the expanding nature of the Universe is modified in the NMC theory. This analysis was originally conducted in Ref. \cite{NMCFriedmann}, with additional results presented in Refs. \cite{NMCAcceleratedExpansion,NMCHubbleTension}. By considering the background metric in comoving coordinates as presented in Eq. (\ref{BackgroundMetric}), we can use the previously discussed field Eqs. (\ref{FieldEquations}) to arrive at the modified Friedmann equation
\begin{equation}\label{ModifiedFriedmann}
    \mathcal{H}^2=\frac{1}{6F}\left[2(1+f_2)a^2\rho-6\mathcal{H}F'-a^2f_1+a^2F R\right].
\end{equation}
The spatial components of the field equations yield the modified Raychaudhuri equation
\begin{equation}\label{ModifiedRaychaudhuri}
\begin{aligned}
\mathcal{H}^2+2\mathcal{H}'=-\frac{1}{2F}&\left[2F''+3\mathcal{H}F'+a^2f_1\right.\\
&\left.-a^2F R+2(1+f_2)a^2 p\right]
\end{aligned}
\end{equation}
and the modified conservation equation of the perfect fluid stress-energy tensor $T^\mu_\nu=\text{diag}(-\rho,p,p,p)$, with $\rho$ and $p$ representing the energy density and pressure respectively, leads to the usual result
\begin{equation}\label{ConservationEq}
    \rho' +3\mathcal{H}(\rho+p)=0,
\end{equation}
where the choice of $\mathcal{L}_m=-\rho$ causes the modifications in Eq. (\ref{NonConservationEq}) to vanish. This greatly simplifies the background cosmological evolution, as we can consider the usual dynamics of the energy density of all kinds of matter with respect to the scale factor $\rho\propto a^{-3(1+\omega)}$, which depends only on the equation of state parameter $\omega=p/\rho$ ($\omega=0$ for non-relativistic matter, $\omega=1/3$ for radiation). \par
At the background level, we thus expect deviations from the standard $\Lambda$CDM model, which is already enough to induce modifications on the propagation of GWs, whose properties are dependent on the dynamics of the expansive nature of the Universe. Based on the research conducted in Refs. \cite{NMCHubbleTension,NMCDarkMatter,NMCAcceleratedExpansion}, we know that observational data from galaxy rotation curves and direct measurements of the Hubble and deceleration parameters can be explained by the NMC theory with the inclusion of negative exponent terms in the power series form of $f_2$ and without need for the inclusion of a cosmological constant. This means that to a reasonable approximation, one can take the background to evolve according to simpler observationally motivated models such as $\Lambda$CDM, especially if the numerical evolution of the perturbations proves too computationally expensive to perform in parallel to the simulation of the FLRW background dynamics in the NMC theory.

\subsection{Perturbed NMC equations}
Considering the general perturbations to the FLRW background metric, for which we maintain the convention introduced at the start of this paper, and the matter content, the NMC-modified field Eqs. (\ref{FieldEquations}) yield 
\begin{widetext}
    \begin{equation}\label{NMCLinearisedFieldEq}
\begin{aligned}
& \left(F_{1,R} \delta R+2  F_{2,R} \mathcal{L}_m \delta R+2  F_2 \delta \mathcal{L}_m\right) R_{\mu \nu}+\left(F_1+2  F_2 \mathcal{L}_m\right) \delta R_{\mu \nu} \\
& \quad-\frac{1}{2} g_{\mu \nu} F_1 \delta R-\frac{1}{2} a^2 h_{\mu \nu} f_1-\left[\delta(\nabla_\mu \nabla_\nu)-a^2 h_{\mu \nu} \square-g_{\mu\nu}\delta(\Box)\right]\left(F_1+2  F_2 \mathcal{L}_m\right) \\
& \quad-\left[\nabla_\mu \nabla_\nu-g_{\mu \nu} \square\right]\left(F_{1,R} \delta R+2  F_{2,R} \mathcal{L}_m \delta R+2  F_2 \delta \mathcal{L}_m\right) \\
& \quad=(1+ f_2) \delta T_{\mu \nu}+ F_2 T_{\mu \nu} \delta R,
\end{aligned}
\end{equation}
\end{widetext}
where we have written $F_{i,R}=dF_i/dR$, with the notation $F_i'=F_{i,R}R'$ still referring to differentiation with respect to $\eta$. Given our previous choice of $f_1=R$, from now on we shall set $F_1=1$. The linear perturbation of the covariant derivatives is due to the perturbation of the Christoffel symbols
\begin{equation}
    \delta\Gamma^{\rho}_{\mu\nu}=\frac{1}{2}\bar{g}^{\rho\gamma}\left[\nabla_\mu \left(a^2 h_{\gamma\nu}\right)+\nabla_\nu \left(a^2 h_{\gamma\mu}\right)-\nabla_\gamma \left(a^2 h_{\mu\nu}\right)\right]
\end{equation}
and so
\begin{equation}
    \delta(\nabla_\mu\nabla_\nu)\varphi=\delta \Gamma^\rho_{\mu\nu} \partial_\rho \varphi
\end{equation}
for any scalar quantity $\varphi$. For reasons that will become clear later, it is convenient to write the perturbed field equations in the form 
\begin{equation}\label{NMCLinearisedFieldEqsDeltaF}
\begin{aligned}
    &F\delta R^\mu_\nu +R_\mu^\nu \delta F-\frac{1}{2}\delta^\mu_\nu\delta R+\left[\delta^\mu_\nu\Box-\nabla^\mu_\nu\right]\delta F\\
    &+\left[\delta^\mu_\nu\delta(\Box)-\delta(\nabla^\mu\nabla_\nu)\right]F=(1+f_2)\delta^\mu_\nu+ F_2 T^\mu_\nu\delta R,
\end{aligned}
\end{equation}
where $F=F_1+2 F_2 \mathcal{L}_m=1+2 F_2 \mathcal{L}_m$ and so $\delta F=2(\mathcal{L}_m F_{2,R}\delta R+F_2 \delta\mathcal{L}_m)$. Additionally, this shows how the explicit dependence on the Lagrangian density is particularly relevant in the context of perturbation theory, as the matter content itself is perturbed, thus leading to a more complex interplay between the evolution of the small changes in the metric and the stress-energy tensor.

\section{Helicity decomposition of perturbations}\label{DecompositionSection}
\subsection{Metric perturbations}
We aim here to analyse general perturbations to an FLRW background metric. The spatial homogeneity of the system means that all background quantities will be functions of comoving time $\eta$ only. This is especially important for $R=R(\eta)$, as the background modified theory functions $f_i$ and their derivatives will be evaluated at this curvature. Most of the assumptions considered in Ref. \cite{NMCGravWaves} are no longer valid, such as $\partial_\mu F_i=\partial_\mu \mathcal{L}_m=0$, as these are now evolving quantities in spacetime. This means that the direct analytic conclusions drawn in that work are no longer possible to reach directly, forcing us to explicitly decompose the gravitational waves into 6 general modes and analyse them individually \cite{GWPolarisations}. Nevertheless, when considering the evolution of the perturbations, we may choose to take the sub-horizon (or high-frequency) approximation, in which the background dynamics may be taken to be static in comparison to the rapidly-varying metric perturbations. We write the general helicity decomposition of $h_{\mu\nu}$ as \cite{GWPolarisations}
\begin{widetext}
    \begin{equation}
\begin{aligned}
    h_{\eta\eta}=&2\phi \\
    h_{\eta i}=&\beta_i+\partial_i\gamma\\
    h_{ij}=&\frac{1}{a}h_{ij}^{TT}+\frac{1}{3}\Theta\delta_{ij}+\partial_{i}\varepsilon_{j}+\partial_{j}\varepsilon_{i}
    +\left(\partial_i\partial_j-\frac{1}{3}\delta_{ij}\nabla^2\right)\lambda,
\end{aligned}
\end{equation}
\end{widetext}
where we assume that $h_{\mu\nu}\rightarrow0$ as we move out to spatial infinity, as one would not expect to feel gravitational effects from an infinitely distant event. We see that we have 6 components from the symmetric $h_{ij}^{TT}$ tensor mode, 3 components for each of the vectors $\varepsilon_i$ and $\beta_i$, and 4 scalars ${\phi,\gamma,\Theta,\lambda}$. This gives a total of 16 modes. Due to the invariance of $a(\eta)$ under spatial rotations, the 4 scalars indeed have helicity 0, the vectors have helicity $\pm1$ and the tensor has helicity $\pm2$ \cite{GWDecompositionBook}. Additionally, we impose the constraints
\begin{equation}
\begin{aligned}
    &\partial_i\beta_i=0\\
    &\partial_i\varepsilon_i=0\\
    &\partial_i h_{ij}^{TT}=0\\
    &\delta^{ij} h_{ij}^{TT}=0,
\end{aligned}
\end{equation}
which give 1/1/3/1 individual equations respectively, totalling 16-6=10 independent variables, consistent with a 4$\times$4 symmetric tensor. Note also that we have written spatial indices in equations like $\partial_i\varepsilon_i=0$ with no indication of being raised or lowered, as these are interpreted as ``helicity" indices \cite{GWDecompositionBook}. These defined quantities are gauge-dependent, allowing for further simplification \cite{GWDecompositionBook}. We thus define the gauge-invariant quantities
\begin{equation}
\begin{aligned}
    &\Phi\equiv-\phi+\frac{1}{a}\partial_\eta\left[a\left(\gamma-\frac{1}{2}\partial_\eta \lambda\right)\right] \\
    &\Psi\equiv\frac{1}{6}\left[-\Theta-\nabla^2\lambda+\mathcal{H}\left(\gamma-\frac{1}{2}\partial_\eta \lambda\right)\right]\\
    &\Xi_i\equiv\beta_i-\partial_\eta{\varepsilon}_i,
    \end{aligned}
\end{equation}
which yields 6 gauge-invariant functions: 1 for $\Phi$, 1 for $\Psi$, 3 for $\Xi_i$ and 6 for $h_{ij}^{TT}$, minus 3 for $\partial_i h_{ij}^{TT}=0$, minus 1 for $\delta^{ij} h_{ij}^{TT}=0$ and minus 1 for $\partial_i \Xi_i=0$. We thus have 6 physical and 4 gauge degrees of freedom \cite{GWPolarisations}. The physical quantities are associated with different modes in the relative motion between two test particles \cite{GWModes} and so introduce detectable modifications to the theory. The remaining gauge freedom can be used to choose the conformal Newtonian gauge \cite{GWDecompositionBook}, for which 
\begin{equation}
    \lambda=\gamma=\beta_i=0
\end{equation}
and so the gauge-invariant quantities can be written as
\begin{equation}
\begin{aligned}
    &\Phi\equiv-\phi \\
    &\Psi\equiv-\frac{1}{6}\Theta\\
    &\Xi_i\equiv-\partial_\eta{\varepsilon}_i,
    \end{aligned}
\end{equation}
which allows the line element to be written as
\begin{equation}\label{LineElement}
\begin{aligned}
    ds^2=a^2(\eta)&\bigg[\left.-(1+2\Phi)d\eta^2
    +\{(1-2\Psi)\delta_{ij}
    +\partial_i\varepsilon_j\right.\\
    +&\left.\partial_j \varepsilon_i
    +\frac{1}{a}h_{ij}^{TT}\}dx^idx^j\right].
\end{aligned}
\end{equation}
The scalar sector resembles the Newtonian limit of General Relativity, aiding in the interpretation of the effects of the scalar perturbations $\Psi$ and $\Phi$ \cite{GWDecompositionBook}.

\subsection{Stress-energy tensor perturbations}
Apart from decomposing the metric perturbations, we may also decompose the stress-energy tensor perturbations into their different helicity components \cite{GWDecompositionBook}. It is convenient to use $\delta T^\mu_\nu$, whose components can be written as
\begin{equation}\label{StressEnergyDecomposition}
    \begin{aligned}
        & \delta T^\eta_\eta=-\delta \rho \\
        & \delta T^i_\eta= S^i+\partial^i S\\
        & \delta T^i_j=\delta p \delta^i_j+\Sigma^i_j,
    \end{aligned}
\end{equation}
where we have defined the anisotropic stress tensor $\Sigma^i_j$. This can in turn be decomposed into scalar, vector and tensor parts as 
\begin{equation}\label{AnisotropicTensor}
    \Sigma_{ij}=\left(\partial_i\partial_j-\frac{1}{3}\delta_{ij}\nabla^2\right)\sigma+(\partial_i\sigma_j+\partial_j\sigma_i)+\sigma_{ij}^{TT},
\end{equation}
with $\sigma_{ij}^{TT}$ being traceless $\partial_i\sigma_{ij}^{TT}=0$ and transverse in the sense that $\delta^{ij}\sigma_{ij}^{TT}=0$. Specifically, in the case of a perfect fluid, such as the one considered in the context of the FLRW metric, the perturbations to the stress-energy tensor can be written as \cite{GWDecompositionBook}
\begin{equation}
    \begin{aligned}
        & \delta T^\eta_\eta=-\delta \rho \\
        & \delta T^i_\eta= -(\rho+ p)v^i\\
        & \delta T^i_j=\delta p \delta^i_j=c_s^2\delta\rho\delta^i_j,
    \end{aligned}
\end{equation}
where we have used the definition of the speed of sound in a perfect fluid $\delta p=c_s^2\delta\rho$ ($c_s^2=0$ for non-relativistic matter and $c_s^2=1/3$ for radiation). Additionally, we have defined the peculiar velocity of particles due to the perturbations, which we can further decompose into its transverse and longitudinal parts as 
\begin{equation}
    v^i=V^i+\partial^i v,
\end{equation}
where $V$ is transverse ($\partial_i V^i=0$). Comparing with Eq. (\ref{StressEnergyDecomposition}), we identify $S=-(\rho+ p)v$, $S^i=-( \rho+ p)V^i$ and $\Sigma^i_j=0$. This means that the tensor part of the stress-energy perturbations vanishes for a perfect fluid. \par
Due to $T^\mu_\nu$ also being a tensor, we expect it to transform under the same gauge transformations as discussed in the previous section. This motivates us to define gauge-invariant stress-energy perturbations in terms of the already defined quantities. These are \cite{GWDecompositionBook} 
\begin{equation}
\begin{aligned}
    &\delta\rho_*=\delta\rho-3\mathcal{H}( \rho+ p)(v+\gamma)\\
    &v_*=v+\frac{1}{2}\lambda'\\
    &V_*^i=V^i+\beta^i
    ,
    \end{aligned}
\end{equation}
which in our gauge ($\lambda=\gamma=\beta_i=0$) simply give
\begin{equation}
    \delta\rho_*=\delta\rho-3\mathcal{H}( \rho+ p)v
\end{equation}
with $v$ and $V^i$ being gauge-invariant.

\subsection{Effects of polarisations on test particles}
\begin{figure}
    \centering
    \includegraphics[width=\linewidth]{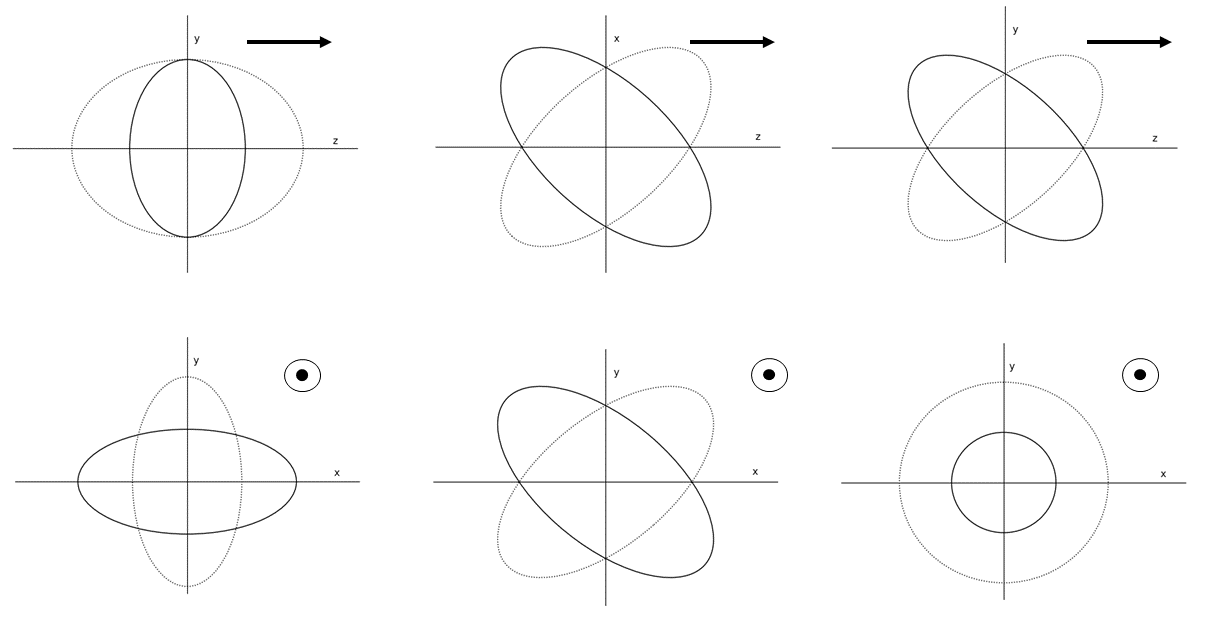}
    \caption{The six polarisation modes of gravitational waves and their effects on a ring of test particles. In the top row, we show the longitudinal, vector-x and vector-y modes respectively. In the bottom row, we show the $+$, $\times$ and breathing modes. The direction of propagation is shown as an arrow pointing left to right in the top plots and out of the page in the bottom plots. We show the configuration of test particles at the two extremes of the periodic oscillation. }
    \label{PolarisationFigure}
\end{figure}

The presence of GWs can be inferred from the relative ``stretching/squeezing" of intervals in spacetime. Specifically, this has been used in the LIGO experiment to obtain the first concrete detection of gravitational waves passing through our planet \cite{GWDetection1st}. This observation focused on the more relevant tensorial modes predicted by perturbations to vacuum backgrounds in GR. However, all polarisations discussed here present their respective effects on spacetime intervals and can be associated with different types of effects. To see this, we can analyse perturbations to the geodesic deviation equation due to the metric fluctuations \cite{WaldGR}
\begin{equation}
    \frac{D^2X^\mu}{d\tau^2}\equiv 
    u^{\alpha}\nabla_{\alpha}(u^{\beta}\nabla_{\beta}X^{\mu})=R^{\mu}_{\ \alpha\beta\sigma}{}^{(1)}u^{\alpha}u^{\beta}X^{\sigma},
\end{equation}
where $u^\alpha$ is the four-velocity of the observer, $X^\mu$ is the displacement of two objects travelling along infinitesimally separated geodesics and the superscript ``$(1)$" refers to taking only first-order contributions to the Riemann tensor. By assuming a non-relativistic observer, we can set $u^\alpha\approx(a^{-1},0,0,0)$ and thus obtain 
\begin{equation}
   \frac{\partial^2X^i}{\partial\eta^2}=-a^{-2}R^i_{\ \eta j\eta}{}^{(1)}X^{j},
\end{equation}
where we have used comoving time $\eta$ to obtain a simpler relation with our previous calculations. Here we have also taken measurements to be made in the local inertial frame (LIF) of the observer, thus simplifying the left-hand side of the equation to a standard partial derivative \cite{WaldGR}. Comoving time is particularly useful for this, as our definition of the full metric in Eq. (\ref{MetricDefinition}) allowed for perturbations to be directly applied to a ``Minkowski" background, as one would observe in its LIF. This allows us to determine the effects of metric perturbations on slowly moving test particles with the use of the perturbed components of $R^i_{\ \eta j\eta}$. Aligning the propagation of any waves with the z-axis, we can separate the Riemann tensor into different modes as \cite{GWModes,GWPolarisationOld}
\begin{equation}
    R^i_{\ \eta j\eta}{}^{(1)}=\begin{pmatrix}
P_4+P_6 & P_5 & P_2\\
P_5 & -P_4+P_6 & P_3 \\
P_2 & P_3 & P_1
\end{pmatrix}_{ij},
\end{equation}
where we have labelled the 6 independent polarisation modes as $P_i$. In order of increasing $i$, these describe the longitudinal, vector-x, vector-y, $+$, $\times$ and breathing modes \cite{GWModes}, which are shown in Figure \ref{PolarisationFigure}. By calculating the respective linearised Riemann tensor components for the chosen gauge-invariant metric perturbations, we obtain 
\begin{equation}
\begin{array}{l}
P_1=\partial_z^2\Phi+\Psi''+\mathcal{H}(\Psi'+\Phi') \\ P_2=\frac{1}{2}\left(\mathcal{H}\partial_z\Xi_1+\partial_z\Xi_1'\right) \\ P_3=\frac{1}{2}\left(\mathcal{H}\partial_z\Xi_2+\partial_z\Xi_2'\right) \\ P_4=\frac{1}{2a}\left(\mathcal{H}'h_++\mathcal{H}h_+'-h_+''\right) \\ P_5=\frac{1}{2a}\left(\mathcal{H}'h_\times+\mathcal{H}h_\times'-h_\times''\right) \\ P_6=\mathcal{H}(\Psi'+\Phi')+\Psi'',
\end{array}
\end{equation}
where the expanding nature of the Universe is present through the comoving Hubble parameter $\mathcal{H}$ and its derivative. The association of the different helicity perturbations with the different polarisation modes serves as yet another reason why the chosen decomposition is useful for observational testing. By considering waves varying much faster than the expansion timescale, we obtain the simpler form
\begin{equation}
\begin{array}{ll}
    P_1=\partial_z^2\Phi+\Psi''&
    P_2=\frac{1}{2}\partial_z\Xi_1'\\
    P_3=\frac{1}{2}\partial_z\Xi_2'&
        P_4=-\frac{1}{2a}h_+''\\
    P_5=-\frac{1}{2a}h_\times''&
    P_6=\Psi'',
    \end{array}
\end{equation}
which provides a clear connection between the investigated metric perturbations and the relative motion of test particles. Any detection (or non-detection) of modified behaviour following from the analysis of the NMC theory could provide means to further restrict the theory's parameters or even test for its presence \cite{LIGOLISAPolarisations,LIGOLISAPolarisations2,LISAPolarisations,FR_GWsLISA}.

\section{Perturbation dynamics in GR}\label{GRDynamicsSection}

\subsection{Stress-energy tensor perturbations}

In General Relativity, the stress-energy perturbations are constrained by the theory's conservation equation
\begin{equation}
    \nabla_\mu T^\mu_\nu=0,
\end{equation}
thus providing us with additional constraints on their behaviour. Considering the metric in the conformal Newtonian gauge as in Eq. (\ref{LineElement}), the conservation equation can be separated into temporal and spatial components. The $\nu=0$ component implies that
\begin{equation}\label{SEConservationEta}
    \delta\rho'+3\mathcal{H}(1+c_s^2)\delta\rho-3(\rho+ p)\Psi'+(\rho+ p)\nabla^2 v=0,
\end{equation}
which only depends on scalar quantities. By defining the relative density perturbations as $\hat{\delta}=\delta\rho/\Bar{\rho}$, with the bar still denoting background quantities, and using the background perfect fluid relation $\Bar p=\omega \Bar \rho$, along with the FLRW conservation equation, this simplifies to 
\begin{equation}
    \hat{\delta}'+3\mathcal{H}(c_s^2-\omega)\hat{\delta}-3(1+\omega)\Psi'+(1+\omega)\nabla^2 v=0,
\end{equation}
where for a perfect fluid $c_s^2=\omega$ and so 
\begin{equation}
    \hat{\delta}'=(1+\omega)(3\Psi'-\nabla^2 v).
\end{equation}
The $\nu=i$ equations can be generally written as
\begin{equation}
    4\mathcal{H}(\rho+ p)v_i+\left((\rho+ p)v_i\right)'+\partial_i\left(c_s^2\delta\rho+(\rho+ p)\Phi\right)=0,
\end{equation}
which we can further decompose using $v_i=V_i+\partial_i v$. This now becomes 
\begin{equation}
\begin{aligned}
    &4\mathcal{H}(\rho+ p)V_i+\left((\rho+ p)V_i\right)'+\partial_i\left[c_s^2\delta\rho+(\rho+ p)\Phi\right.\\
    &\left.+4\mathcal{H}(\rho+ p)v+\left((\rho+ p)v\right)'\right]=0,
\end{aligned}
\end{equation}
where we now notice the helicity decomposition of the equation allows us to set both 
\begin{equation}
4\mathcal{H}(\rho+ p)V_i+\left((\rho+ p)V_i\right)'=0
\end{equation}
and 
\begin{equation}
    c_s^2\delta\rho+(\rho+ p)\Phi+4\mathcal{H}(\rho+ p)v+\left((\rho+ p)v\right)'=0.
\end{equation}
Here we have used the condition that all perturbations go to 0 at infinity when assuming that the constant term within the spatial derivative must be 0 and not any other constant. The vector equation indicates that we may separate the temporal and spatial evolutions of $V_i$ as 
\begin{equation}
    V_i=\frac{V_i^{(s)}(\Vec{x})}{(\rho+ p)a^4},
\end{equation}
where we have denoted the spatial part of the perturbation with a superscript $(s)$. However, the scalar equation shows how the same separation can not be assumed for $v$, due to the coupling with the density and metric perturbations.

\subsection{Scalar perturbations}
The scalar sector describes the effects of the $\Psi$ and $\Phi$ perturbations. The non-vanishing components of the perturbed Einstein tensor are 
\begin{equation}
    \delta G^\eta_\eta=\frac{6}{a^2}\mathcal{H}^2\Phi+\frac{2}{a^2}(3\mathcal{H}\Psi'-\nabla^2\Psi)
\end{equation}
\begin{equation}
    \delta G^i_\eta=\frac{2}{a^2}(\mathcal{H}\partial_i\Phi+\partial_i\Psi')
\end{equation}
\begin{equation}
\begin{aligned}
    \delta G^i_j=&\frac{1}{a^2}\left[\delta^i_j\left(2(\mathcal{H}^2+2\mathcal{H}' )\Phi +2\mathcal{H}(\Phi'+2\Psi')+2\Psi''\right)\right.\\
    &+\left.\left(\delta^i_j\nabla^2-\partial_i\partial_j\right)(\Phi-\Psi)\right]
\end{aligned}
\end{equation}
with the respective scalar stress-energy perturbations 
\begin{equation}
    \begin{aligned}
        & \delta T^\eta_\eta=-\delta \rho \\
        & \delta T^i_\eta= -(\rho+ p)\partial_i v\\
        & \delta T^i_j=\delta p \delta^i_j=c_s^2\delta\rho\delta^i_j.
    \end{aligned}
\end{equation}
The $(i,j)$ equations can be split into a $\delta^i_j$ term and a $\partial_i\partial_j$ term. These two kinds of terms behave as different helicity components and can thus be separated into independent equations \cite{GWDecompositionBook}:
\begin{equation}
    2(\mathcal{H}^2+2\mathcal{H}' )\Phi +2\mathcal{H}(\Phi'+2\Psi')+2\Psi''=a^2 c_s^2\delta\rho
\end{equation}
and 
\begin{equation}
    \Phi-\Psi=0\Rightarrow\Phi=\Psi,
\end{equation}
respectively, where we see that the two scalar perturbations are exactly equal due to the chosen convention in their initial definition in terms of the original perturbation variables. This equality is a direct consequence of the vanishing of $\sigma$ in the anisotropic stress tensor $\Sigma^i_j$ in Eq. (\ref{AnisotropicTensor}) when assuming a perfect fluid \cite{GWDecompositionBook}. \par
We can now analyse the $(i,\eta)$ components of the perturbed field equations with $\Phi=\Psi$. These give
\begin{equation}
\begin{aligned}
    \mathcal{H}\partial_i\Phi+\partial_i\Phi'&=-\frac{a^2}{2}( \rho+ p)\partial_i v \\
    &\Rightarrow \mathcal{H}\Phi+\Phi'=-\frac{a^2}{2}( \rho+ p) v,
\end{aligned}
\end{equation}
which we can then apply to the $(\eta,\eta)$ equation
\begin{equation}
\begin{aligned}
    &\frac{2}{a^2}\left(3\mathcal{H}^2\Phi+3\mathcal{H}\Phi'-\nabla^2\Phi\right)\\
    &=\frac{6}{a^2}\mathcal{H}\left(\mathcal{H}\Phi+\Phi'\right)-\frac{2}{a^2}\nabla^2\Phi\\
    &=-3\mathcal{H}(\rho+ p)v-\frac{2}{a^2}\nabla^2\Phi=-\delta\rho
\end{aligned}
\end{equation}
or in the more illuminating form
\begin{equation}\label{PerturbedPoisson}
    \nabla^2\Phi=\frac{a^2}{2}\left(\delta\rho-3\mathcal{H}(\rho+ p)v\right)=\frac{a^2}{2}\delta\rho_*.
\end{equation}
This equation is written fully in terms of gauge-invariant quantities and so is valid in any gauge. It is analogous to the classical Newtonian potential equation $\nabla^2\Phi=4\pi G \rho$ \cite{GWDecompositionBook}, here written as $\rho/2$ due to the chosen convention $8\pi G=1$ .\par
Considering the $\delta^i_j$ component of the $(i,j)$ equation together with the relation $\Phi=\Psi$, we get
\begin{equation}\label{PhiTimeEq}
2(\mathcal{H}^2+2\mathcal{H}')\Phi+6\mathcal{H}\Phi'+2\Phi''=a^2c_s^2\delta\rho,    
\end{equation}
which we combine with the $(i,\eta)$ equation to obtain a gauge-invariant form 
\begin{equation}
\begin{aligned}
&\Phi''+3(1+c_s^2)\mathcal{H}\Phi'+\left(2\mathcal{H}'+(1+3c_s^2)\mathcal{H}^2\right)\Phi\\
&= \frac{a^2}{2}c_s^2\left(\delta\rho-3\mathcal{H}(\rho+ p)v\right)=\frac{a^2}{2}c_s^2\delta\rho_* .
\end{aligned}
\end{equation}
By observing that the right-hand side of this equation matches that of Eq. (\ref{PerturbedPoisson}) with an additional factor of $c_s^2$, we obtain the full independent ``master" equation for $\Phi$
\begin{equation}\label{ScalarWaveEq}
    \Phi''-c_s^2\nabla^2\Phi+3(1+c_s^2)\mathcal{H}\Phi'+\left(2\mathcal{H}'+(1+3c_s^2)\mathcal{H}^2\right)\Phi=0,
\end{equation}
where we see that $\Phi$ obeys a wave-like equation with propagation speed $c_s^2$. The $\Phi'$ term is a friction term that indicates that the wave has the form $(\Phi a^{3(1+c_s^2)/2})$ instead of simply $\Phi$. The physical perturbation $\Phi$ thus propagates as a wave for $c_s^2\neq0$, but also decays with the expansion of the Universe as
\begin{equation}
    \Phi=\Psi\propto a^{-\frac{3}{2}(1+c_s^2)}\sim\sqrt{\Bar{\rho}} \ \ \ (c_s^2\neq0).
\end{equation}
During the radiation-dominated epoch, we see that the perturbations oscillate while decaying at the same rate as the square root of the background matter density, to which they are inherently connected. After solving for the evolution of $\Phi$ we may then determine the behaviour of $\delta\rho$ and $v$ from the previous field and conservation equations \cite{GWDecompositionBook}. Even though we have found a wave-like equation for the scalar perturbations, this radiative behaviour is only present for $c_s^2\neq0$, meaning that in a matter-dominated epoch, such as the recent past of our Universe, this wave-like evolution of scalar perturbations would not be present, with these instead having a time evolution given by the decoupled Eq. (\ref{PhiTimeEq}) with $c_s^2\delta\rho=0$. This equation further simplifies during matter domination, as $\mathcal{H}^2+2\mathcal{H}'=0$ and $c_s^2=0$ during this epoch, leading to the temporal equation for the scalar perturbation $\Phi''+3\mathcal{H}\Phi'=0$, which has a solution that is constant in comoving time and one obeying $\Phi\propto\eta^{-5}\propto a^{-5/2}$ \cite{GWDecompositionBook}. \par
Additionally, by combining the perturbed field and conservation equations for the scalar sector, we may determine the evolution of the matter perturbations $\delta\rho$ in more detail. This is particularly simple in the sub-Hubble limit, with the time scale of perturbations being much smaller than that of the expanding Universe. In this limit, combining the conservation of the stress-energy tensor with the $(\eta,\eta)$ component of the field equations gives 
\begin{equation}
    \delta\rho''-c_s^2\nabla^2\delta\rho=0,
\end{equation}
which shows how the matter fluctuations also evolve with wave-like behaviour, as one would expect from their intrinsic relation to the scalar metric perturbations. The propagation speed is given by the speed of sound in the corresponding type of perfect fluid, meaning that these matter perturbations propagate at the same speed as the scalar potential $\Phi$ (or equivalently $\Psi$), while also not exhibiting wave-like properties in a matter-dominated background ($c_s^2=0$). \par
Although we find the scalar perturbations obey wave-like equations in the presence of matter, the same is no longer true in vacuum ($\rho=\delta\rho=0$). In this case, the metric perturbations have a spatial profile given by the Laplace equation $\nabla^2\Phi=0$ and time behaviour given by the unsourced version of Eq. (\ref{PhiTimeEq}). This is expected, as GR only predicts the existence of 2 radiative degrees of freedom \cite{GWPolarisations}, which will be discussed in the context of the tensor sector below.

\subsection{Vector perturbations}
The vector sector consists of contributions from $\epsilon_i$ and the associated $\Xi_i=-\epsilon_i'$. The non-zero components of the perturbed Einstein tensor are 
\begin{equation}
    \delta G^\eta_i=\frac{1}{2a^2}\nabla^2\Xi_i
\end{equation}
\begin{equation}
    \delta G^i_j=-\frac{1}{2a^2}\left[2\mathcal{H}(\partial_i\Xi_j+\partial_j\Xi_i)+(\partial_i\Xi_j+\partial_j\Xi_i)'\right]
\end{equation}
with the corresponding vector stress-energy perturbations
\begin{equation}
    \begin{aligned}
        & \delta T^\eta_\eta=0 \\
        & \delta T^\eta_i= (\rho+p)V^i\\
        & \delta T^i_j=0,
    \end{aligned}
\end{equation}
where the transverse $V^i$ obeys $\partial_i V^i=0$. The $(i,j)$ components with $i=j$ then give
\begin{equation}
\begin{aligned}
&2\mathcal{H}\partial_i\Xi_j+\partial_i\Xi_j'=0\\
&\Rightarrow \Xi_j'=-2\mathcal{H}\Xi_j\Rightarrow \Xi_j=R_j(\Vec{x})a^{-2}
\end{aligned}
\end{equation}
while the $(\eta,i)$ components imply
\begin{equation}
\begin{aligned}
    \nabla^2\Xi_i&=2a^2(\rho+ p)V_i \\
    &\Rightarrow \nabla^2 R_j=2a^4(\rho+ p)V_i=2V_i^{(s)},
\end{aligned}
\end{equation}
which has the form of a sourced Poisson equation for the vector perturbations $\Xi_i$. It thus becomes clear that these perturbations do not exhibit radiative behaviour in an FLRW background in GR, i.e. they do not behave as propagating waveforms. We also observe that the $\eta$-dependence is removed for both first-order quantities in the equation, again showing how we can separate their respective temporal and spatial dependencies \cite{GWDecompositionBook}.

\subsection{Tensor perturbations}
The tensor sector of the field equations is made up of the traceless-transverse $h_{ij}^{TT}$, which we have normalised as $h_{ij}=h_{ij}^{TT}/a$ for simplicity of the final equations. These have the same form as in the standard GR derivation \cite{WaldGR}, with $h_{xx}^{TT}=-h_{yy}^{TT}=h_{+}$ and $h_{yx}^{TT}=h_{xy}^{TT}=h_{\times}$. The non-zero Einstein tensor perturbations are
\begin{equation}
    \delta G^x_x=-\delta G^y_y=-\frac{1}{2a^2}\left[(\mathcal{H}^2+\mathcal{H}')h_+ +\Box_\eta h_+\right]
\end{equation}

\begin{equation}
    \delta G^x_y=\delta G^y_x=-\frac{1}{2a^2}\left[(\mathcal{H}^2+\mathcal{H}')h_\times +\Box_\eta h_\times \right],
\end{equation}
where we have defined $\Box_\eta=-\partial^2_\eta+\nabla^2$. As discussed previously, for a perfect fluid there are no anisotropic perturbations and so the tensor sector has $\delta T^\mu_\nu=0$, leading to the final equations
\begin{equation}
\begin{aligned}
    \Box_\eta h_{\times/+}=-(\mathcal{H}^2+\mathcal{H}')h_{\times/+}&=-\frac{a^2R}{6} h_{\times/+}\\
    &=-\frac{a''}{a}h_{\times/+},
\end{aligned}
\end{equation}
which represent a wave-like propagation of the tensor modes with luminal speed $c_{gw}^2=1$. Note that we can think of the right-hand side of the wave equations as an ``effective mass" term with $m^2_{gw}=-(\mathcal{H}^2+\mathcal{H}')$. The sign of this quantity depends on the Ricci scalar $R$. This curvature is 0 during a radiation-dominated epoch ($a(\eta)\propto \eta$) and $>0$ in a matter-dominated one ($a(\eta)\propto \eta^2$). Thus it seems that in these stages of the evolution of the Universe we could have a negative graviton mass. However, as discussed in Ref. \cite{SpeedOfGravity}, the presence of such a term merely indicates an effective mass, with a vanishing actual mass of the graviton. This distinction is also relevant when considering the causal structure of these waves, where one can typically take the sub-horizon approximation $\partial_z^2 h\sim k^2 h>>(a''/a) h$ and retrieve the same luminal speed discussed above \cite{SpeedOfGravity}.

\section{Perturbation dynamics in nonminimally coupling model}\label{NMCDynamicsSection}

For simplicity, when analysing GWs in the modified theory we assume all perturbations to be only functions of $\eta$ and $z$, as that should not affect any possible radiative behaviour, which can always be aligned with the $z$-axis without loss of generality. With this in mind, any second-order spatial derivative terms in scalar quantities like $\delta R$ choose no preferred direction and thus one can replace $\partial_z^2$ for $\nabla^2$ when extending to more general coordinate dependence. For each sector, we find that the corresponding equations are analogous to their GR counterparts, allowing for a simple reconciliation with general dependence on all spatial coordinates. \par
Some analysis of the scalar sector perturbations has been conducted in Ref. \cite{NMCCosmologicalPerturbations} in the context of cosmological perturbation theory and the formation of the large-scale structure of the Universe. However, in that work, only approximate behaviour was obtained due to neglecting any time derivatives in comparison with spatial derivatives. While we will sometimes invoke the sub-Hubble regime, we do not make this quasi-static approximation, as we aim to investigate the presence of dynamical gravitational wave-like behaviour in the NMC theory. Another notable difference is the inclusion of vector and tensor sector perturbations, leading to a more complete picture of the evolution of metric fluctuations in this modified theory.

\subsection{Stress-energy tensor perturbations}
As discussed in Section \ref{NMCSection}, the conservation equation in the NMC theory is modified to give 
\begin{equation}
    \nabla_\mu T^\mu_\nu=\frac{F_2}{1+f_2}(\delta^\mu_\nu\mathcal{L}_m-T^\mu_\nu)\partial_\mu R,
\end{equation}
where we take $\mathcal{L}_m=-\rho$. Perturbing this equation to linear order gives 
\begin{equation}
\begin{aligned}
    \delta\left( \nabla_\mu T^\mu_\nu\right)=&\frac{F_2}{1+f_2}\left[(\delta^\mu_\nu \delta\mathcal{L}_m-\delta T^\mu_\nu)\partial_\mu R\right.\\
    &+\left.(\delta^\mu_\nu\mathcal{L}_m-T^\mu_\nu)\partial_\mu (\delta R)\right]\\
    &+\delta\left(\frac{F_2}{1+f_2}\right)(\delta^\mu_\nu\mathcal{L}_m-T^\mu_\nu)\partial_\mu R,
\end{aligned}
\end{equation}
where the left-hand side is unaltered from GR. Noting that in FLRW all background quantities depend only on $\eta$, we see that to zeroth order
\begin{equation}
\begin{aligned}
    (\delta^\mu_\nu\mathcal{L}_m-T^\mu_\nu)\partial_\mu R&= (\delta^\eta_\nu\mathcal{L}_m-T^\eta_\nu)\partial_\eta R\\
    &=\delta^\eta_\nu(-\rho+\rho)\partial_\eta R=0,
\end{aligned}
\end{equation}
where we have used our chosen form of $\mathcal{L}_m=-\rho$ and the $\eta\eta$ component of the diagonal background stress-energy tensor $T^\eta_\eta=-\rho$. We thus see that the final term in the perturbed conservation equation vanishes for all values of the free index $\nu$. Similarly, by analysing the $\nu=\eta$ component we see that 
 \begin{equation}
\begin{aligned}
    \delta\left( \nabla_\mu T^\mu_\eta\right)&=\frac{F_2}{1+f_2}\left[(\delta^\mu_\eta \delta\mathcal{L}_m-\delta T^\mu_\eta)\partial_\mu R\right.\\
    &\left.+(\delta^\mu_\eta\mathcal{L}_m-T^\mu_\eta)\partial_\mu (\delta R)\right]\\
    &=\frac{F_2}{1+f_2}\left[(\delta\mathcal{L}_m-\delta T^\eta_\eta)\partial_\eta R\right.\\
    &\left.+(\mathcal{L}_m-T^\eta_\eta)\partial_\eta(\delta R)\right]=0,
\end{aligned}
\end{equation}
as $\delta T^\eta_\eta=\delta\mathcal{L}_m=-\delta\rho$. This shows that the $\nu=\eta$ component of the conservation equation is unaltered by the NMC theory and is thus identical to Eq. (\ref{SEConservationEta}). \par
For $\nu=i$ we have 
\begin{equation}
\begin{aligned}
     \delta\left( \nabla_\mu T^\mu_i\right)&=\frac{F_2}{1+f_2}\left[-\delta T^\eta_i\partial_\eta R+(\delta^\mu_i\mathcal{L}_m-T^\mu_i)\partial_\mu (\delta R)\right]\\
     &=\frac{F_2}{1+f_2}\left[-\delta T^\eta_i\partial_\eta R+(\mathcal{L}_m-p)\partial_i (\delta R)\right]\\
     &=-\frac{F_2}{1+f_2}\left[v_i\partial_\eta R+\partial_i (\delta R)\right](\rho+p),
\end{aligned}
\end{equation}
where we again use $v_i=V_i+\partial_i v$ and so 
\begin{equation}
\begin{aligned}
    \delta\left( \nabla_\mu T^\mu_i\right)=&-\frac{F_2}{1+f_2}(\rho+p)V_i\partial_\eta R\\
    &-\partial_i\left[\frac{F_2}{1+f_2}(\rho+p)(v\partial_\eta R+\delta R)\right].
\end{aligned}
\end{equation}
Here we have used the fact that $f_2$, $\rho$ and $p$ only depend on $\eta$ at background level. Similarly to the left-hand side of this equation, we have 2 different terms with different helicities. This means we again have 2 separate equations for each spatial component of the conservation equation. The first of these, associated with the vector sector, is
\begin{equation}
    \begin{aligned}
        4\mathcal{H}(\rho+p)V_i+&\left[(\rho+p)V_i\right]'=-\frac{F_2}{1+f_2}R'(\rho+p)V_i\\
        &=-\ln{(1+f_2)}'(\rho+p)V_i \\
        &\Rightarrow V_i=\frac{V_i^{(s)}(z)}{(1+f_2)(\rho+p)a^4},
    \end{aligned}
\end{equation}
where we see that as in GR we can separate the spatial dependence of the vector matter perturbations (here again written with a superscript $(s)$) from their temporal evolution, with the latter decaying with an additional factor of $(1+f_2)$ due to the nonminimal coupling. \par
For the scalar sector we have 
\begin{equation}
\begin{aligned}
    &c_s^2\delta\rho+(\rho+p)\Phi+\left[(\rho+p)v\right]'\\
    &=-\left[4\mathcal{H}+\ln{(1+f_2)}'\right](\rho+p)v-\frac{F_2}{1+f_2}(\rho+p)\delta R,
\end{aligned}
\end{equation}
where again we find a contribution of a $(1+f_2)$ factor from the NMC on the evolution of $v$, along with additional coupling to the scalar metric perturbations via the purely scalar-dependent $\delta R$. This is explicitly given by
\begin{equation}
\begin{aligned}
    \delta R=\frac{2}{a^2}&\left[\nabla^2(2\Psi-\Phi)-3\mathcal{H}(3\Psi+\Phi)'\right.\\
    &\left.-6(\mathcal{H}^2+\mathcal{H}')\Phi-3\Psi''\right],
\end{aligned} 
\end{equation}
showing the added complexity introduced by the NMC theory to the coupling between the scalar quantities $\delta\rho$, $v$, $\Phi$ and $\Psi$, which is not present in minimally coupled $f(R)$ theories, as these present no alterations to the stress-energy conservation equation \cite{FR_GWs,ExtraForce}.

\subsection{Tensor perturbations} \label{NMCDynamicsTensorSubsection}
The tensor sector of the field equations is relatively simple due to both the zeroth and first-order stress-energy tensor having no tensorial helicity terms. As with the perturbations in GR, the equation for the $\times$ term comes from the $(x,y)$ component of the field equations, while for the $+$ term an identical equation follows from the $(x,x)=-(y,y)$ components. The evolution of the tensor perturbations is described by 
\begin{equation}
\begin{aligned}
    &(1-2F_2\rho)\Box_\eta h_{+/\times}+2(F_2 \rho)'h_{+/\times}'\\
    &=\left[2\mathcal{H}(F_2\rho)'-(\mathcal{H}^2+\mathcal{H}')(1-2F_2\rho)\right]h_{+/\times},
\end{aligned}
\end{equation}
where we note that the $\Box_\eta$ operator indicates a luminal propagation speed $c_{tensor}^2=1$. This equation clearly reduces to the GR equivalent when we set $f_2=0$. However, we now also have a friction term due to the nonminimal coupling term $(F_2 \rho)'$. To simplify this, we may use the fact that 
\begin{equation}
\begin{aligned}
    &(1-2F_2\rho)h''-2(F_2\rho)'h'=Fh''+F'h'\\
    &=\sqrt{F}\left(\sqrt{F}h''+2\frac{F'}{2\sqrt{F}}h'\right)\\
&=\sqrt{F}\left(\sqrt{F}h''+2\frac{F'}{2\sqrt{F}}h'+(\sqrt{F})''h-(\sqrt{F})''h\right)\\
&=\sqrt{F}(\sqrt{F}h)''-\sqrt{F}(\sqrt{F})''h
\end{aligned}
\end{equation}
to write the equation as 
\begin{equation}
\Box_\eta(\tilde{h}_{+/\times})=-\left(\mathcal{H}\frac{F'}{F}+\frac{F''}{2F}-\frac{F'^2}{4F^2}+(\mathcal{H}^2+\mathcal{H}')\right)\tilde{h}_{+/\times},
\end{equation}
where we have used
\begin{equation}
    \frac{(\sqrt{F})''}{\sqrt{F}}=\frac{F''}{2F}-\frac{F'^2}{4F^2}
\end{equation}
and defined $\tilde{h}_{+/\times}=\sqrt{F}h_{+/\times}$.
We thus see that the gravitational wave amplitude will be scaled by a $\eta$-dependent factor of $F^{-1/2}$. For late-time modifications with $f_2=(R/R_n)^n$ ($n<0$), as used in the context of removing the Hubble tension \cite{NMCHubbleTension} or mimicking dark matter profiles \cite{NMCDarkMatter}, we see that $F_2\rho<0$ and so $F=1-2F_2\rho>1$, meaning that gravitational waves decay faster as $F$ increases into the future and $f_2$ dominates. If we take the background to be slowly evolving in comparison to the waves and introduce a stress-energy source, we obtain the modified gravitational wave equation
\begin{equation}\label{TensorWaveEq}
    \Box h_{ij}^{TT}=-16\pi G\frac{1+f_2}{F}\delta T_{ij},
\end{equation}
where we have temporarily restored the gravitational constant. This differs from the GR result by a factor of $\frac{1+f_2}{F}$ stemming from the non-minimal coupling. We will later identify this as a parameter ($\Sigma=\frac{1+f_2}{F}$) that will also modify the weak lensing predictions from the scalar sector perturbations, which may also be thought of as a rescaling of the gravitational constant $\tilde{G}=G\Sigma $ \cite{NMCCosmologicalPerturbations}.

\subsubsection{Generation of gravitational waves in NMC theory}
The production of GWs in GR can be analysed by solving the unmodified version of wave Eq. (\ref{TensorWaveEq}) using the Green's function of the $\Box$ operator. For the remainder of this section, we shall simplify the equations by writing $\delta T_{ij}$ as $T_{ij}$ and consider perturbations varying faster than the background, allowing us to approximate $t\approx\eta$ and $a(t)\approx 1$. By applying the GR conservation equation $\partial_\mu T^{\mu\nu}=0$, one can relate the time derivative of the temporal components of the stress-energy tensor to the spatial derivatives of its spatial components. This leads to the well known quadrupole formula
\begin{equation}
    h^{ij}(t,\Vec{x})=\frac{2G}{r}\ddot{I}^{ij}(t_r,r)=\frac{2G}{r}\frac{d^2}{dt^2}\int d^3x' \ x'^i x'^j T^{00},
\end{equation}
which describes the waves generated by a source at a distance $r$ away and at retarded time $t_r=t-r$. This formula indicates that in GR we do not expect monopoles and dipoles to produce gravitational radiation \cite{WaldGR}. \par
However, in the NMC theory, there are two significant changes. These are the modified dynamical value of the gravitational constant $\tilde{G}$ and the non-conservation of the stress-energy tensor given in Eq. (\ref{NonConservationEq}). This means that there are additional steps we must take to derive the form of the waves $h^{ij}$ from the matter content that originates them. We start by applying the Green's function \cite{WaldGR} 
\begin{equation}
\begin{aligned}
    h^{ij}(t,\Vec{x})&=4G\int dt' \int d^3x' \ \frac{\delta(t'-t_r)}{|\Vec{x}-\Vec{x}'|} \left(\frac{1+f_2}{F}\right)T^{ij}\\
    &\approx\frac{4G}{r}\int d^3x' \ \left[\left(\frac{1+f_2}{F}\right)T^{ij}\right](t_r,\Vec{x}') ,
\end{aligned}
\end{equation}

where we have assumed a source at a distance $|\Vec{x}|=r$ away with $\Vec{x}\gg\Vec{x}'$. To simplify this equation, we assume that the background $R(t,\Vec{x})\approx R(t)$ such that we may move the scaling of the modified constant outside the spatial integral. This means we now have
\begin{equation}
\begin{aligned}
     h^{ij}(t,\Vec{x})&=\frac{4G}{r}\left.\left(\frac{1+f_2}{F}\right)\right\vert_{t=t_r}\int d^3x' \ T^{ij}(t_r,\Vec{x}')\\
     &=\frac{4\tilde{G}(t_r)}{r}\int d^3x' \ T^{ij}(t_r,\Vec{x}'),
\end{aligned}
\end{equation}
which is what one would expect by considering the rescaled gravitational constant as a ``true" constant. We then integrate by parts and neglect boundary terms as we expect no contributions at infinity, while using the conservation equation to substitute $T^{ij}$ terms by $T^{00}$ terms. If the conservation equation was unaltered with respect to GR, as is the case in minimally coupled $f(R)$ theories, then the quadrupole formula for the spin-2 sector would be recovered as 
\begin{equation}
    h^{ij}(t,\Vec{x})= \frac{2\tilde{G}}{r}\ddot{I}^{ij}(t_r,\Vec{x}),
\end{equation}
where in that case we would have $\tilde{G}=G/F_1$. However, the conservation equation in the NMC theory gives
\begin{equation}
\begin{aligned}
    \partial_k T^{kj}=&-\partial_t T^{0j}+\frac{F_2}{1+f_2}(g^{\mu j}\mathcal{L}_m-T^{\mu j})\partial_\mu R\\
    =&-\partial_t T^{0j}+\frac{F_2}{1+f_2}(g^{0 j}\mathcal{L}_m-T^{0 j})\dot{R}\\
    =&-\frac{1}{1+f_2}\partial_t[(1+f_2)T^{0j}],
\end{aligned}
\end{equation}
where we have again taken $R\approx R(t)$ and taken the background metric to be flat such that $g^{0j}=0$. We thus have

\begin{equation}
\begin{aligned}
     h^{ij}(t,\Vec{x}) &=-\frac{4\tilde{G}}{r}\int d^3x' \ x'^i \partial_k T^{kj}\\
     &= \frac{4\tilde{G}}{r(1+f_2)}\int d^3x' \ x'^i \partial_t \left[(1+f_2)T^{0j}\right].
\end{aligned}
\end{equation}
Following the same steps we develop this expression as
\begin{equation}
\begin{aligned}
     h^{ij}(t,\Vec{x}) &=-\frac{2\tilde{G}}{r(1+f_2)}\int d^3x' \ x'^i x'^j \partial_t \left[(1+f_2)\partial_k T^{0k}\right]\\
     &=\frac{2\tilde{G}}{r(1+f_2)}\int d^3x' \ x'^i x'^j \partial_t\left[(1+f_2)\partial_t T^{00}\right]\\
     &=\frac{2\tilde{G}}{r}\left[\ddot{I}^{ij}+\frac{F_2}{1+f_2}\dot{R}\dot{I}^{ij}\right],
    \end{aligned} 
\end{equation}
where we have used the conservation equation again to simplify $\partial_k T^{k0}$ as
\begin{equation}
\begin{aligned}
     \partial_k T^{k0}&=-\partial_t T^{00}+\frac{F_2}{1+f_2}(g^{0 0}\mathcal{L}_m-T^{00})\dot{R}\\
     &=-\partial_t T^{00}+\frac{F_2}{1+f_2}(\rho-T^{00})\dot{R}=-\partial_t T^{00},
\end{aligned}
\end{equation}
as we have chosen the matter Lagrangian to be $\mathcal{L}_m=-\rho=-T^{00}$. We note that the $\dot R$ term should only be considered when its dynamics are comparable to those of $I_{ij}$, as it should otherwise be ignored under the sub-horizon approximation. Therefore, under the stated assumptions, we find that the production of gravitational waves is still quadrupolar, obeying a similar equation to the one in GR with a rescaled gravitational constant $\tilde{G}$ and an additional term due to the modified conservation law, which follows from the nonminimal coupling between matter and curvature. As expected, by setting $f_1=R$ and $f_2=F_2=0$, we recover the GR quadrupole formula, thus ensuring a smooth reconciliation with GR in the weak modification regime. However, as discussed later in this work, the NMC theory admits scalar gravitational wave degrees of freedom, as previously discussed in Ref. \cite{NMCGravWaves}. This additional degree of freedom predicts gravitational radiation from all multipoles, down to monopoles and dipoles, as in most alternative gravitational theories \cite{ScalarWavesf_R}. Nevertheless, showing that the traceless-transverse modes see no effects from monopoles and dipoles at leading order provides clarity on another fundamental property of what we still expect to be the dominant modes of gravitational radiation in the NMC theory.

\subsection{Vector perturbations}
As we consider $\Xi_i=\Xi_i(\eta,z)$, the condition $\partial_i\Xi_i=0$ implies $\Xi_3=0$. As in the GR analysis, the fully spatial components of the perturbed field equations have no vector contribution from the stress-energy terms and thus give us the temporal evolution of the vector sector as
\begin{equation}
\begin{aligned}
F\Xi_i'-2(F_2\rho)'\Xi_i+2\mathcal{H}F\Xi_i=&\left[F\Xi_i\right]'+2\mathcal{H}F\Xi_i=0\\
&\Rightarrow \Xi_i=\frac{R_i(z)}{a^2 F},
\end{aligned}
\end{equation}
which shows that the vector perturbations evolve in comoving time $\eta$ similarly to how they did in GR, with an additional decay due to the nonminimal coupling term in $F$. \par
The $(\eta,i)$ components again determine the spatial evolution of the perturbations
\begin{equation}
\begin{aligned}
    \frac{1}{2a^2}F\nabla^2\Xi_i&=(1+f_2)\delta T^\eta_i+F_2 \bar T^\eta_i\delta R=(1+f_2)(\rho+p)V_i \\
    &\Rightarrow \nabla^2 R_i=2 V^{(s)}_i(z),
\end{aligned}
\end{equation}
where in the final step we have removed the $\eta$-dependence from the equations determined above and from the conservation equations. This is precisely the same spatial equation as in GR, leading us to the conclusion that the spatial dependence of the vector perturbations seems to be unaltered by the NMC theory. Additionally, the separation of the $\eta$ and spatial evolution means that we have no radiative behaviour and thus the NMC theory does not predict wave-like evolution for the vector terms in the metric perturbation. The modification of the temporal evolution of the vector sector could provide an opportunity to observationally test the presence of a nonminimal coupling in the gravitational theory \cite{LIGOLISAPolarisations,LISAPolarisations}. However, this would involve obtaining observations at a considerable variety of redshifts and throughout a wide variety of systems in order to draw any solid conclusions.

\subsection{Scalar perturbations}
When considering scalar perturbations in GR, we found the equality $\Psi=\Phi$ from the purely spatial components of the field equations, leading to a greatly simplified analysis of the remaining components. Analysing the same purely spatial components again reveals the possibility of separating these into $\delta^i_j$ and $\partial_i\partial_j$ terms, as expected from the helicity decomposition. For simplicity, we show these in the high-frequency limit, which gives 
\begin{equation}
\begin{aligned}
    &\delta^i_j\left[2(1+2F_2 p)\Psi''+(1+2 F_2 p)\nabla^2\left(\Phi-\Psi\right)\right.\\
    &\left.-(1+2F_2 p-F)\Box_\eta \Psi+\Box_\eta \delta F-a^2(1+f_2)\delta p\right]\\
    &-\partial_i\partial_j\left[F(\Phi-\Psi)+\delta F\right]=0,
\end{aligned} 
\end{equation}
where we have not expanded $\delta F$ into its full dependence on scalar sector quantities for simplicity and again assumed a perfect fluid with $p=c_s^2\rho$. By setting $F=1$, $F_2=0$ and $\delta F=0$, we recover the high-frequency GR result, as expected. There is a strong dependence on the quantity $F_p\equiv1+2F_2 p$, which interestingly would be precisely the value of $F$ if we had defined the matter Lagrangian in terms of pressure ($\mathcal{L}_m=p$). We can again set both of these terms equal to zero independently, with the latter giving
\begin{equation}
    \Psi-\Phi=\delta\ln{F}=\frac{\delta F}{F}=\frac{-2F_2\delta\rho-2\rho F_{2,R}\delta R}{F},
\end{equation}
which shows that the NMC, present through $F$, breaks the equality between the different scalar metric perturbations \cite{NMCCosmologicalPerturbations,NMCPerfectFluidDynamics}. This significantly complicates the remaining equations, as we are unable to replace all $\Psi$-dependence with $\Phi$-dependence like in GR. Additionally, the $\nabla\nabla\delta F$ terms in the perturbed field equations lead to first and second-order derivatives of the density fluctuations $\delta\rho$, present due to the explicit presence of the matter Lagrangian density in $F$. However, by rewriting the previous relation as $\delta F=(\Psi-\Phi)F$ and inserting this into the perturbed field Eqs. (\ref{NMCLinearisedFieldEqsDeltaF}), we may remove most of the complexity from the equations, with stress-energy perturbations now only present in $\delta T^\mu_\nu$, analogously to GR.\par
The remaining non-trivial equations are obtained from the $(\eta,\eta)$ component, the $(i,\eta)$ component and the $\delta^i_j$ term in the purely spatial components of the field equations. The choice of only $\eta$ and $z$-dependence means that only the $i=3$ component of $(i,\eta)$ is non-trivial, with the remaining $(i,\eta)$ components yielding the same equation in the general case and thus not causing any loss of generality. The $\delta^i_j$ term is the same for all $i=j$ components, as expected. This means we are left with 3 non-trivial field equations, along with 2 stress-energy conservation equations for the scalar sector. This would be consistent with the 5 degrees of freedom from $\Phi,\Psi,v,\delta\rho$ and $\delta p$. However, as we have assumed a perfect fluid with $\delta p=c_s^2\delta\rho$, we are only left with 4 independent degrees of freedom and expect one of the equations to be obtained from the others, as is the case with the equations in GR \cite{GWDecompositionBook}. \par
The $(i,\eta)$ components of the field equations can again be written as $\partial_i(\cdots)=0$, which together with the condition of vanishing perturbations at infinity leads us to set $(\cdots)=0$. This can be written as 
\begin{equation}
\begin{aligned}
    F\left[\mathcal{H}(\Phi+\Psi)+(\Phi+\Psi)'\right]&+F'(2\Phi-\Psi)\\
    &=-a^2(1+f_2)(\rho+p)v,
\end{aligned}
\end{equation}
which reduces to the GR result obtained previously when $f_2=0$, as expected. The $(\eta,\eta)$ component gives 
\begin{equation}
\begin{aligned}
    -a^2(1+f_2)\delta\rho=&-F\nabla^2(\Phi+\Psi)+3F'\Psi'+3\mathcal{H}F'\Phi\\
    &+3F(\mathcal{H}'-\mathcal{H}^2)(\Psi-\Phi)\\
    &+3\mathcal{H}\left(F\left[\mathcal{H}(\Phi+\Psi)+(\Phi+\Psi)'\right]\right.\\
    &\left.+F'(2\Phi-\Psi)\right),
\end{aligned}
\end{equation}
where the last term can be simplified with the $(i,\eta)$ equation to give
\begin{equation}
\begin{aligned}
F\nabla^2(\Phi+\Psi)-3F'(\Psi'+\mathcal{H}\Phi)&+3F(\mathcal{H}^2-\mathcal{H}')(\Psi-\Phi)\\
&=a^2(1+f_2)\delta\rho_*,
\end{aligned}
\end{equation}
with $\delta\rho_*=\delta\rho-3\mathcal{H}(\rho+p)v$ still gauge-invariant in the NMC theory. This is the modified form of the ``Poisson-like" Eq. (\ref{PerturbedPoisson}), with additional metric terms proportional to $F'$ and an added factor of $(1+f_2)$ multiplying the matter content of the equation due to the nonminimal coupling. The final term on the left-hand side is proportional to $F$, which does not vanish in GR ($F=1$). However, in that scenario, we know that $\Psi=\Phi$ and the term will still be removed, thus ensuring a reconciliation with GR. In the sub-horizon regime, where we take the perturbations to vary much faster than the rate expansion of the Universe ($\partial_z\sim k>>\mathcal{H}$), this equation simplifies to 
\begin{equation} \label{ScalarWeakLensingEq}
\begin{aligned}
\nabla^2(\Phi_{WL})\equiv\nabla^2(\Phi+\Psi)&=\frac{1+f_2}{F}a^2\delta\rho_*\\
&\equiv \Sigma(a,k)a^2\delta\rho_*, 
\end{aligned}
\end{equation}
which matches the expression found in Ref. \cite{NMCCosmologicalPerturbations}. Note that we have taken all background quantities (such as $F$) to evolve with the expansion of the Universe as $F'\sim \mathcal{H}F$ and thus neglect them when compared to derivatives of perturbative quantities. Similarly to what was done in the tensor sector, we have again defined $\Sigma$ in analogy with the formalism proposed in Ref. \cite{PerturbationParameters}, where the importance of this parameter in the context of weak lensing observations was discussed. This definition comes from the relationship between the so-called ``weak lensing potential" $\Phi_{WL}$ and the density perturbations. As previously pointed out, this effect can be encapsulated as a rescaling of the gravitational constant \cite{NMCCosmologicalPerturbations}. \par
All scalar equations presented so far have had explicit dependence on the perturbed matter content. To remove this, we may recognise that to first-order the stress-energy tensor for a perfect fluid obeys $T^z_z+c_s^2 T^\eta_\eta=0$, meaning that doing the same to the corresponding components of the perturbed field equations would remove all matter perturbation content. As before, we choose the high-frequency limit for simplicity, which gives
\begin{equation}\label{FullNMCScalarMasterEqs}
\begin{aligned}
0=&\left[3(1+c_s^2)-F(2+3c_s^2)\right]\Psi''\\
&+\left[F(2+c_s^2)-2(1+c_s^2)\right]\nabla^2\Psi\\
&+F\Phi''+\left[1+c_s^2-F(1+2c_s^2)\right]\nabla^2 \Phi.\\
\end{aligned}
\end{equation}
This is the NMC-modified version of the scalar ``wave" Eq. (\ref{ScalarWaveEq}). The main difference between these is the dependence on both scalars $\Phi$ and $\Psi$, which are no longer equivalent. It looks like the combination of two wave-like equations for these scalars. Unlike in GR, the assumption of a matter-dominated Universe with $c_s^2=0$ no longer fully removes the $\partial_z^2$ part of the equation. This is due to the presence of the NMC terms proportional to $F_2\rho$. The complexity of the coupled equations means we cannot find concrete analytic solutions even under the high-frequency approximation. Nevertheless, one could consider the weak modification regime, which allows for an additional linear treatment of the NMC effects. However, this would only hold in specific circumstances and is thus left as a discussion in Appendix \ref{WeakNMCRegime}. 

\subsubsection{Detection of scalar polarisations}
Even without an analytical solution for the separate metric perturbations, the analysis conducted here provides a possible method for distinguishing between the standard theory and the $f(R)$ modified theory. This follows from the initial discussion on the scalar sector in the NMC theory, where we determined that the presence of the nonminimal coupling breaks the equality $\Psi=\Phi$ with the introduction of a $\delta\ln{F}$ term. When introducing the scalar polarisation components in the tidal tensor ($P_1$ and $P_6$), we found that in the high-frequency (sub-horizon) limit these depend on the metric perturbations as 
\begin{equation}
\begin{array}{ll}
    P_1=\partial_z^2\Phi+\Psi''\sim- k^2\Phi-\omega^2\Psi\\
    P_6=\Psi''\sim -\omega^2\Psi,
    \end{array}
\end{equation}
where we note the mixed functional dependence in $P_1$. We have also assumed a monochromatic plane wave decomposition of the scalar functions for simplicity in the final step, although one could of course consider more complex Fourier space forms with a mixed frequency spectrum in general.  These terms correspond to longitudinal and breathing modes respectively and lead to distinct behaviour from the tensorial $+$ and $\times$ modes \cite{GWPolarisations}. In GR, only the massless spin-2 modes are expected to be present, while some modified theories of gravity predict the existence of the aforementioned scalar modes \cite{LIGOLISAPolarisations,LISAPolarisations}. Particularly, this was already established in the context of minimally and nonminimally coupled $f(R)$ theories \cite{FR_GWs,FRGWs2,NMCGravWaves}. This means that the detection of such polarisations would serve as a direct test of modified gravity, although their detection alone would not necessarily distinguish between different modified theories without further determination of their properties. With the above results, we can go further in our predictions, as we not only find the presence of breathing and scalar modes, but we also obtain a clear distinction between their properties, as we no longer have equivalent temporal and spatial metric perturbations in the scalar sector. If we detect both scalar and longitudinal gravitational wave polarisations and are able to compare their effects, one could theoretically use these observations to isolate the effects of the $\Psi$ perturbation present in both $P_1$ and $P_6$ and consequently determine the $\Phi$ contribution to $P_1$. If these are found to be equivalent, then the nonminimally coupled theory could be ruled out, while evidence of a ``decoherence" of $\Phi$ and $\Psi$ would indicate the possible presence of the NMC. Of course, this would not necessarily provide definitive proof of its validity over other alternative theories, but it would nevertheless serve as a remarkable step toward obtaining a more general gravitational theory. \par
Detecting additional polarisations of GWs would require careful synchronization of multiple interferometers, as we would need data from different orientations to be able to disentangle the 6 possible independent polarisations \cite{PolarisationTestsStochastic,BreathingLongitudinalDegeneracy}. As discussed in Refs. \cite{LISAPolarisations,ScalarDegenerateGroundResponses}, the responses of ground-based laser interferometers to breathing and longitudinal modes are completely degenerate, meaning that these can not be distinguished by present experiments. This follows from the equal pattern functions obtained for these polarisations in the context of laser interferometer arrays \cite{BreathingLongitudinalDegeneracy}. The emergence of space-based GW detectors, such as LISA \cite{LISA}, leads to the possibility of probing wider ranges of frequencies with minimal interference from the atmosphere \cite{FRGWs2}. This means that the distinction and independent analysis of the breathing and longitudinal modes, along with the consequent comparison of the two scalar metric perturbations, could become feasible in the coming years \cite{PolarisationTestsStochastic,BeyondLISA,ExtraPolarisationDetection,SpaceBasedPolarisation}. Additionally, there are many recent developments on the detection methods for high-frequency gravitational waves (HFGWs) beyond the range of interferometry experiments like LIGO and LISA. These HFGWs can be generated in the very early Universe (see, for instance, Ref. \cite{GWBubbles}) or high-energy astrophysical events \cite{CosmographyGWs,HFGW_Interferometer}, therefore providing a direct connection to extreme conditions in which modified gravity would be indispensable. They would also be in an optimal frequency range for an electromagnetic response \cite{HFGWDetection_Review,GW_EMCavity,HFGW_Recent}, whose observation may be used to detect and separate the six polarisations, as discussed in Ref. \cite{HighFrequencyGWs}, where it was also found that there is a good complementarity between the proposed electromagnetic response data analysis and existing ground-based telescopes. \par

Another notable avenue for observational tests of modified gravity at the other end of the GW frequency spectrum is the analysis of pulsar timing arrays (PTAs) \cite{BreathingLongitudinalDegeneracy}. This method utilises the highly regular electromagnetic pulsations of distant rotating neutron stars, known as pulsars, to detect low-frequency GWs from the cosmological stochastic background or isolated strong-field events like those caused by super-massive black holes \cite{PTA_SKAStrongFieldTests}. As light is emitted from the pulsars and received on Earth, both its start and end points are met with the effects of passing gravitational waves from different points in the Universe, leading to path differences that may be observed as measurable alterations to the otherwise periodic pulsation \cite{PTA1}. Each Earth-Pulsar system then serves as an effective interferometer arm of cosmological scale, allowing for the detection of GWs with frequencies ($\sim$nHz) well below those detectable by LIGO ($\sim$100Hz) or LISA ($\sim$mHz) \cite{PTA_SKAStrongFieldTests}. Of course, the large distances to these objects mean that we expect a combination of different gravitational waves to affect the observed result. This stochastic gravitational wave background (SGWB) is a complex composition of different wave polarisations, especially if one considers a general metric theory with 6 possible wave modes. Thankfully, this has been thoroughly researched in past literature, where it has been found that the correlation between data from arrays of many different pulsars can be used to decompose this background into its respective polarisations \cite{PTA1,PTA2,PTADataConstraints,PTAMapping}. With the constantly increasing number of pulsars under permanent observation and improvements in the respective data analysis, this method provides yet another chance to individually detect the longitudinal and breathing modes, thus allowing for the separation of the effects from $\Psi$ and $\Phi$ \cite{PTAPulsarAmount}. This is especially relevant due to the possible early origin of parts of the SGWB, which could provide a window to epochs with some of the more intense NMC presence. Additionally, the longitudinal nature of the $P_1$ polarisation leads to an increased effect on the irregularity of the pulsar radiation, with sensitivities up to two orders of magnitude greater than those of tensor modes \cite{PTA1}.

\subsection{Perturbation dynamics in models with \texorpdfstring{$f_2(R)=(R_n/R)^n$}{f2=(Rn/R)n}} \label{WorkedExampleSection}
The discussion so far has focused on an agnostic form of $f_2(R)$, chosen to keep all arguments as general as possible. However, it is useful to analyse the dynamics resulting from the modified perturbation equations. For this, we choose an inverse power law form given by $f_2=(R_n/R)^n$, where $n$ is a positive integer. This choice ensures the decoupling of curvature and matter for large values of $R$, which from a cosmological point of view corresponds to the early-time Universe, while allowing for the emergence of modified behaviour near the present. Notably, this kind of model has been shown to mimic the effects of dark matter in galaxy rotation curves \cite{NMCDarkMatter}, yield the observed accelerated expansion of the Universe \cite{NMCAcceleratedExpansion} and resolve the Hubble tension \cite{NMCHubbleTension}. In the latter case, it was found that the NMC model can approximately recreate the late-time behaviour expansion of the $\Lambda$CDM model with $H_0\sim73.2 \ \text{km/s/Mpc}$. This means that for small redshifts we can take the evolution of the background quantities from the results of that work, which greatly simplifies the generation of visualisations for the worked example shown here. \par
In the remainder of this section, we will focus on models with $n=4$ and $n=10$, as these were shown to capture different kinds of characteristics of the NMC model in Ref. \cite{NMCHubbleTension}. The most important function to calculate is $F=1+2F_2\mathcal{L}_m$, which for these models is given by 
\begin{equation}
    F=1+2n\left(\frac{R_n}{R}\right)^n\frac{\rho}{R}\geq 1,
\end{equation}
where we can consider suitable values for the NMC ``characteristic scale" constants $R_n$ from the results on the Hubble tension presented in Ref. \cite{NMCHubbleTension}, namely $R_4=4.1\times 10^4$ and $R_{10}=4.4\times10^4$ in the chosen unit convention ($c=1$). 

\subsubsection{Vector perturbations}
\begin{figure}[ht!]
    \centering
    \includegraphics[width=0.95\linewidth]{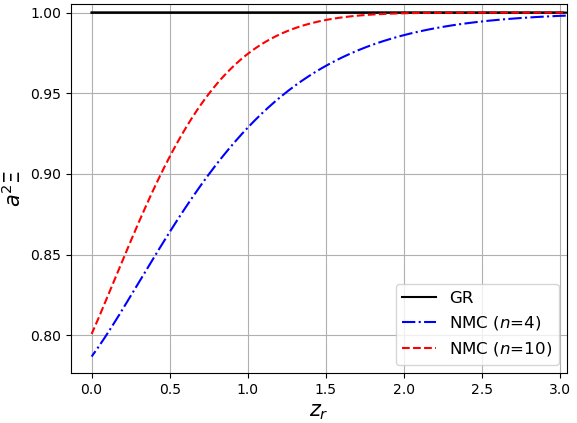}
    \includegraphics[width=0.95\linewidth]{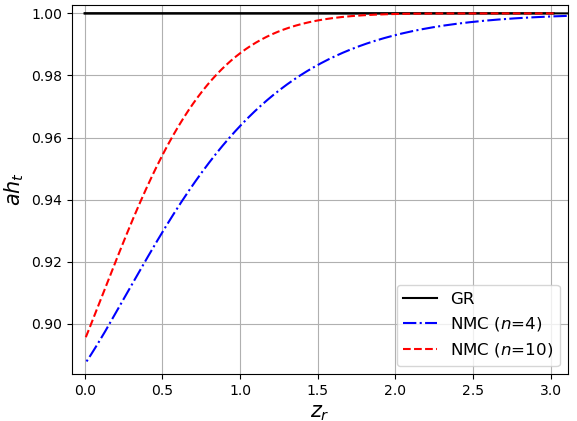}
    \caption{Time evolution of the amplitude of vector (top) and tensor (bottom) metric perturbations in GR and in the NMC modified theory with $f_2=(R_n/R)^n$. Here the time dependence is shown in terms of the corresponding redshift $z_r$, with any temporal dependence present in both GR and the modified theories being factored out for easier comparison, as pointed out in the y-axis label.}
    \label{WorkedExamplePlots}
\end{figure}

As discussed earlier, the vector perturbations have a temporal evolution given by 
\begin{equation}
    \Xi_i\propto\frac{1}{a^2F}=\frac{(1+z_r)^2}{F(z_r)},
\end{equation}
where we have defined the redshift as $a\sim(1+z_r)^{-1}$ in order to avoid confusion with the spatial coordinate $z$ used before. Due to the inverse nature of $f_2$, we have $F_2<0$, such that $F>1$, thus leading to an overall faster temporal decrease of the perturbations due to the expansion of the Universe, as shown in the left panel in Figure \ref{WorkedExamplePlots}. Note that the $n=10$ model behaves approximately like GR up until $z_r\approx1.75$, and presents larger perturbation values than $n=4$ for all redshifts, with both models predicting a similar $\sim20\%$ decrease in the present vector perturbation magnitude when compared to GR. \par

\subsubsection{Tensor perturbations}
The tensor perturbations not only decay with the expansion of the Universe, but also oscillate with their own characteristic frequency, which is redshifted similarly to what is observed for frequencies of light originating in the far past and detected on Earth ($\omega_{\text{obs}}=\omega_i \frac{a(t_i)}{a(t_{\text{obs}})}$). However, this decreasing frequency is not affected by the NMC model, instead following from the general expanding features of spacetime in the chosen coordinate system. Similarly, we will factor out the decay over distance ($\sim1/r$), an effect present in both GR and in the NMC model, and regard this as due to the dissipation of the waves as they spread over larger distances. With this in mind, we neglect both of these behaviours and focus on the remaining characteristics of the evolution of the amplitude of these waves. From the previously derived equations, we see that even in GR the amplitude of the tensor perturbations will decay as $a^{-1}\sim(1+z_r)$, which has been factored out from the start in the definition of the perturbed metric. However, in the modified model we obtain oscillatory behaviour for the combination $\sqrt{F}h$, which means that the perturbation $h$ will have an additional temporal decrease given by $F^{-1/2}$. Putting all of this together, we see that
\begin{equation}
    h_{\text{t}}\propto\frac{1}{a\sqrt{F}}=\frac{1+z_r}{\sqrt{F(z_r)}},
\end{equation}
where $h_{\text{t}}$ denotes the amplitude of the observed tensorial perturbations. The results for the chosen example models are shown in the right panel of Figure \ref{WorkedExamplePlots}. Again both models are in agreement with GR for larger redshifts ($z_r\gtrapprox3$), with the $n=10$ model maintaining an approximate GR-like behaviour until $z_r\approx1.75$, after which we see that both models predict a $\sim10\%$ decrease in the presently observed gravitational wave amplitudes due to the expansion of the Universe in the NMC model. 

\subsubsection{Scalar perturbations}
\begin{figure}[ht!]
    \centering
    \includegraphics[width=0.95\linewidth]{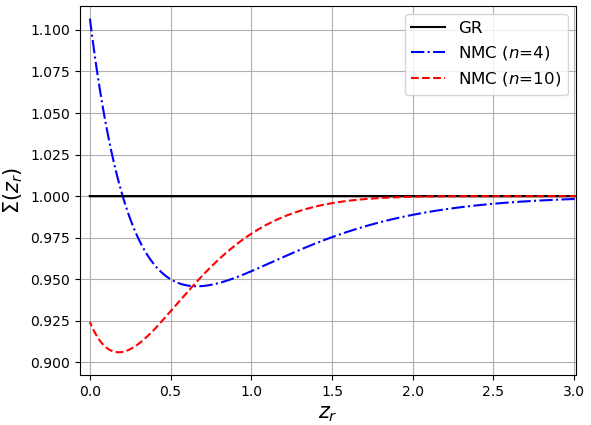}
    \caption{The evolution of the ``weak lensing parameter" $\Sigma$ in terms of redshift $z_r$ in the NMC modified theory with $n=4$ and $n=10$. The GR prediction is shown for comparison. }
    \label{WeakLensingPlot}
\end{figure}

The scalar perturbations present more complex equations than the remaining sectors, as seen by the lack of decoupling between the equations for the scalar metric fluctuations $\Phi$ and $\Psi$ discussed in the sections above. In fact, even when considering a specific example such as the previously chosen model, we are still unable to obtain analytical solutions for the evolution of these quantities. However, one particular parameter can be predicted in the sub-Hubble limit of Eq. (\ref{ScalarWeakLensingEq}), where we defined the ``weak lensing parameter" ($\Sigma$) in terms of the relation between the weak lensing potential $\Phi_{WL}\equiv\Phi+\Psi$ and the density perturbations, similarly to what has been done in past works on cosmological parameters (see Ref. \cite{PerturbationParameters} for a review). Distinctly from the vector and tensor sectors, the evolution of this quantity depends not only on $F$ and therefore $F_2\rho$, but also on the original NMC function $f_2(R)$. Again applying the generated simulation data from previous work on the Hubble tension \cite{NMCHubbleTension}, we can make predictions on the late-time behaviour of this parameter for different NMC models, as shown in Figure \ref{WeakLensingPlot}. \par

We find that $\Sigma$ approaches its GR value ($\Sigma_{\text{GR}}=1$) for large redshifts, as expected. In the $n=10$ model, this parameter departs from unity later than in the $n=4$ model, due to the higher suppression of $f_2$ with $n=10$ for large curvatures. Additionally, $\Sigma$ consistently decreases almost until the present ($z_r=0$), reaching a minimum value of $\Sigma\approx 0.9$ before increasing to $\sim0.925$, with the turning point occurring around $z_r=0.2$. The $n=4$ model is similar in many aspects, such as the initial decrease and consequent turning point. However, there are some notable differences, namely the earlier departure from GR and the earlier turning point around $z_r=0.7$. This latter distinction is of particular importance, as the increase of $\Sigma$ starts early enough for its value to pass the GR prediction around the same time that it starts increasing in the $n=10$ model, reaching an increase of $\sim 10\%$ at present. These are noteworthy results, not only because they provide us with predictions for weak lensing properties of the modified theory which could potentially be observationally tested, but also because they offer a clear distinction between the lower-power $n=4$ model and its higher-power $n=10$ counterpart, based on their effect on weak lensing at lower redshifts. \par 
Recent attempts to constrain the present value of $\Sigma$ in the context of testing modified gravity theories typically obtain values around 1, with $1\sigma$ regions in the $\sim\left[0.9,1.2\right]$ range at best \cite{WLMeasurement2}, which places both of the NMC models' predictions within error of observational constraints \cite{WeakLensingMeasurement,WeakLensingMeasurement2,WLMeasurement2,WLMeasurement3}. However, in Ref. \cite{DESYear3Lensing} it was found from the Dark Energy Survey (DES) year 3 results that the present value of the weak lensing parameter, analysed by its deviation from GR, i.e. $\Sigma_0\equiv\Sigma(z_r=0)-1$, can be constrained to $\Sigma_0=0.6\pm0.4$ from DES alone and to $\Sigma_0=0.04\pm0.05$ from DES combined with external data, both of which show a greater tension with the prediction from the $n=10$ NMC model than with that of the $n=4$ model. Similar conclusions were reached in the Planck experiment's 2018 results \cite{WLPlanck2018}. This is particularly relevant when considering that these constraints were obtained when taking a parametrisation for $\Sigma$ of the form
\begin{equation}
    \Sigma_{\text{par}}(z_r)=1+\Sigma_0\frac{\Omega_\Lambda(z_r)}{\Omega_\Lambda(z_r=0)},
\end{equation}
where $\Omega_\Lambda$ represents the dark energy density, which does not need to be constant in general. Although this is not equivalent to the predictions of the NMC theories considered here, taking the deviations of $\Sigma$ from its GR value to be caused by dark energy is in some ways analogous to what happens in late-time NMC models with dependence on inverse powers of $R$, which have been shown to yield the same accelerated expansion with no need for the presence of dark energy \cite{NMCAcceleratedExpansion,NMCHubbleTension}. Improvements to the accuracy of observational constraints on growth parameters in modified gravity could provide crucial evidence to confirm or rule out the presence of modified theories such as the NMC theory considered in this work. 

\section{Gravitational waves in scalar-tensor theory representation} \label{ScalarTensorSection}

An alternative method to determine the spectrum of polarisations expected in an $f(R)$ NMC gravity model is to look at an equivalent scalar-tensor formulation of the same theory. In fact, it has been shown in Ref. \cite{NMCScalarTensorAnalogy} that a two-field scalar-tensor model provides a suitable choice. This is given in the form of a Jordan-Brans-Dicke theory with a potential 
\begin{equation}
    S=\int d^4x \ \sqrt{-g} \left[\Omega R-V(\gamma,\Omega)+2\left(1+f_2(\gamma)\right)\mathcal{L}_m\right],
\end{equation}
where $\gamma$ and $\Omega$ are scalar fields and we define 
\begin{equation}
    V(\gamma,\Omega)=\gamma\Omega-f_1(\gamma)=\gamma(\Omega-1)
\end{equation}
as the potential. Note that we have already assumed the form of $f_1(x)=x$ as our work focuses on the NMC theory. Varying the action with respect to the fields yields $\gamma=R$, along with
\begin{equation}
\Omega=F_1(\gamma)+2F_2(\gamma)\mathcal{L}_m=1+2F_2(\gamma)\mathcal{L}_m,
\end{equation}
while varying with respect to the ``physical" metric gives the field equations
\begin{equation}
\begin{aligned}
    \Omega\left(R_{\mu\nu}-\frac{1}{2}g_{\mu\nu}R\right)=&(\nabla_\mu\nabla_\nu-g_{\mu\nu}\Box)\Omega-\frac{1}{2}g_{\mu\nu} V(\gamma,\Omega)\\
    &+\left[1+f_2(\gamma)\right]T_{\mu\nu},
\end{aligned}
\end{equation}
which reduce to the result shown in Section \ref{NMCSection} upon substitution of $\gamma$ and $\Omega$ from the other equations. From now on we will identify $\gamma=R$, while keeping $\Omega$ as an independent field, as its dependence on the matter Lagrangian $\mathcal{L}_m$ does not allow us to solve for $\Omega=\Omega(\gamma)$. Taking the trace of the field equations provides a useful relation
\begin{equation}
    3\Box\Omega+2V(R,\Omega)-R=3\Box\Omega+R(\Omega-2)=(1+f_2)T,
\end{equation}
which shows that $\Omega$ obeys a Klein-Gordon equation. Perturbing the trace equation around some slowly evolving background curvature $R_0\approx 0$ and field $\Omega_0$, we obtain 
\begin{equation}
    3\Box\delta\Omega+(\Omega_0-2)\delta R=(1+f_2)\delta T+F_2 T \delta R,
\end{equation}
while the field equations now read
\begin{equation}
\begin{aligned}
    \Omega_0\delta G_{\mu\nu}=&(\nabla_\mu\nabla_\nu-g_{\mu\nu}\Box)\delta\Omega-\frac{1}{2}\eta_{\mu\nu}(\Omega_0-1)\delta R\\
    &+F_2 T_{\mu\nu}\delta R+ (1+f_2)\delta T_{\mu\nu},
\end{aligned}
\end{equation}
where we have written the metric perturbations over an approximately flat background as $g_{\mu\nu}=\eta_{\mu\nu}+h_{\mu\nu}$. We then choose a gauge in which \cite{NMCGravWaves}
\begin{equation}
    \partial_\mu\left(h^{\mu\nu}-\frac{1}{2}\eta^{\mu\nu}h-\eta^{\mu\nu}\frac{\delta\Omega}{\Omega_0}\right)=0,
\end{equation}
which leads to
\begin{equation}
\begin{array}{ll}
    \delta R_{\mu\nu}=\partial_\mu\partial_\nu\delta\tilde{\Omega}-\frac{1}{2}\Box h_{\mu\nu}\\
    \delta R=\Box\delta\tilde{\Omega}-\frac{1}{2}\Box h,
    \end{array}
\end{equation}
where we have defined $\delta\tilde{\Omega}=\delta\Omega/\Omega_0$ for simplicity. Note that this is equivalent to the quantity $\delta(\ln{F})=\delta F/F_0$ presented in the previous section, where we found that this is the source of the inequality between the two scalar metric perturbations $\Phi$ and $\Psi$. The field equations are then
\begin{equation}
\begin{aligned}
    \Box&\left[\Omega_0 h_{\mu\nu}+\left(2\Omega_0F_2T_{\mu\nu}+(\Omega_0-2)\eta_{\mu\nu}\right)\delta\tilde{\Omega}\right.\\
    &\left.-\left(F_2T_{\mu\nu}+\frac{1}{2}\eta_{\mu\nu}\right)h\right]=-2(1+f_2)\delta T_{\mu\nu},
\end{aligned}
\end{equation}
while the trace equation gives
\begin{equation}
\begin{aligned}
    \left(\Omega_0F_2 T+2\Omega_0-4\right)\Box\delta\tilde{\Omega}&+\frac{1}{2}\left(\Omega_0-2-F_2 T\right)\Box h\\
    &=-(1+f_2)\delta T,
\end{aligned}
\end{equation}
where moving background quantities inside of the derivative operators is justified when considering a slowly evolving background, which may be taken to be approximately constant in comparison to the rapidly evolving perturbations. Note that by considering only the tensor sector, i.e. by neglecting the scalar degree of freedom and the trace of the perturbations, we get the same result as in  Section \ref{NMCDynamicsTensorSubsection}, with a modified gravitational constant $\tilde{G}=(1+f_2)G/\Omega_0=(1+f_2)G/F$, as expected. Even when assuming the more complex case of the presence of background matter content, it is clear that the NMC theory introduces at least one additional radiative degree of freedom in the form of $\delta\tilde{\Omega}$, which is associated with the non-zero trace of the perturbations as we would expect for a scalar degree of freedom.  \par
Indeed, when considering the simpler case where no background matter is present ($\Bar{T}_{\mu\nu}=0$ and $\Omega_0=1$), while still allowing for matter perturbations to occur due to the interplay of curvature and matter in the NMC model, the field equations give
\begin{equation}   \Box\tilde{h}_{\mu\nu}\equiv\Box\left[h_{\mu\nu}-\frac{1}{2}\eta_{\mu\nu}h-\eta_{\mu\nu}\delta{\Omega}\right]=-2(1+f_2)\delta T_{\mu\nu},
\end{equation}
where the redefined perturbation $\tilde{h}_{\mu\nu}$ is transverse ($\partial_\mu \tilde{h}^{\mu\nu}=0$) due to the previously chosen gauge. In the absence of sources, we thus have a wave equation for the quantity $\tilde{h}_{\mu\nu}$, which is initially sourced by the matter perturbations at the origin of the GW signal, while the trace equation directly shows the relation between $\delta\Omega$ and $h$ as $\Box\delta\Omega=-\frac{1}{4}\Box h$. This means that we may solve this equation with the same traceless-transverse solution used in GR. We then invert the definition of $\tilde{h}_{\mu\nu}$ by using $h=-\tilde{h}-4\delta\Omega$ to find the physical metric perturbations  
\begin{equation}
h_{\mu\nu}=\tilde{h}_{\mu\nu}-\frac{1}{2}\eta_{\mu\nu}\tilde{h}-\eta_{\mu\nu}\delta{\Omega}=\tilde{h}_{\mu\nu}-\eta_{\mu\nu}\delta{\Omega},
\end{equation}
where we have taken $\tilde{h}=0$ due to the traceless property taken for the solution. The physical metric perturbation is then given by  
\begin{equation}
    h_{\mu\nu}=\begin{pmatrix}
\delta\Omega & 0 & 0 & 0\\
0 & h_+-\delta \Omega & h_{\times} &0 \\
0 & h_{\times} & -h_+-\delta \Omega &0 \\
0 & 0 & 0 &-\delta \Omega
\end{pmatrix}_{\mu\nu},
\end{equation}
which indicates the presence of luminally propagating breathing and longitudinal polarisation modes induced by the scalar perturbation $\delta\Omega$ along the standard $+$ and $\times$ polarisations present in GR \cite{ScalarWavesf_R,GWPolarisations}. Note that in terms of the scalar perturbations discussed in Section \ref{NMCDynamicsSection}, this would mean $\Psi=-\Phi=\delta\Omega/2$, leading to the exact relation we found from using a gauge-invariant formalism $\Psi-\Phi=\delta\Omega=\delta F$. We also see that this perturbation obeys $h=-4\Omega$, which agrees with the trace equation in vacuum.   \par
However, it is important to keep in mind that the NMC contribution to $\delta\Omega$ follows from two distinct terms, namely $F_2\delta\mathcal{L}_m$ and $F_{2,R}\mathcal{L}_m\delta R$, meaning that the existence of these waves would depend on the presence of background matter to allow for their propagation. When such waves pass through cosmic voids ($\mathcal{L}_m=\delta\mathcal{L}_m=0$), as will inevitably be the case in a Universe where dark energy is a purely gravitational phenomenon, both of these contributions are negligible or even zero. The specific case of NMC GW perturbations over dark energy-like fluids has been analysed in Ref. \cite{NMCGravWaves}, but we do not consider these in this work as the NMC theory has no need for dark energy to generate the accelerated expansion of the Universe \cite{NMCAcceleratedExpansion,NMCHubbleTension}. Of course, the prior reasoning only applies if we consider a purely NMC theory with no modification to the minimal sector through $f_1$, as minimally coupled $f(R)$ theories introduce scalar gravitational radiation polarisations even in vacuum \cite{ScalarWavesf_R}, as these do not vanish by taking $\mathcal{L}_m=\delta\mathcal{L}_m=0$. This means that, although the non-detection of scalar polarisations in future experiments could rule out the validity of minimally coupled $f(R)$ theories, the same cannot be said about their NMC counterparts, as to leading order the scalar polarisations may have been ``washed out" by the passage of the GWs through cosmic voids.

\section{Conclusions}\label{ConclusionsSection}
In this work, we have analysed the properties of metric and matter perturbations in the context of an expanding Universe in a nonminimally coupled theory of gravity. These same perturbations have been thoroughly studied in the context of GR\cite{WaldGR,GWDecompositionBook}, establishing the evolution of small deviations in homogeneous FLRW spacetimes and matter backgrounds and their role in the formation of large-scale structure in the Universe. The latter studies allow for the prediction of the propagation properties of (tensorial) gravitational waves, which radiate at luminal speeds and decay with the expansion of the Universe, as one would expect \cite{SpeedOfGravity,GWDecompositionBook}. GR also predicts the presence of 2 scalar and 2 vectorial degrees of freedom. While the latter do not exhibit any radiative behaviour, evolving independently with respect to time and spatial coordinates, the former behave as one equivalent term with different epoch-dependent properties \cite{GWDecompositionBook}. The corresponding scalar sector term obeys a wave-like equation during the radiation-dominated epoch of the Universe, propagating at the speed of sound in that medium ($c_s^2=1/3$). However, during matter domination ($c_s^2=0$) the time and spatial dependencies separate, leading to a decaying evolution with no oscillatory behaviour. This means that the only radiative behaviour one would expect to detect in a GR-ruled Universe would come from the usual tensorial degrees of freedom, typically separated into $+$ and $\times$ modes \cite{WaldGR}.    \par
When considering the corresponding field equations in the NMC theory, several alterations arise \cite{NMCGravWaves}. Although the tensorial modes still propagate luminally, their effective mass is altered and the gravitational waves decay with an additional factor of $F^{-1/2}$ while oscillating. These changes are in line with the analogous result presented in Ref. \cite{NMCGravWaves}, where the additional time dependence was not obtained due to the assumption of simple constant curvature backgrounds. In late-time conditions, we expect negative exponent terms in a full power series expansion of $f_2$ to dominate \cite{NMCHubbleTension,NMCAcceleratedExpansion}, thus leading to a faster decrease in the wave magnitude as it propagates. We then analysed the production of these waves, which under our assumptions maintain their quadrupolar nature at leading order, while monopole or dipole sources would only induce the scalar polarisations \cite{ScalarWavesf_R}. \par
The evolution of vector perturbations is mostly unaltered, with spatial and temporal dependencies separating as in GR, except for the latter being modified by a factor of $F^{-1}$, which accelerates their attenuation over time. Even though we cannot classify this sector as containing gravitational waves, any effects stemming from these kinds of metric fluctuations would be affected by the nonminimally coupled theory, thus providing means for future testing of the presence of such modifications to GR \cite{LIGOLISAPolarisations,LISAPolarisations}. \par

The scalar sector has the most non-trivial changes, mostly due to its inherent connection with the scalar density perturbations, which directly enter the field and conservation equations via the nonminimal coupling and due to our choice of Lagrangian density $\mathcal{L}_m=-\rho$ \cite{ExtraForce,LagrangianChoice}. This is particularly relevant when considering the relation between the scalar metric perturbations $\Psi$ and $\Phi$, which are no longer simply equivalent as in GR, but rather related by a more complex coupled equation sourced by the variation in $F=F_1+2F_2\mathcal{L}_m$ \cite{NMCCosmologicalPerturbations}. This quantity depends on the matter density and the Ricci scalar, which are affected by the perturbations to the stress-energy content and metric components respectively. However, the same equation can be used to remove all density perturbation derivative terms from the field equations, thus allowing for a better comparison between the GR and NMC dynamics. We derived the modified ``Poisson-like" equation, where the Laplacian term is now related to the gauge-invariant density perturbations via an ``effective gravitational constant" $\tilde{G}=G(1+f_2)/F$ which is unique to the NMC theory \cite{NMCCosmologicalPerturbations}. We analysed the effect of this modification on the weak lensing parameter ($\Sigma$) in examples where $f_2=(R_n/R)^n$, with the $n=4,10$ models chosen from their previously determined behaviour in Ref. \cite{NMCHubbleTension}. We found our predictions to be in good agreement with recent observational constraints.  \par 
A particularity of the modified theory considered in this work is the aforementioned inequality of the scalar metric fluctuations $\Phi$ and $\Psi$, as initially introduced in Ref. \cite{NMCCosmologicalPerturbations} in the context of large-scale structure formation. Although this adds considerable complexity to the analytical determination of properties of the scalar modes, it provides us with a general prediction which could be used in tests of the validity of the NMC theory. We have outlined how this hinges on the detection and distinction of the associated breathing and longitudinal modes, which depend on future space-based gravitational wave interferometry experiments, as the two scalar polarisations produce degenerate responses in ground-based laser interferometer experiments \cite{LISAPolarisations,ScalarDegenerateGroundResponses}. With the predicted creation of future experiments such as LISA, the status of modified theories like the one addressed in this work could thus be put to the test \cite{LISAPolarisations}. Recent interest in the detection of lower and higher frequency signals of the SGWB spectrum could also bring about additional tests of modified gravity theories. Among some of the possibilities discussed in this work are the analysis of PTA data ($\sim$nHz) \cite{PTA_SKAStrongFieldTests} and the proposed adaptation of microwave cavity experiments ($\sim$MHz) \cite{HFGWDetection_Review}. This is a fundamental step in the analysis of the NMC $f(R)$ theory, which has already been shown to fit several observational effects (see, for example, Refs. \cite{NMCAcceleratedExpansion,NMCDarkMatter,NMCHubbleTension}), while only now providing predictions that can be tested without being led by previous results, as any consistent theory must do. \par
We then analysed an alternative method to determine the polarisation spectrum in the NMC theory by making use of the representation of this model as a two-field scalar-tensor theory \cite{NMCScalarTensorAnalogy}. We find that the scalar field associated with the nonminimal coupling induces an additional degree of freedom that mixes curvature and matter perturbations. The scalar degree of freedom is associated with the trace of the metric perturbations and presents itself in the form of breathing and longitudinal polarisation modes propagating luminally, in agreement with what has been found in $f(R)$ theories \cite{ScalarWavesf_R} and in other work on the NMC theory \cite{NMCGravWaves}. However, we discussed the particularities of the propagation of these scalar modes through empty regions such as cosmic voids \cite{CosmicVoids}, in which the absence of background matter to perturb would remove any NMC effects, in particular the scalar polarisations. This only applies to pure NMC models with no minimal $f(R)$ component, meaning that a future lack of detection of scalar GW polarisations from distant sources could provide insight into an experimental distinction between mixed (MC+NMC) and pure nonminimally coupled $f(R)$ theories.   \par
Extensions of the work described here are varied, with perhaps the clearest direction being performing the analysis of the scalar sector under a general regime. Particularly, research on the evolution of the density perturbations has been conducted in Ref. \cite{NMCCosmologicalPerturbations}, where the sub-horizon regime was assumed for simplification of the coupled equations. Once a clearer picture of the evolution of scalar quantities is obtained, a logical extension would be to apply those results, together with those presented here, to the prediction of observable primordial imprints of the different perturbation sectors in cosmological data, such as the CMB power spectrum \cite{PrimordialGWs}. Additionally, applying the Newman-Penrose tetrad formalism to the perturbed NMC theory in a cosmological context could provide further conclusions on the predicted polarisations and their properties, as done in Refs. \cite{NMCGravWaves,ProbingFR_GWs}. 

\begin{acknowledgments}
    The work of one of us (O.B.) is partially supported by FCT (Fundação para a Ciência e Tecnologia, Portugal) through the project CERN/FIS-PAR/0027/2021, with DOI identifier  10.54499/CERN/FIS-PAR/0027/2021.
\end{acknowledgments}

\appendix
\section{Scalar perturbations in the weak NMC regime}\label{WeakNMCRegime}
Decoupling the scalar field equations analytically in an abstract NMC theory with an unspecified $f_2(R)$ is extremely convoluted. Employing a numerical method could be necessary to extract more quantitative results in general scenarios \cite{GWDecompositionBook}, which with the results of this work would simply consist of evolving the coupled differential equations with a robust numerical integrator with sensible initial conditions and a set chosen NMC theory parameters. With this in mind, some results on a concrete example were presented in Section \ref{WorkedExampleSection}. Nevertheless, some additional conclusions on the scalar sector can be reached in the weak regime, in which we take all modifications to be treated perturbatively to linear order in a separate expansion to that of the metric perturbations.  This assumption is not expected to hold under general conditions, as can be seen by considering the hypothetical complete power expansion form of the nonminimal coupling function $f_2$ discussed in Section \ref{NMCSection}. The behaviour of negative and positive exponents of the curvature scalar is such that we expect the strength of the NMC to be dominant in spacetime regions with significantly small (late-time Universe) \cite{NMCAcceleratedExpansion,NMCHubbleTension} or large (early-time Universe) \cite{NMCInflation,NMCInflation2,NMCInflation3} curvatures respectively. However, intermediary situations, such as those expected around the creation of the cosmic microwave background, could provide regions in which the NMC might be treated in the weak regime, as discussed in Ref. \cite{NMCHubbleTension}. The following perturbative analysis, although not general, is therefore still a relevant example to consider when describing some fundamental points of the evolution of the Universe. \par
In order to simplify the distinction between expansion orders, we introduce a new factor $\chi$ multiplying $f_2$. As discussed in Refs. \cite{GWSchwarzschild,GWExtremal}, this method allows us to analyse any modifications to the equations under the light of their GR counterpart, as they are already first-order terms and thus can be evaluated with the zeroth-order equations in mind. This is particularly useful, as the scalar perturbations $\Psi$ and $\Phi$ are equivalent in GR. More specifically, in the perturbative regime we may write
\begin{equation}
    \Psi-\Phi=\frac{\delta F}{F}\approx\delta F=-2\chi(\rho F_{2,R}\delta R+F_2\delta\rho)=\mathcal{O}(\chi),
\end{equation}
which may be introduced into the field Eqs. (\ref{NMCLinearisedFieldEqsDeltaF}). To decouple the equations, we need to remove the $\delta\rho$ dependence, which can be done by using the $(\eta,\eta)$ component of the GR perturbed field equations, as the $\delta\rho$ term above is already $\mathcal{O}(\chi)$. By taking a Fourier wave form for each scalar sector function with $\nabla^2\sim -k^2$ and the explicit expression for $\delta R$ given previously, we can rearrange the equation above as 
\begin{equation}
    \Psi=\Phi+\chi q(\Phi,\Phi',\Phi'')
\end{equation}
or equivalently 
\begin{equation}
    \Phi=\Psi-\chi q(\Psi,\Psi',\Psi''),
\end{equation}
where we have introduced the function $q(\Phi,\Phi',\Phi'')$ for simplicity. This can be applied to the coupled master Eq. (\ref{FullNMCScalarMasterEqs}) by writing $\Psi'=\Phi'+\chi\tilde q(\Phi,\Phi',\Phi'',\Phi^{(3)})$ and equivalently for $\Psi''$, hence removing all $\Psi$ dependence, while now obtaining fourth-order derivative terms in the decoupled master equation. \par
However, any third or fourth-order derivatives are necessarily $\mathcal{O}(\chi)$ as the zeroth-order master equation only included derivatives of up to second order. We may thus apply Eq.(\ref{ScalarWaveEq}) to lower any higher-order derivatives and write the master equation in the form \cite{SpeedOfGravity}
\begin{equation}
\partial_\eta^2\Phi+\beta_1(\eta)\partial_\eta\Phi+\beta_0(\eta,k^2)\Phi+\alpha_0(\eta)\Phi=0,
\end{equation}
from which we can read off the propagation speed $c_{\Phi}^2=\beta_0/k^2$, with equivalent reasoning leading to analogous equations for $\Psi$. This is clear when disregarding the friction term, which simply leads to a modified decay of the wave, after which one may take a wave-like time dependence of the form $e^{i\omega\eta}$ and solve the equation in the large-$k$ limit with $\omega^2=\beta_0+\alpha_0$. This implies the previously asserted propagation speed $c_\Phi^2=\lim_{k\rightarrow\infty}\frac{\omega^2}{k^2}=\beta_0/k^2$, as the $\alpha_0/k^2$ term is negligible in this limit. \par
As expected, these speeds are of the form $c_{\Phi/\Psi}^2=c_s^2+\chi\Delta c_{\Phi/\Psi}^2$, where these are given by
\begin{widetext}
    \begin{equation}
\begin{aligned}
c_{\Phi}^2=c_s^2&-\frac{\chi}{a^2}\left[-2 \mathcal{H}^2 \rho F_{2,R}+2 \left(\mathcal{H}^2+\mathcal{H}'\right)\rho F_{2,R}-6 \mathcal{H}^2  c_s^2 \rho F_{2,R}-12 \left(\mathcal{H}^2+\mathcal{H}'\right)c_s^2 \rho F_{2,R}\right. \\
& +36 \mathcal{H}^2  c_s^4 \rho F_{2,R}-18 \left(\mathcal{H}^2+\mathcal{H}'\right)c_s^4 \rho F_{2,R}-2 \mathcal{H}^2  F_2+2 \left(\mathcal{H}^2+\mathcal{H}'\right)F_2 \\
& -18 \mathcal{H}^2  c_s^2 F_2+6 \left(\mathcal{H}^2+\mathcal{H}'\right)c_s^2 F_2-a^2 \rho F_2+2 a^2 c_s^2 \rho F_2+3 a^2 c_s^4 \rho F_2+ \\
& +2 \mathcal{H}  F_{2,R} \rho^{\prime}-18 \mathcal{H}  c_s^4 F_{2,R} \rho^{\prime}+2 \mathcal{H}  \rho F_{2,R}^{\prime}-18 \mathcal{H}  c_{s}^4 \rho F_{2,R}^{\prime} \\
& -4  \rho^{\prime} F_{2,R}^{\prime}+12  c_{s}^2 \rho^{\prime} F_{2,R}^{\prime}+2 \mathcal{H}  F_2{ }^{\prime}+6 \mathcal{H}  c_{s}^2 F_2^{\prime}-2  F_{2,R} \rho^{\prime \prime} \\
& \left.+6  c_s^2 F_{2,R} \rho^{\prime \prime}-2  \rho F_{2,R}^{\prime \prime}+6  c_{s}^2 \rho F_{2,R}^{\prime \prime}-2  F_2^{\prime \prime}\right] 
\end{aligned}
\end{equation}
\begin{equation}
\begin{aligned}
c_{\Psi}^2=c_s^2&-\frac{\chi}{a^2}\left[-6 \mathcal{H}^2  F_2+2 \left(\mathcal{H}^2+\mathcal{H}'\right)F_2+42 \mathcal{H}^2  c_s^2 F_2-18 \left(\mathcal{H}^2+\mathcal{H}'\right)c_s^2 F_2-a^2 \rho F_2\right. \\
& +2 a^2 c_s^2\rho F_2+3 a^2 c_s^4 \rho F_2-6 \mathcal{H}^2  \rho F_{2, R}+2 \left(\mathcal{H}^2+\mathcal{H}'\right)\rho F_{2, R}+54 \mathcal{H}^2  c_s^2 \rho F_{2, R} \\
& -12 \left(\mathcal{H}^2+\mathcal{H}'\right)c_s^2 \rho F_{2, R}-108 \mathcal{H}^2  c_{s}^4 \rho F_{2, R}+54 \left(\mathcal{H}^2+\mathcal{H}'\right)c_s^4 \rho F_{2, R} \\
& -2 \mathcal{H}  F_{2, R} \rho^{\prime}+18 \mathcal{H}  c_{s}^4 F_{2, R} \rho^{\prime}-2 \mathcal{H}  F_2^{\prime}-6 \mathcal{H}  c_{s}^2 F_2^{\prime}-2 \mathcal{H}  \rho F_{2,R}' \\
& +18 \mathcal{H}  c_{s}^4 \rho F_{2,R}'+4  \rho^{\prime} F_{2,R}'-12  c_{s}^2 \rho^{\prime} F_{2,R}'+2  F_{2, R} \rho^{\prime \prime} \\
& -\left.6  c_s^2 F_{2, R} \rho^{\prime \prime}+2  F_2^{\prime \prime}+2  \rho F_{2, R}{ }^{\prime \prime}-6  c_s^2 \rho F_{2, R}''\right]. 
\end{aligned}
\end{equation}
\end{widetext}

Not only do these speeds match at zeroth order, but they still depend on the speed of sound from the perfect fluid considered for the FLRW background. Although in an RD Universe ($c_s^2=1/3$) the weak NMC corrections would be relatively small in comparison to the GR speed, this is no longer the case in the more recent MD Universe ($c_s^2=0$), where the radiative behaviour would be fully due to the NMC corrections, even when considering the weak regime. This means that we could in principle use the detection of this behaviour of scalar polarisations of GWs via their effects on test particles to test the presence of a nonminimal coupling in the gravitational theory. We should note that we have ignored the possibility of a dark energy-dominated ($\Lambda$D) Universe, as the source of the currently observed accelerated expansion has been shown to be accounted for in the NMC theory with no need for the presence of a cosmological constant \cite{NMCAcceleratedExpansion,NMCHubbleTension}. The difference between the scalar perturbation speeds in the NMC theory ($c_\Psi^2\neq c_\Phi^2$) could provide additional avenues of testing \cite{GWPolarisationSpeeds}.

\bibliography{References}

\begin{thebibliography}{74}%
\makeatletter
\providecommand \@ifxundefined [1]{%
 \@ifx{#1\undefined}
}%
\providecommand \@ifnum [1]{%
 \ifnum #1\expandafter \@firstoftwo
 \else \expandafter \@secondoftwo
 \fi
}%
\providecommand \@ifx [1]{%
 \ifx #1\expandafter \@firstoftwo
 \else \expandafter \@secondoftwo
 \fi
}%
\providecommand \natexlab [1]{#1}%
\providecommand \enquote  [1]{``#1''}%
\providecommand \bibnamefont  [1]{#1}%
\providecommand \bibfnamefont [1]{#1}%
\providecommand \citenamefont [1]{#1}%
\providecommand \href@noop [0]{\@secondoftwo}%
\providecommand \href [0]{\begingroup \@sanitize@url \@href}%
\providecommand \@href[1]{\@@startlink{#1}\@@href}%
\providecommand \@@href[1]{\endgroup#1\@@endlink}%
\providecommand \@sanitize@url [0]{\catcode `\\12\catcode `\$12\catcode `\&12\catcode `\#12\catcode `\^12\catcode `\_12\catcode `\%12\relax}%
\providecommand \@@startlink[1]{}%
\providecommand \@@endlink[0]{}%
\providecommand \url  [0]{\begingroup\@sanitize@url \@url }%
\providecommand \@url [1]{\endgroup\@href {#1}{\urlprefix }}%
\providecommand \urlprefix  [0]{URL }%
\providecommand \Eprint [0]{\href }%
\providecommand \doibase [0]{https://doi.org/}%
\providecommand \selectlanguage [0]{\@gobble}%
\providecommand \bibinfo  [0]{\@secondoftwo}%
\providecommand \bibfield  [0]{\@secondoftwo}%
\providecommand \translation [1]{[#1]}%
\providecommand \BibitemOpen [0]{}%
\providecommand \bibitemStop [0]{}%
\providecommand \bibitemNoStop [0]{.\EOS\space}%
\providecommand \EOS [0]{\spacefactor3000\relax}%
\providecommand \BibitemShut  [1]{\csname bibitem#1\endcsname}%
\let\auto@bib@innerbib\@empty
\bibitem [{\citenamefont {Abbott}\ \emph {et~al.}(2016)\citenamefont {Abbott} \emph {et~al.}}]{GWDetection1st}%
  \BibitemOpen
  \bibfield  {author} {\bibinfo {author} {\bibfnamefont {B.~P.}\ \bibnamefont {Abbott}} \emph {et~al.} (\bibinfo {collaboration} {LIGO Scientific, Virgo}),\ }\bibfield  {title} {\bibinfo {title} {{Observation of Gravitational Waves from a Binary Black Hole Merger}},\ }\href {https://doi.org/10.1103/PhysRevLett.116.061102} {\bibfield  {journal} {\bibinfo  {journal} {Phys. Rev. Lett.}\ }\textbf {\bibinfo {volume} {116}},\ \bibinfo {pages} {061102} (\bibinfo {year} {2016})},\ \Eprint {https://arxiv.org/abs/1602.03837} {arXiv:1602.03837 [gr-qc]} \BibitemShut {NoStop}%
\bibitem [{\citenamefont {Sakellariadou}(2022)}]{GWProbing}%
  \BibitemOpen
  \bibfield  {author} {\bibinfo {author} {\bibfnamefont {M.}~\bibnamefont {Sakellariadou}},\ }\bibfield  {title} {\bibinfo {title} {{Gravitational Waves: The Theorist\textquoteright{}s Swiss Knife}},\ }\href {https://doi.org/10.3390/universe8020132} {\bibfield  {journal} {\bibinfo  {journal} {Universe}\ }\textbf {\bibinfo {volume} {8}},\ \bibinfo {pages} {132} (\bibinfo {year} {2022})},\ \Eprint {https://arxiv.org/abs/2202.00735} {arXiv:2202.00735 [astro-ph.CO]} \BibitemShut {NoStop}%
\bibitem [{\citenamefont {Xu}(2015)}]{GWProbing2}%
  \BibitemOpen
  \bibfield  {author} {\bibinfo {author} {\bibfnamefont {L.}~\bibnamefont {Xu}},\ }\bibfield  {title} {\bibinfo {title} {{Gravitational Waves: A Test for Modified Gravity}},\ }\href {https://doi.org/10.1103/PhysRevD.91.103520} {\bibfield  {journal} {\bibinfo  {journal} {Phys. Rev. D}\ }\textbf {\bibinfo {volume} {91}},\ \bibinfo {pages} {103520} (\bibinfo {year} {2015})},\ \Eprint {https://arxiv.org/abs/1410.6977} {arXiv:1410.6977 [astro-ph.CO]} \BibitemShut {NoStop}%
\bibitem [{\citenamefont {Bogdanos}\ \emph {et~al.}(2010)\citenamefont {Bogdanos}, \citenamefont {Capozziello}, \citenamefont {De~Laurentis},\ and\ \citenamefont {Nesseris}}]{MassiveMasslessGWsHigherOrderGravity}%
  \BibitemOpen
  \bibfield  {author} {\bibinfo {author} {\bibfnamefont {C.}~\bibnamefont {Bogdanos}}, \bibinfo {author} {\bibfnamefont {S.}~\bibnamefont {Capozziello}}, \bibinfo {author} {\bibfnamefont {M.}~\bibnamefont {De~Laurentis}},\ and\ \bibinfo {author} {\bibfnamefont {S.}~\bibnamefont {Nesseris}},\ }\bibfield  {title} {\bibinfo {title} {{Massive, massless and ghost modes of gravitational waves from higher-order gravity}},\ }\href {https://doi.org/10.1016/j.astropartphys.2010.08.001} {\bibfield  {journal} {\bibinfo  {journal} {Astropart. Phys.}\ }\textbf {\bibinfo {volume} {34}},\ \bibinfo {pages} {236} (\bibinfo {year} {2010})},\ \Eprint {https://arxiv.org/abs/0911.3094} {arXiv:0911.3094 [gr-qc]} \BibitemShut {NoStop}%
\bibitem [{\citenamefont {de~Rham}\ \emph {et~al.}(2020)\citenamefont {de~Rham}, \citenamefont {Francfort},\ and\ \citenamefont {Zhang}}]{GWSchwarzschild}%
  \BibitemOpen
  \bibfield  {author} {\bibinfo {author} {\bibfnamefont {C.}~\bibnamefont {de~Rham}}, \bibinfo {author} {\bibfnamefont {J.}~\bibnamefont {Francfort}},\ and\ \bibinfo {author} {\bibfnamefont {J.}~\bibnamefont {Zhang}},\ }\bibfield  {title} {\bibinfo {title} {{Black Hole Gravitational Waves in the Effective Field Theory of Gravity}},\ }\href {https://doi.org/10.1103/PhysRevD.102.024079} {\bibfield  {journal} {\bibinfo  {journal} {Phys. Rev. D}\ }\textbf {\bibinfo {volume} {102}},\ \bibinfo {pages} {024079} (\bibinfo {year} {2020})},\ \Eprint {https://arxiv.org/abs/2005.13923} {arXiv:2005.13923 [hep-th]} \BibitemShut {NoStop}%
\bibitem [{\citenamefont {Barroso~Varela}\ and\ \citenamefont {Rauch}(2024)}]{GWExtremal}%
  \BibitemOpen
  \bibfield  {author} {\bibinfo {author} {\bibfnamefont {M.}~\bibnamefont {Barroso~Varela}}\ and\ \bibinfo {author} {\bibfnamefont {H.}~\bibnamefont {Rauch}},\ }\bibfield  {title} {\bibinfo {title} {{Gravitational waves on charged black hole backgrounds in modified gravity}},\ }\href {https://doi.org/10.1007/s10714-024-03198-9} {\bibfield  {journal} {\bibinfo  {journal} {Gen. Rel. Grav.}\ }\textbf {\bibinfo {volume} {56}},\ \bibinfo {pages} {16} (\bibinfo {year} {2024})},\ \Eprint {https://arxiv.org/abs/2311.07376} {arXiv:2311.07376 [gr-qc]} \BibitemShut {NoStop}%
\bibitem [{\citenamefont {de~Rham}\ and\ \citenamefont {Tolley}(2020)}]{SpeedOfGravity}%
  \BibitemOpen
  \bibfield  {author} {\bibinfo {author} {\bibfnamefont {C.}~\bibnamefont {de~Rham}}\ and\ \bibinfo {author} {\bibfnamefont {A.~J.}\ \bibnamefont {Tolley}},\ }\bibfield  {title} {\bibinfo {title} {{Speed of gravity}},\ }\href {https://doi.org/10.1103/PhysRevD.101.063518} {\bibfield  {journal} {\bibinfo  {journal} {Phys. Rev. D}\ }\textbf {\bibinfo {volume} {101}},\ \bibinfo {pages} {063518} (\bibinfo {year} {2020})},\ \Eprint {https://arxiv.org/abs/1909.00881} {arXiv:1909.00881 [hep-th]} \BibitemShut {NoStop}%
\bibitem [{\citenamefont {de~Rham}\ \emph {et~al.}(2017)\citenamefont {de~Rham}, \citenamefont {Deskins}, \citenamefont {Tolley},\ and\ \citenamefont {Zhou}}]{GravitonMassBounds}%
  \BibitemOpen
  \bibfield  {author} {\bibinfo {author} {\bibfnamefont {C.}~\bibnamefont {de~Rham}}, \bibinfo {author} {\bibfnamefont {J.~T.}\ \bibnamefont {Deskins}}, \bibinfo {author} {\bibfnamefont {A.~J.}\ \bibnamefont {Tolley}},\ and\ \bibinfo {author} {\bibfnamefont {S.-Y.}\ \bibnamefont {Zhou}},\ }\bibfield  {title} {\bibinfo {title} {{Graviton Mass Bounds}},\ }\href {https://doi.org/10.1103/RevModPhys.89.025004} {\bibfield  {journal} {\bibinfo  {journal} {Rev. Mod. Phys.}\ }\textbf {\bibinfo {volume} {89}},\ \bibinfo {pages} {025004} (\bibinfo {year} {2017})},\ \Eprint {https://arxiv.org/abs/1606.08462} {arXiv:1606.08462 [astro-ph.CO]} \BibitemShut {NoStop}%
\bibitem [{\citenamefont {Cornish}\ \emph {et~al.}(2017)\citenamefont {Cornish}, \citenamefont {Blas},\ and\ \citenamefont {Nardini}}]{GWSpeedBound}%
  \BibitemOpen
  \bibfield  {author} {\bibinfo {author} {\bibfnamefont {N.}~\bibnamefont {Cornish}}, \bibinfo {author} {\bibfnamefont {D.}~\bibnamefont {Blas}},\ and\ \bibinfo {author} {\bibfnamefont {G.}~\bibnamefont {Nardini}},\ }\bibfield  {title} {\bibinfo {title} {{Bounding the speed of gravity with gravitational wave observations}},\ }\href {https://doi.org/10.1103/PhysRevLett.119.161102} {\bibfield  {journal} {\bibinfo  {journal} {Phys. Rev. Lett.}\ }\textbf {\bibinfo {volume} {119}},\ \bibinfo {pages} {161102} (\bibinfo {year} {2017})},\ \Eprint {https://arxiv.org/abs/1707.06101} {arXiv:1707.06101 [gr-qc]} \BibitemShut {NoStop}%
\bibitem [{\citenamefont {Abbott}\ \emph {et~al.}(2018)\citenamefont {Abbott} \emph {et~al.}}]{LIGOLISAPolarisations}%
  \BibitemOpen
  \bibfield  {author} {\bibinfo {author} {\bibfnamefont {B.~P.}\ \bibnamefont {Abbott}} \emph {et~al.} (\bibinfo {collaboration} {LIGO Scientific, Virgo}),\ }\bibfield  {title} {\bibinfo {title} {{Search for Tensor, Vector, and Scalar Polarizations in the Stochastic Gravitational-Wave Background}},\ }\href {https://doi.org/10.1103/PhysRevLett.120.201102} {\bibfield  {journal} {\bibinfo  {journal} {Phys. Rev. Lett.}\ }\textbf {\bibinfo {volume} {120}},\ \bibinfo {pages} {201102} (\bibinfo {year} {2018})},\ \Eprint {https://arxiv.org/abs/1802.10194} {arXiv:1802.10194 [gr-qc]} \BibitemShut {NoStop}%
\bibitem [{\citenamefont {Isi}\ \emph {et~al.}(2015)\citenamefont {Isi}, \citenamefont {Weinstein}, \citenamefont {Mead},\ and\ \citenamefont {Pitkin}}]{LIGOLISAPolarisations2}%
  \BibitemOpen
  \bibfield  {author} {\bibinfo {author} {\bibfnamefont {M.}~\bibnamefont {Isi}}, \bibinfo {author} {\bibfnamefont {A.~J.}\ \bibnamefont {Weinstein}}, \bibinfo {author} {\bibfnamefont {C.}~\bibnamefont {Mead}},\ and\ \bibinfo {author} {\bibfnamefont {M.}~\bibnamefont {Pitkin}},\ }\bibfield  {title} {\bibinfo {title} {{Detecting Beyond-Einstein Polarizations of Continuous Gravitational Waves}},\ }\href {https://doi.org/10.1103/PhysRevD.91.082002} {\bibfield  {journal} {\bibinfo  {journal} {Phys. Rev. D}\ }\textbf {\bibinfo {volume} {91}},\ \bibinfo {pages} {082002} (\bibinfo {year} {2015})},\ \Eprint {https://arxiv.org/abs/1502.00333} {arXiv:1502.00333 [gr-qc]} \BibitemShut {NoStop}%
\bibitem [{\citenamefont {Mastrogiovanni}\ \emph {et~al.}(2020)\citenamefont {Mastrogiovanni}, \citenamefont {Steer},\ and\ \citenamefont {Barsuglia}}]{CosmographyGWs}%
  \BibitemOpen
  \bibfield  {author} {\bibinfo {author} {\bibfnamefont {S.}~\bibnamefont {Mastrogiovanni}}, \bibinfo {author} {\bibfnamefont {D.~A.}\ \bibnamefont {Steer}},\ and\ \bibinfo {author} {\bibfnamefont {M.}~\bibnamefont {Barsuglia}},\ }\bibfield  {title} {\bibinfo {title} {Probing modified gravity theories and cosmology using gravitational-waves and associated electromagnetic counterparts},\ }\href {https://doi.org/10.1103/PhysRevD.102.044009} {\bibfield  {journal} {\bibinfo  {journal} {Phys. Rev. D}\ }\textbf {\bibinfo {volume} {102}},\ \bibinfo {pages} {044009} (\bibinfo {year} {2020})}\BibitemShut {NoStop}%
\bibitem [{\citenamefont {Dong}\ \emph {et~al.}(2024)\citenamefont {Dong}, \citenamefont {Liua},\ and\ \citenamefont {Liu}}]{GWModes}%
  \BibitemOpen
  \bibfield  {author} {\bibinfo {author} {\bibfnamefont {Y.-Q.}\ \bibnamefont {Dong}}, \bibinfo {author} {\bibfnamefont {Y.-Q.}\ \bibnamefont {Liua}},\ and\ \bibinfo {author} {\bibfnamefont {Y.-X.}\ \bibnamefont {Liu}},\ }\bibfield  {title} {\bibinfo {title} {{Polarization modes of gravitational waves in general modified gravity: General metric theory and general scalar-tensor theory}},\ }\href {https://doi.org/10.1103/PhysRevD.109.044013} {\bibfield  {journal} {\bibinfo  {journal} {Phys. Rev. D}\ }\textbf {\bibinfo {volume} {109}},\ \bibinfo {pages} {044013} (\bibinfo {year} {2024})},\ \Eprint {https://arxiv.org/abs/2310.11336} {arXiv:2310.11336 [gr-qc]} \BibitemShut {NoStop}%
\bibitem [{\citenamefont {Schumacher}\ \emph {et~al.}(2023)\citenamefont {Schumacher}, \citenamefont {Yunes},\ and\ \citenamefont {Yagi}}]{GWPolarisationSpeeds}%
  \BibitemOpen
  \bibfield  {author} {\bibinfo {author} {\bibfnamefont {K.}~\bibnamefont {Schumacher}}, \bibinfo {author} {\bibfnamefont {N.}~\bibnamefont {Yunes}},\ and\ \bibinfo {author} {\bibfnamefont {K.}~\bibnamefont {Yagi}},\ }\bibfield  {title} {\bibinfo {title} {Gravitational wave polarizations with different propagation speeds},\ }\href {https://doi.org/10.1103/PhysRevD.108.104038} {\bibfield  {journal} {\bibinfo  {journal} {Phys. Rev. D}\ }\textbf {\bibinfo {volume} {108}},\ \bibinfo {pages} {104038} (\bibinfo {year} {2023})}\BibitemShut {NoStop}%
\bibitem [{\citenamefont {Maggiore}(2018)}]{GWDecompositionBook}%
  \BibitemOpen
  \bibfield  {author} {\bibinfo {author} {\bibfnamefont {M.}~\bibnamefont {Maggiore}},\ }\href@noop {} {\emph {\bibinfo {title} {{Gravitational Waves. Vol. 2: Astrophysics and Cosmology}}}}\ (\bibinfo  {publisher} {Oxford University Press},\ \bibinfo {year} {2018})\BibitemShut {NoStop}%
\bibitem [{\citenamefont {Flanagan}\ and\ \citenamefont {Hughes}(2005)}]{GWPolarisations}%
  \BibitemOpen
  \bibfield  {author} {\bibinfo {author} {\bibfnamefont {E.~E.}\ \bibnamefont {Flanagan}}\ and\ \bibinfo {author} {\bibfnamefont {S.~A.}\ \bibnamefont {Hughes}},\ }\bibfield  {title} {\bibinfo {title} {{The Basics of gravitational wave theory}},\ }\href {https://doi.org/10.1088/1367-2630/7/1/204} {\bibfield  {journal} {\bibinfo  {journal} {New J. Phys.}\ }\textbf {\bibinfo {volume} {7}},\ \bibinfo {pages} {204} (\bibinfo {year} {2005})},\ \Eprint {https://arxiv.org/abs/gr-qc/0501041} {arXiv:gr-qc/0501041} \BibitemShut {NoStop}%
\bibitem [{\citenamefont {Newman}\ and\ \citenamefont {Penrose}(1962)}]{NPFormalism}%
  \BibitemOpen
  \bibfield  {author} {\bibinfo {author} {\bibfnamefont {E.}~\bibnamefont {Newman}}\ and\ \bibinfo {author} {\bibfnamefont {R.}~\bibnamefont {Penrose}},\ }\bibfield  {title} {\bibinfo {title} {{An Approach to gravitational radiation by a method of spin coefficients}},\ }\href {https://doi.org/10.1063/1.1724257} {\bibfield  {journal} {\bibinfo  {journal} {J. Math. Phys.}\ }\textbf {\bibinfo {volume} {3}},\ \bibinfo {pages} {566} (\bibinfo {year} {1962})}\BibitemShut {NoStop}%
\bibitem [{\citenamefont {Bertolami}\ \emph {et~al.}(2018)\citenamefont {Bertolami}, \citenamefont {Gomes},\ and\ \citenamefont {Lobo}}]{NMCGravWaves}%
  \BibitemOpen
  \bibfield  {author} {\bibinfo {author} {\bibfnamefont {O.}~\bibnamefont {Bertolami}}, \bibinfo {author} {\bibfnamefont {C.}~\bibnamefont {Gomes}},\ and\ \bibinfo {author} {\bibfnamefont {F.~S.~N.}\ \bibnamefont {Lobo}},\ }\bibfield  {title} {\bibinfo {title} {{Gravitational waves in theories with a non-minimal curvature-matter coupling}},\ }\href {https://doi.org/10.1140/epjc/s10052-018-5781-5} {\bibfield  {journal} {\bibinfo  {journal} {Eur. Phys. J. C}\ }\textbf {\bibinfo {volume} {78}},\ \bibinfo {pages} {303} (\bibinfo {year} {2018})},\ \Eprint {https://arxiv.org/abs/1706.06826} {arXiv:1706.06826 [gr-qc]} \BibitemShut {NoStop}%
\bibitem [{\citenamefont {Alves}\ \emph {et~al.}(2009)\citenamefont {Alves}, \citenamefont {Miranda},\ and\ \citenamefont {de~Araujo}}]{ProbingFR_GWs}%
  \BibitemOpen
  \bibfield  {author} {\bibinfo {author} {\bibfnamefont {M.~E.~S.}\ \bibnamefont {Alves}}, \bibinfo {author} {\bibfnamefont {O.~D.}\ \bibnamefont {Miranda}},\ and\ \bibinfo {author} {\bibfnamefont {J.~C.~N.}\ \bibnamefont {de~Araujo}},\ }\bibfield  {title} {\bibinfo {title} {{Probing the f(R) formalism through gravitational wave polarizations}},\ }\href {https://doi.org/10.1016/j.physletb.2009.08.005} {\bibfield  {journal} {\bibinfo  {journal} {Phys. Lett. B}\ }\textbf {\bibinfo {volume} {679}},\ \bibinfo {pages} {401} (\bibinfo {year} {2009})},\ \Eprint {https://arxiv.org/abs/0908.0861} {arXiv:0908.0861 [gr-qc]} \BibitemShut {NoStop}%
\bibitem [{\citenamefont {Liang}\ \emph {et~al.}(2017)\citenamefont {Liang}, \citenamefont {Gong}, \citenamefont {Hou},\ and\ \citenamefont {Liu}}]{FR_GWs}%
  \BibitemOpen
  \bibfield  {author} {\bibinfo {author} {\bibfnamefont {D.}~\bibnamefont {Liang}}, \bibinfo {author} {\bibfnamefont {Y.}~\bibnamefont {Gong}}, \bibinfo {author} {\bibfnamefont {S.}~\bibnamefont {Hou}},\ and\ \bibinfo {author} {\bibfnamefont {Y.}~\bibnamefont {Liu}},\ }\bibfield  {title} {\bibinfo {title} {{Polarizations of gravitational waves in $f(R)$ gravity}},\ }\href {https://doi.org/10.1103/PhysRevD.95.104034} {\bibfield  {journal} {\bibinfo  {journal} {Phys. Rev. D}\ }\textbf {\bibinfo {volume} {95}},\ \bibinfo {pages} {104034} (\bibinfo {year} {2017})},\ \Eprint {https://arxiv.org/abs/1701.05998} {arXiv:1701.05998 [gr-qc]} \BibitemShut {NoStop}%
\bibitem [{\citenamefont {Rizwana~Kausar}\ \emph {et~al.}(2017)\citenamefont {Rizwana~Kausar}, \citenamefont {Philippoz},\ and\ \citenamefont {Jetzer}}]{FRGWs}%
  \BibitemOpen
  \bibfield  {author} {\bibinfo {author} {\bibfnamefont {H.}~\bibnamefont {Rizwana~Kausar}}, \bibinfo {author} {\bibfnamefont {L.}~\bibnamefont {Philippoz}},\ and\ \bibinfo {author} {\bibfnamefont {P.}~\bibnamefont {Jetzer}},\ }\bibfield  {title} {\bibinfo {title} {{Gravitational wave polarization modes in $f(R)$ theories}},\ }in\ \href {https://doi.org/10.1142/9789813226609_0088} {\emph {\bibinfo {booktitle} {{14th Marcel Grossmann Meeting on Recent Developments in Theoretical and Experimental General Relativity, Astrophysics, and Relativistic Field Theories}}}},\ Vol.~\bibinfo {volume} {2}\ (\bibinfo {year} {2017})\ pp.\ \bibinfo {pages} {1220--1226}\BibitemShut {NoStop}%
\bibitem [{\citenamefont {Katsuragawa}\ \emph {et~al.}(2019)\citenamefont {Katsuragawa}, \citenamefont {Nakamura}, \citenamefont {Ikeda},\ and\ \citenamefont {Capozziello}}]{FRGWs2}%
  \BibitemOpen
  \bibfield  {author} {\bibinfo {author} {\bibfnamefont {T.}~\bibnamefont {Katsuragawa}}, \bibinfo {author} {\bibfnamefont {T.}~\bibnamefont {Nakamura}}, \bibinfo {author} {\bibfnamefont {T.}~\bibnamefont {Ikeda}},\ and\ \bibinfo {author} {\bibfnamefont {S.}~\bibnamefont {Capozziello}},\ }\bibfield  {title} {\bibinfo {title} {{Gravitational Waves in $F(R)$ Gravity: Scalar Waves and the Chameleon Mechanism}},\ }\href {https://doi.org/10.1103/PhysRevD.99.124050} {\bibfield  {journal} {\bibinfo  {journal} {Phys. Rev. D}\ }\textbf {\bibinfo {volume} {99}},\ \bibinfo {pages} {124050} (\bibinfo {year} {2019})},\ \Eprint {https://arxiv.org/abs/1902.02494} {arXiv:1902.02494 [gr-qc]} \BibitemShut {NoStop}%
\bibitem [{\citenamefont {Bertolami}\ \emph {et~al.}(2007)\citenamefont {Bertolami}, \citenamefont {Boehmer}, \citenamefont {Harko},\ and\ \citenamefont {Lobo}}]{ExtraForce}%
  \BibitemOpen
  \bibfield  {author} {\bibinfo {author} {\bibfnamefont {O.}~\bibnamefont {Bertolami}}, \bibinfo {author} {\bibfnamefont {C.~G.}\ \bibnamefont {Boehmer}}, \bibinfo {author} {\bibfnamefont {T.}~\bibnamefont {Harko}},\ and\ \bibinfo {author} {\bibfnamefont {F.~S.~N.}\ \bibnamefont {Lobo}},\ }\bibfield  {title} {\bibinfo {title} {{Extra force in f(R) modified theories of gravity}},\ }\href {https://doi.org/10.1103/PhysRevD.75.104016} {\bibfield  {journal} {\bibinfo  {journal} {Phys. Rev. D}\ }\textbf {\bibinfo {volume} {75}},\ \bibinfo {pages} {104016} (\bibinfo {year} {2007})},\ \Eprint {https://arxiv.org/abs/0704.1733} {arXiv:0704.1733 [gr-qc]} \BibitemShut {NoStop}%
\bibitem [{\citenamefont {Bertolami}\ \emph {et~al.}(2012)\citenamefont {Bertolami}, \citenamefont {Fraz\~ao},\ and\ \citenamefont {P\'aramos}}]{NMCDarkMatter}%
  \BibitemOpen
  \bibfield  {author} {\bibinfo {author} {\bibfnamefont {O.}~\bibnamefont {Bertolami}}, \bibinfo {author} {\bibfnamefont {P.}~\bibnamefont {Fraz\~ao}},\ and\ \bibinfo {author} {\bibfnamefont {J.}~\bibnamefont {P\'aramos}},\ }\bibfield  {title} {\bibinfo {title} {{Mimicking dark matter in galaxy clusters through a non-minimal gravitational coupling with matter}},\ }\href {https://doi.org/10.1103/PhysRevD.86.044034} {\bibfield  {journal} {\bibinfo  {journal} {Phys. Rev. D}\ }\textbf {\bibinfo {volume} {86}},\ \bibinfo {pages} {044034} (\bibinfo {year} {2012})},\ \Eprint {https://arxiv.org/abs/1111.3167} {arXiv:1111.3167 [gr-qc]} \BibitemShut {NoStop}%
\bibitem [{\citenamefont {Bertolami}\ and\ \citenamefont {P\'aramos}(2010)}]{NMCDarkMatter2}%
  \BibitemOpen
  \bibfield  {author} {\bibinfo {author} {\bibfnamefont {O.}~\bibnamefont {Bertolami}}\ and\ \bibinfo {author} {\bibfnamefont {J.}~\bibnamefont {P\'aramos}},\ }\bibfield  {title} {\bibinfo {title} {{Mimicking dark matter through a non-minimal gravitational coupling with matter}},\ }\href {https://doi.org/10.1088/1475-7516/2010/03/009} {\bibfield  {journal} {\bibinfo  {journal} {JCAP}\ }\textbf {\bibinfo {volume} {03}},\ \bibinfo {pages} {009}},\ \Eprint {https://arxiv.org/abs/0906.4757} {arXiv:0906.4757 [astro-ph.GA]} \BibitemShut {NoStop}%
\bibitem [{\citenamefont {March}\ \emph {et~al.}(2019)\citenamefont {March}, \citenamefont {Bertolami}, \citenamefont {P\'aramos},\ and\ \citenamefont {Dell'Agnello}}]{NMCSolarSystem}%
  \BibitemOpen
  \bibfield  {author} {\bibinfo {author} {\bibfnamefont {R.}~\bibnamefont {March}}, \bibinfo {author} {\bibfnamefont {O.}~\bibnamefont {Bertolami}}, \bibinfo {author} {\bibfnamefont {J.}~\bibnamefont {P\'aramos}},\ and\ \bibinfo {author} {\bibfnamefont {S.}~\bibnamefont {Dell'Agnello}},\ }\bibfield  {title} {\bibinfo {title} {{Nonminimally coupled curvature-matter gravity models and Solar System constraints}},\ }in\ \href {https://doi.org/10.1142/9789811258251_0104} {\emph {\bibinfo {booktitle} {{15th Marcel Grossmann Meeting on Recent Developments in Theoretical and Experimental General Relativity, Astrophysics, and Relativistic Field Theories}}}}\ (\bibinfo {year} {2019})\ \Eprint {https://arxiv.org/abs/1903.07059} {arXiv:1903.07059 [gr-qc]} \BibitemShut {NoStop}%
\bibitem [{\citenamefont {March}\ \emph {et~al.}(2017)\citenamefont {March}, \citenamefont {P\'aramos}, \citenamefont {Bertolami},\ and\ \citenamefont {Dell'Agnello}}]{NMCSolarSystem2}%
  \BibitemOpen
  \bibfield  {author} {\bibinfo {author} {\bibfnamefont {R.}~\bibnamefont {March}}, \bibinfo {author} {\bibfnamefont {J.}~\bibnamefont {P\'aramos}}, \bibinfo {author} {\bibfnamefont {O.}~\bibnamefont {Bertolami}},\ and\ \bibinfo {author} {\bibfnamefont {S.}~\bibnamefont {Dell'Agnello}},\ }\bibfield  {title} {\bibinfo {title} {{Nonminimally coupled gravity and planetary motion}},\ }\href@noop {} {\bibfield  {journal} {\bibinfo  {journal} {Frascati Phys. Ser.}\ }\textbf {\bibinfo {volume} {64}},\ \bibinfo {pages} {33} (\bibinfo {year} {2017})}\BibitemShut {NoStop}%
\bibitem [{\citenamefont {March}\ \emph {et~al.}(2024)\citenamefont {March}, \citenamefont {Bertolami}, \citenamefont {Muccino},\ and\ \citenamefont {Dell'Agnello}}]{NMCSolarSystem3}%
  \BibitemOpen
  \bibfield  {author} {\bibinfo {author} {\bibfnamefont {R.}~\bibnamefont {March}}, \bibinfo {author} {\bibfnamefont {O.}~\bibnamefont {Bertolami}}, \bibinfo {author} {\bibfnamefont {M.}~\bibnamefont {Muccino}},\ and\ \bibinfo {author} {\bibfnamefont {S.}~\bibnamefont {Dell'Agnello}},\ }\bibfield  {title} {\bibinfo {title} {{Equivalence principle violation in nonminimally coupled gravity and constraints from lunar laser ranging}},\ }\href {https://doi.org/10.1103/PhysRevD.109.124013} {\bibfield  {journal} {\bibinfo  {journal} {Phys. Rev. D}\ }\textbf {\bibinfo {volume} {109}},\ \bibinfo {pages} {124013} (\bibinfo {year} {2024})},\ \Eprint {https://arxiv.org/abs/2312.14618} {arXiv:2312.14618 [gr-qc]} \BibitemShut {NoStop}%
\bibitem [{\citenamefont {Bertolami}\ \emph {et~al.}(2016)\citenamefont {Bertolami}, \citenamefont {Bessa},\ and\ \citenamefont {P\'aramos}}]{NMCInflation}%
  \BibitemOpen
  \bibfield  {author} {\bibinfo {author} {\bibfnamefont {O.}~\bibnamefont {Bertolami}}, \bibinfo {author} {\bibfnamefont {V.}~\bibnamefont {Bessa}},\ and\ \bibinfo {author} {\bibfnamefont {J.}~\bibnamefont {P\'aramos}},\ }\bibfield  {title} {\bibinfo {title} {{Inflation with a massive vector field nonminimally coupled to gravity}},\ }\href {https://doi.org/10.1103/PhysRevD.93.064002} {\bibfield  {journal} {\bibinfo  {journal} {Phys. Rev. D}\ }\textbf {\bibinfo {volume} {93}},\ \bibinfo {pages} {064002} (\bibinfo {year} {2016})},\ \Eprint {https://arxiv.org/abs/1511.03520} {arXiv:1511.03520 [gr-qc]} \BibitemShut {NoStop}%
\bibitem [{\citenamefont {Bertolami}\ \emph {et~al.}(2011)\citenamefont {Bertolami}, \citenamefont {Fraz\~ao},\ and\ \citenamefont {Páramos}}]{NMCInflation2}%
  \BibitemOpen
  \bibfield  {author} {\bibinfo {author} {\bibfnamefont {O.}~\bibnamefont {Bertolami}}, \bibinfo {author} {\bibfnamefont {P.}~\bibnamefont {Fraz\~ao}},\ and\ \bibinfo {author} {\bibfnamefont {J.}~\bibnamefont {Páramos}},\ }\bibfield  {title} {\bibinfo {title} {{Reheating via a generalized non-minimal coupling of curvature to matter}},\ }\href {https://doi.org/10.1103/PhysRevD.83.044010} {\bibfield  {journal} {\bibinfo  {journal} {Phys. Rev. D}\ }\textbf {\bibinfo {volume} {83}},\ \bibinfo {pages} {044010} (\bibinfo {year} {2011})},\ \Eprint {https://arxiv.org/abs/1010.2698} {arXiv:1010.2698 [gr-qc]} \BibitemShut {NoStop}%
\bibitem [{\citenamefont {Gomes}\ \emph {et~al.}(2017)\citenamefont {Gomes}, \citenamefont {Rosa},\ and\ \citenamefont {Bertolami}}]{NMCInflation3}%
  \BibitemOpen
  \bibfield  {author} {\bibinfo {author} {\bibfnamefont {C.}~\bibnamefont {Gomes}}, \bibinfo {author} {\bibfnamefont {J.~a.~G.}\ \bibnamefont {Rosa}},\ and\ \bibinfo {author} {\bibfnamefont {O.}~\bibnamefont {Bertolami}},\ }\bibfield  {title} {\bibinfo {title} {{Inflation in non-minimal matter-curvature coupling theories}},\ }\href {https://doi.org/10.1088/1475-7516/2017/06/021} {\bibfield  {journal} {\bibinfo  {journal} {JCAP}\ }\textbf {\bibinfo {volume} {06}},\ \bibinfo {pages} {021}},\ \Eprint {https://arxiv.org/abs/1611.02124} {arXiv:1611.02124 [gr-qc]} \BibitemShut {NoStop}%
\bibitem [{\citenamefont {Bertolami}\ \emph {et~al.}(2013)\citenamefont {Bertolami}, \citenamefont {Fraz\~ao},\ and\ \citenamefont {P\'aramos}}]{NMCCosmologicalPerturbations}%
  \BibitemOpen
  \bibfield  {author} {\bibinfo {author} {\bibfnamefont {O.}~\bibnamefont {Bertolami}}, \bibinfo {author} {\bibfnamefont {P.}~\bibnamefont {Fraz\~ao}},\ and\ \bibinfo {author} {\bibfnamefont {J.}~\bibnamefont {P\'aramos}},\ }\bibfield  {title} {\bibinfo {title} {{Cosmological perturbations in theories with non-minimal coupling between curvature and matter}},\ }\href {https://doi.org/10.1088/1475-7516/2013/05/029} {\bibfield  {journal} {\bibinfo  {journal} {JCAP}\ }\textbf {\bibinfo {volume} {05}},\ \bibinfo {pages} {029}},\ \Eprint {https://arxiv.org/abs/1303.3215} {arXiv:1303.3215 [gr-qc]} \BibitemShut {NoStop}%
\bibitem [{\citenamefont {Di~Valentino}\ \emph {et~al.}(2021)\citenamefont {Di~Valentino}, \citenamefont {Mena}, \citenamefont {Pan}, \citenamefont {Visinelli}, \citenamefont {Yang}, \citenamefont {Melchiorri}, \citenamefont {Mota}, \citenamefont {Riess},\ and\ \citenamefont {Silk}}]{HubbleTensionReview}%
  \BibitemOpen
  \bibfield  {author} {\bibinfo {author} {\bibfnamefont {E.}~\bibnamefont {Di~Valentino}}, \bibinfo {author} {\bibfnamefont {O.}~\bibnamefont {Mena}}, \bibinfo {author} {\bibfnamefont {S.}~\bibnamefont {Pan}}, \bibinfo {author} {\bibfnamefont {L.}~\bibnamefont {Visinelli}}, \bibinfo {author} {\bibfnamefont {W.}~\bibnamefont {Yang}}, \bibinfo {author} {\bibfnamefont {A.}~\bibnamefont {Melchiorri}}, \bibinfo {author} {\bibfnamefont {D.~F.}\ \bibnamefont {Mota}}, \bibinfo {author} {\bibfnamefont {A.~G.}\ \bibnamefont {Riess}},\ and\ \bibinfo {author} {\bibfnamefont {J.}~\bibnamefont {Silk}},\ }\bibfield  {title} {\bibinfo {title} {{In the realm of the Hubble tension\textemdash{}a review of solutions}},\ }\href {https://doi.org/10.1088/1361-6382/ac086d} {\bibfield  {journal} {\bibinfo  {journal} {Class. Quant. Grav.}\ }\textbf {\bibinfo {volume} {38}},\ \bibinfo {pages} {153001} (\bibinfo {year} {2021})},\ \Eprint {https://arxiv.org/abs/2103.01183} {arXiv:2103.01183 [astro-ph.CO]} \BibitemShut {NoStop}%
\bibitem [{\citenamefont {Barroso~Varela}\ and\ \citenamefont {Bertolami}(2024)}]{NMCHubbleTension}%
  \BibitemOpen
  \bibfield  {author} {\bibinfo {author} {\bibfnamefont {M.}~\bibnamefont {Barroso~Varela}}\ and\ \bibinfo {author} {\bibfnamefont {O.}~\bibnamefont {Bertolami}},\ }\bibfield  {title} {\bibinfo {title} {{Hubble tension in a nonminimally coupled curvature-matter gravity model}},\ }\href {https://doi.org/10.1088/1475-7516/2024/06/025} {\bibfield  {journal} {\bibinfo  {journal} {JCAP}\ }\textbf {\bibinfo {volume} {06}},\ \bibinfo {pages} {025}},\ \Eprint {https://arxiv.org/abs/2403.11683} {arXiv:2403.11683 [gr-qc]} \BibitemShut {NoStop}%
\bibitem [{\citenamefont {Bertolami}\ and\ \citenamefont {P\'aramos}(2015)}]{LagrangianChoice}%
  \BibitemOpen
  \bibfield  {author} {\bibinfo {author} {\bibfnamefont {O.}~\bibnamefont {Bertolami}}\ and\ \bibinfo {author} {\bibfnamefont {J.}~\bibnamefont {P\'aramos}},\ }\bibfield  {title} {\bibinfo {title} {{Homogeneous spherically symmetric bodies with a non-minimal coupling between curvature and matter: the choice of the Lagrangian density for matter}},\ }\href {https://doi.org/10.1007/s10714-014-1835-7} {\bibfield  {journal} {\bibinfo  {journal} {Gen. Rel. Grav.}\ }\textbf {\bibinfo {volume} {47}},\ \bibinfo {pages} {1835} (\bibinfo {year} {2015})},\ \Eprint {https://arxiv.org/abs/1306.1177} {arXiv:1306.1177 [gr-qc]} \BibitemShut {NoStop}%
\bibitem [{\citenamefont {Bertolami}\ \emph {et~al.}(2008)\citenamefont {Bertolami}, \citenamefont {Lobo},\ and\ \citenamefont {P\'aramos}}]{LagrangianForm}%
  \BibitemOpen
  \bibfield  {author} {\bibinfo {author} {\bibfnamefont {O.}~\bibnamefont {Bertolami}}, \bibinfo {author} {\bibfnamefont {F.~S.~N.}\ \bibnamefont {Lobo}},\ and\ \bibinfo {author} {\bibfnamefont {J.}~\bibnamefont {P\'aramos}},\ }\bibfield  {title} {\bibinfo {title} {Nonminimal coupling of perfect fluids to curvature},\ }\href {https://doi.org/10.1103/PhysRevD.78.064036} {\bibfield  {journal} {\bibinfo  {journal} {Phys. Rev. D}\ }\textbf {\bibinfo {volume} {78}},\ \bibinfo {pages} {064036} (\bibinfo {year} {2008})}\BibitemShut {NoStop}%
\bibitem [{\citenamefont {Brown}(1993)}]{LagrangianChoice2}%
  \BibitemOpen
  \bibfield  {author} {\bibinfo {author} {\bibfnamefont {J.~D.}\ \bibnamefont {Brown}},\ }\bibfield  {title} {\bibinfo {title} {Action functionals for relativistic perfect fluids},\ }\href {https://doi.org/10.1088/0264-9381/10/8/017} {\bibfield  {journal} {\bibinfo  {journal} {Classical and Quantum Gravity}\ }\textbf {\bibinfo {volume} {10}},\ \bibinfo {pages} {1579} (\bibinfo {year} {1993})}\BibitemShut {NoStop}%
\bibitem [{\citenamefont {Bertolami}\ and\ \citenamefont {P\'aramos}(2014)}]{NMCFriedmann}%
  \BibitemOpen
  \bibfield  {author} {\bibinfo {author} {\bibfnamefont {O.}~\bibnamefont {Bertolami}}\ and\ \bibinfo {author} {\bibfnamefont {J.}~\bibnamefont {P\'aramos}},\ }\bibfield  {title} {\bibinfo {title} {{Modified Friedmann Equation from Nonminimally Coupled Theories of Gravity}},\ }\href {https://doi.org/10.1103/PhysRevD.89.044012} {\bibfield  {journal} {\bibinfo  {journal} {Phys. Rev. D}\ }\textbf {\bibinfo {volume} {89}},\ \bibinfo {pages} {044012} (\bibinfo {year} {2014})},\ \Eprint {https://arxiv.org/abs/1311.5615} {arXiv:1311.5615 [gr-qc]} \BibitemShut {NoStop}%
\bibitem [{\citenamefont {Bertolami}\ \emph {et~al.}(2010)\citenamefont {Bertolami}, \citenamefont {Fraz\~ao},\ and\ \citenamefont {P\'aramos}}]{NMCAcceleratedExpansion}%
  \BibitemOpen
  \bibfield  {author} {\bibinfo {author} {\bibfnamefont {O.}~\bibnamefont {Bertolami}}, \bibinfo {author} {\bibfnamefont {P.}~\bibnamefont {Fraz\~ao}},\ and\ \bibinfo {author} {\bibfnamefont {J.}~\bibnamefont {P\'aramos}},\ }\bibfield  {title} {\bibinfo {title} {{Accelerated expansion from a non-minimal gravitational coupling to matter}},\ }\href {https://doi.org/10.1103/PhysRevD.81.104046} {\bibfield  {journal} {\bibinfo  {journal} {Phys. Rev. D}\ }\textbf {\bibinfo {volume} {81}},\ \bibinfo {pages} {104046} (\bibinfo {year} {2010})},\ \Eprint {https://arxiv.org/abs/1003.0850} {arXiv:1003.0850 [gr-qc]} \BibitemShut {NoStop}%
\bibitem [{\citenamefont {Wald}(1984)}]{WaldGR}%
  \BibitemOpen
  \bibfield  {author} {\bibinfo {author} {\bibfnamefont {R.~M.}\ \bibnamefont {Wald}},\ }\href {https://doi.org/10.7208/chicago/9780226870373.001.0001} {\emph {\bibinfo {title} {{General Relativity}}}}\ (\bibinfo  {publisher} {Chicago Univ. Pr.},\ \bibinfo {address} {Chicago, USA},\ \bibinfo {year} {1984})\BibitemShut {NoStop}%
\bibitem [{\citenamefont {Eardley}\ \emph {et~al.}(1973)\citenamefont {Eardley}, \citenamefont {Lee},\ and\ \citenamefont {Lightman}}]{GWPolarisationOld}%
  \BibitemOpen
  \bibfield  {author} {\bibinfo {author} {\bibfnamefont {D.~M.}\ \bibnamefont {Eardley}}, \bibinfo {author} {\bibfnamefont {D.~L.}\ \bibnamefont {Lee}},\ and\ \bibinfo {author} {\bibfnamefont {A.~P.}\ \bibnamefont {Lightman}},\ }\bibfield  {title} {\bibinfo {title} {{Gravitational-wave observations as a tool for testing relativistic gravity}},\ }\href {https://doi.org/10.1103/PhysRevD.8.3308} {\bibfield  {journal} {\bibinfo  {journal} {Phys. Rev. D}\ }\textbf {\bibinfo {volume} {8}},\ \bibinfo {pages} {3308} (\bibinfo {year} {1973})}\BibitemShut {NoStop}%
\bibitem [{\citenamefont {Hu}\ \emph {et~al.}(2024)\citenamefont {Hu}, \citenamefont {Wang}, \citenamefont {Tan},\ and\ \citenamefont {Shao}}]{LISAPolarisations}%
  \BibitemOpen
  \bibfield  {author} {\bibinfo {author} {\bibfnamefont {Y.}~\bibnamefont {Hu}}, \bibinfo {author} {\bibfnamefont {P.-P.}\ \bibnamefont {Wang}}, \bibinfo {author} {\bibfnamefont {Y.-J.}\ \bibnamefont {Tan}},\ and\ \bibinfo {author} {\bibfnamefont {C.-G.}\ \bibnamefont {Shao}},\ }\bibfield  {title} {\bibinfo {title} {Testing the polarization of gravitational-wave background with the lisa-tianqin network},\ }\href {https://doi.org/10.3847/1538-4357/ad0cef} {\bibfield  {journal} {\bibinfo  {journal} {The Astrophysical Journal}\ }\textbf {\bibinfo {volume} {961}},\ \bibinfo {pages} {116} (\bibinfo {year} {2024})}\BibitemShut {NoStop}%
\bibitem [{\citenamefont {Capozziello}\ \emph {et~al.}(2008)\citenamefont {Capozziello}, \citenamefont {Corda},\ and\ \citenamefont {De~Laurentis}}]{FR_GWsLISA}%
  \BibitemOpen
  \bibfield  {author} {\bibinfo {author} {\bibfnamefont {S.}~\bibnamefont {Capozziello}}, \bibinfo {author} {\bibfnamefont {C.}~\bibnamefont {Corda}},\ and\ \bibinfo {author} {\bibfnamefont {M.~F.}\ \bibnamefont {De~Laurentis}},\ }\bibfield  {title} {\bibinfo {title} {{Massive gravitational waves from f(R) theories of gravity: Potential detection with LISA}},\ }\href {https://doi.org/10.1016/j.physletb.2008.10.001} {\bibfield  {journal} {\bibinfo  {journal} {Phys. Lett. B}\ }\textbf {\bibinfo {volume} {669}},\ \bibinfo {pages} {255} (\bibinfo {year} {2008})},\ \Eprint {https://arxiv.org/abs/0812.2272} {arXiv:0812.2272 [astro-ph]} \BibitemShut {NoStop}%
\bibitem [{\citenamefont {Naf}\ and\ \citenamefont {Jetzer}(2011)}]{ScalarWavesf_R}%
  \BibitemOpen
  \bibfield  {author} {\bibinfo {author} {\bibfnamefont {J.}~\bibnamefont {Naf}}\ and\ \bibinfo {author} {\bibfnamefont {P.}~\bibnamefont {Jetzer}},\ }\bibfield  {title} {\bibinfo {title} {{On Gravitational Radiation in Quadratic $f(R)$ Gravity}},\ }\href {https://doi.org/10.1103/PhysRevD.84.024027} {\bibfield  {journal} {\bibinfo  {journal} {Phys. Rev. D}\ }\textbf {\bibinfo {volume} {84}},\ \bibinfo {pages} {024027} (\bibinfo {year} {2011})},\ \Eprint {https://arxiv.org/abs/1104.2200} {arXiv:1104.2200 [gr-qc]} \BibitemShut {NoStop}%
\bibitem [{\citenamefont {Bertolami}\ and\ \citenamefont {Martins}(2012)}]{NMCPerfectFluidDynamics}%
  \BibitemOpen
  \bibfield  {author} {\bibinfo {author} {\bibfnamefont {O.}~\bibnamefont {Bertolami}}\ and\ \bibinfo {author} {\bibfnamefont {A.}~\bibnamefont {Martins}},\ }\bibfield  {title} {\bibinfo {title} {{On the dynamics of perfect fluids in non-minimally coupled gravity}},\ }\href {https://doi.org/10.1103/PhysRevD.85.024012} {\bibfield  {journal} {\bibinfo  {journal} {Phys. Rev. D}\ }\textbf {\bibinfo {volume} {85}},\ \bibinfo {pages} {024012} (\bibinfo {year} {2012})},\ \Eprint {https://arxiv.org/abs/1110.2379} {arXiv:1110.2379 [gr-qc]} \BibitemShut {NoStop}%
\bibitem [{\citenamefont {Amendola}\ \emph {et~al.}(2018)\citenamefont {Amendola} \emph {et~al.}}]{PerturbationParameters}%
  \BibitemOpen
  \bibfield  {author} {\bibinfo {author} {\bibfnamefont {L.}~\bibnamefont {Amendola}} \emph {et~al.},\ }\bibfield  {title} {\bibinfo {title} {{Cosmology and fundamental physics with the Euclid satellite}},\ }\href {https://doi.org/10.1007/s41114-017-0010-3} {\bibfield  {journal} {\bibinfo  {journal} {Living Rev. Rel.}\ }\textbf {\bibinfo {volume} {21}},\ \bibinfo {pages} {2} (\bibinfo {year} {2018})},\ \Eprint {https://arxiv.org/abs/1606.00180} {arXiv:1606.00180 [astro-ph.CO]} \BibitemShut {NoStop}%
\bibitem [{\citenamefont {Callister}\ \emph {et~al.}(2017)\citenamefont {Callister}, \citenamefont {Biscoveanu}, \citenamefont {Christensen}, \citenamefont {Isi}, \citenamefont {Matas}, \citenamefont {Minazzoli}, \citenamefont {Regimbau}, \citenamefont {Sakellariadou}, \citenamefont {Tasson},\ and\ \citenamefont {Thrane}}]{PolarisationTestsStochastic}%
  \BibitemOpen
  \bibfield  {author} {\bibinfo {author} {\bibfnamefont {T.}~\bibnamefont {Callister}}, \bibinfo {author} {\bibfnamefont {A.~S.}\ \bibnamefont {Biscoveanu}}, \bibinfo {author} {\bibfnamefont {N.}~\bibnamefont {Christensen}}, \bibinfo {author} {\bibfnamefont {M.}~\bibnamefont {Isi}}, \bibinfo {author} {\bibfnamefont {A.}~\bibnamefont {Matas}}, \bibinfo {author} {\bibfnamefont {O.}~\bibnamefont {Minazzoli}}, \bibinfo {author} {\bibfnamefont {T.}~\bibnamefont {Regimbau}}, \bibinfo {author} {\bibfnamefont {M.}~\bibnamefont {Sakellariadou}}, \bibinfo {author} {\bibfnamefont {J.}~\bibnamefont {Tasson}},\ and\ \bibinfo {author} {\bibfnamefont {E.}~\bibnamefont {Thrane}},\ }\bibfield  {title} {\bibinfo {title} {Polarization-based tests of gravity with the stochastic gravitational-wave background},\ }\href {https://doi.org/10.1103/PhysRevX.7.041058} {\bibfield  {journal} {\bibinfo  {journal} {Phys. Rev. X}\ }\textbf {\bibinfo {volume} {7}},\ \bibinfo {pages} {041058} (\bibinfo {year} {2017})}\BibitemShut {NoStop}%
\bibitem [{\citenamefont {Will}(2014)}]{BreathingLongitudinalDegeneracy}%
  \BibitemOpen
  \bibfield  {author} {\bibinfo {author} {\bibfnamefont {C.~M.}\ \bibnamefont {Will}},\ }\bibfield  {title} {\bibinfo {title} {{The Confrontation between General Relativity and Experiment}},\ }\href {https://doi.org/10.12942/lrr-2014-4} {\bibfield  {journal} {\bibinfo  {journal} {Living Rev. Rel.}\ }\textbf {\bibinfo {volume} {17}},\ \bibinfo {pages} {4} (\bibinfo {year} {2014})},\ \Eprint {https://arxiv.org/abs/1403.7377} {arXiv:1403.7377 [gr-qc]} \BibitemShut {NoStop}%
\bibitem [{\citenamefont {Nishizawa}\ \emph {et~al.}(2009)\citenamefont {Nishizawa}, \citenamefont {Taruya}, \citenamefont {Hayama}, \citenamefont {Kawamura},\ and\ \citenamefont {Sakagami}}]{ScalarDegenerateGroundResponses}%
  \BibitemOpen
  \bibfield  {author} {\bibinfo {author} {\bibfnamefont {A.}~\bibnamefont {Nishizawa}}, \bibinfo {author} {\bibfnamefont {A.}~\bibnamefont {Taruya}}, \bibinfo {author} {\bibfnamefont {K.}~\bibnamefont {Hayama}}, \bibinfo {author} {\bibfnamefont {S.}~\bibnamefont {Kawamura}},\ and\ \bibinfo {author} {\bibfnamefont {M.-a.}\ \bibnamefont {Sakagami}},\ }\bibfield  {title} {\bibinfo {title} {Probing nontensorial polarizations of stochastic gravitational-wave backgrounds with ground-based laser interferometers},\ }\href {https://doi.org/10.1103/PhysRevD.79.082002} {\bibfield  {journal} {\bibinfo  {journal} {Phys. Rev. D}\ }\textbf {\bibinfo {volume} {79}},\ \bibinfo {pages} {082002} (\bibinfo {year} {2009})}\BibitemShut {NoStop}%
\bibitem [{\citenamefont {Berti}\ \emph {et~al.}(2006)\citenamefont {Berti}, \citenamefont {Cardoso},\ and\ \citenamefont {Will}}]{LISA}%
  \BibitemOpen
  \bibfield  {author} {\bibinfo {author} {\bibfnamefont {E.}~\bibnamefont {Berti}}, \bibinfo {author} {\bibfnamefont {V.}~\bibnamefont {Cardoso}},\ and\ \bibinfo {author} {\bibfnamefont {C.~M.}\ \bibnamefont {Will}},\ }\bibfield  {title} {\bibinfo {title} {{On gravitational-wave spectroscopy of massive black holes with the space interferometer LISA}},\ }\href {https://doi.org/10.1103/PhysRevD.73.064030} {\bibfield  {journal} {\bibinfo  {journal} {Phys. Rev. D}\ }\textbf {\bibinfo {volume} {73}},\ \bibinfo {pages} {064030} (\bibinfo {year} {2006})},\ \Eprint {https://arxiv.org/abs/gr-qc/0512160} {arXiv:gr-qc/0512160} \BibitemShut {NoStop}%
\bibitem [{\citenamefont {Crowder}\ and\ \citenamefont {Cornish}(2005)}]{BeyondLISA}%
  \BibitemOpen
  \bibfield  {author} {\bibinfo {author} {\bibfnamefont {J.}~\bibnamefont {Crowder}}\ and\ \bibinfo {author} {\bibfnamefont {N.~J.}\ \bibnamefont {Cornish}},\ }\bibfield  {title} {\bibinfo {title} {{Beyond LISA: Exploring future gravitational wave missions}},\ }\href {https://doi.org/10.1103/PhysRevD.72.083005} {\bibfield  {journal} {\bibinfo  {journal} {Phys. Rev. D}\ }\textbf {\bibinfo {volume} {72}},\ \bibinfo {pages} {083005} (\bibinfo {year} {2005})},\ \Eprint {https://arxiv.org/abs/gr-qc/0506015} {arXiv:gr-qc/0506015} \BibitemShut {NoStop}%
\bibitem [{\citenamefont {Philippoz}\ and\ \citenamefont {Jetzer}(2017)}]{ExtraPolarisationDetection}%
  \BibitemOpen
  \bibfield  {author} {\bibinfo {author} {\bibfnamefont {L.}~\bibnamefont {Philippoz}}\ and\ \bibinfo {author} {\bibfnamefont {P.}~\bibnamefont {Jetzer}},\ }\bibfield  {title} {\bibinfo {title} {{Detecting additional polarization modes with LISA}},\ }\href {https://doi.org/10.1088/1742-6596/840/1/012057} {\bibfield  {journal} {\bibinfo  {journal} {J. Phys. Conf. Ser.}\ }\textbf {\bibinfo {volume} {840}},\ \bibinfo {pages} {012057} (\bibinfo {year} {2017})}\BibitemShut {NoStop}%
\bibitem [{\citenamefont {Nishizawa}\ \emph {et~al.}(2010)\citenamefont {Nishizawa}, \citenamefont {Taruya},\ and\ \citenamefont {Kawamura}}]{SpaceBasedPolarisation}%
  \BibitemOpen
  \bibfield  {author} {\bibinfo {author} {\bibfnamefont {A.}~\bibnamefont {Nishizawa}}, \bibinfo {author} {\bibfnamefont {A.}~\bibnamefont {Taruya}},\ and\ \bibinfo {author} {\bibfnamefont {S.}~\bibnamefont {Kawamura}},\ }\bibfield  {title} {\bibinfo {title} {Cosmological test of gravity with polarizations of stochastic gravitational waves around 0.1--1 hz},\ }\href {https://doi.org/10.1103/PhysRevD.81.104043} {\bibfield  {journal} {\bibinfo  {journal} {Phys. Rev. D}\ }\textbf {\bibinfo {volume} {81}},\ \bibinfo {pages} {104043} (\bibinfo {year} {2010})}\BibitemShut {NoStop}%
\bibitem [{\citenamefont {Mantziris}\ and\ \citenamefont {Bertolami}(2024)}]{GWBubbles}%
  \BibitemOpen
  \bibfield  {author} {\bibinfo {author} {\bibfnamefont {A.}~\bibnamefont {Mantziris}}\ and\ \bibinfo {author} {\bibfnamefont {O.}~\bibnamefont {Bertolami}},\ }\bibfield  {title} {\bibinfo {title} {{Gravitational waves from a curvature-induced phase transition of a Higgs-portal dark matter sector}},\ }\Eprint {https://arxiv.org/abs/2407.18845} {arXiv:2407.18845 [astro-ph.CO]}  (\bibinfo {year} {2024})\BibitemShut {NoStop}%
\bibitem [{\citenamefont {Nishizawa}\ \emph {et~al.}(2008)\citenamefont {Nishizawa}, \citenamefont {Kawamura}, \citenamefont {Akutsu}, \citenamefont {Arai}, \citenamefont {Yamamoto}, \citenamefont {Tatsumi}, \citenamefont {Nishida}, \citenamefont {Sakagami}, \citenamefont {Chiba}, \citenamefont {Takahashi},\ and\ \citenamefont {Sugiyama}}]{HFGW_Interferometer}%
  \BibitemOpen
  \bibfield  {author} {\bibinfo {author} {\bibfnamefont {A.}~\bibnamefont {Nishizawa}}, \bibinfo {author} {\bibfnamefont {S.}~\bibnamefont {Kawamura}}, \bibinfo {author} {\bibfnamefont {T.}~\bibnamefont {Akutsu}}, \bibinfo {author} {\bibfnamefont {K.}~\bibnamefont {Arai}}, \bibinfo {author} {\bibfnamefont {K.}~\bibnamefont {Yamamoto}}, \bibinfo {author} {\bibfnamefont {D.}~\bibnamefont {Tatsumi}}, \bibinfo {author} {\bibfnamefont {E.}~\bibnamefont {Nishida}}, \bibinfo {author} {\bibfnamefont {M.-a.}\ \bibnamefont {Sakagami}}, \bibinfo {author} {\bibfnamefont {T.}~\bibnamefont {Chiba}}, \bibinfo {author} {\bibfnamefont {R.}~\bibnamefont {Takahashi}},\ and\ \bibinfo {author} {\bibfnamefont {N.}~\bibnamefont {Sugiyama}},\ }\bibfield  {title} {\bibinfo {title} {Laser-interferometric detectors for gravitational wave backgrounds at 100 mhz: Detector design and sensitivity},\ }\href {https://doi.org/10.1103/PhysRevD.77.022002} {\bibfield  {journal} {\bibinfo  {journal} {Phys. Rev. D}\ }\textbf {\bibinfo {volume}
  {77}},\ \bibinfo {pages} {022002} (\bibinfo {year} {2008})}\BibitemShut {NoStop}%
\bibitem [{\citenamefont {Aggarwal}\ \emph {et~al.}(2021)\citenamefont {Aggarwal} \emph {et~al.}}]{HFGWDetection_Review}%
  \BibitemOpen
  \bibfield  {author} {\bibinfo {author} {\bibfnamefont {N.}~\bibnamefont {Aggarwal}} \emph {et~al.},\ }\bibfield  {title} {\bibinfo {title} {{Challenges and opportunities of gravitational-wave searches at MHz to GHz frequencies}},\ }\href {https://doi.org/10.1007/s41114-021-00032-5} {\bibfield  {journal} {\bibinfo  {journal} {Living Rev. Rel.}\ }\textbf {\bibinfo {volume} {24}},\ \bibinfo {pages} {4} (\bibinfo {year} {2021})},\ \Eprint {https://arxiv.org/abs/2011.12414} {arXiv:2011.12414 [gr-qc]} \BibitemShut {NoStop}%
\bibitem [{\citenamefont {Navarro}\ \emph {et~al.}(2024)\citenamefont {Navarro}, \citenamefont {Gimeno}, \citenamefont {Monz\'o-Cabrera}, \citenamefont {D\'{\i}az-Morcillo},\ and\ \citenamefont {Blas}}]{GW_EMCavity}%
  \BibitemOpen
  \bibfield  {author} {\bibinfo {author} {\bibfnamefont {P.}~\bibnamefont {Navarro}}, \bibinfo {author} {\bibfnamefont {B.}~\bibnamefont {Gimeno}}, \bibinfo {author} {\bibfnamefont {J.}~\bibnamefont {Monz\'o-Cabrera}}, \bibinfo {author} {\bibfnamefont {A.}~\bibnamefont {D\'{\i}az-Morcillo}},\ and\ \bibinfo {author} {\bibfnamefont {D.}~\bibnamefont {Blas}},\ }\bibfield  {title} {\bibinfo {title} {Study of a cubic cavity resonator for gravitational waves detection in the microwave frequency range},\ }\href {https://doi.org/10.1103/PhysRevD.109.104048} {\bibfield  {journal} {\bibinfo  {journal} {Phys. Rev. D}\ }\textbf {\bibinfo {volume} {109}},\ \bibinfo {pages} {104048} (\bibinfo {year} {2024})}\BibitemShut {NoStop}%
\bibitem [{\citenamefont {Gatti}\ \emph {et~al.}(2024)\citenamefont {Gatti}, \citenamefont {Visinelli},\ and\ \citenamefont {Zantedeschi}}]{HFGW_Recent}%
  \BibitemOpen
  \bibfield  {author} {\bibinfo {author} {\bibfnamefont {C.}~\bibnamefont {Gatti}}, \bibinfo {author} {\bibfnamefont {L.}~\bibnamefont {Visinelli}},\ and\ \bibinfo {author} {\bibfnamefont {M.}~\bibnamefont {Zantedeschi}},\ }\bibfield  {title} {\bibinfo {title} {Cavity detection of gravitational waves: Where do we stand?},\ }\href {https://doi.org/10.1103/PhysRevD.110.023018} {\bibfield  {journal} {\bibinfo  {journal} {Phys. Rev. D}\ }\textbf {\bibinfo {volume} {110}},\ \bibinfo {pages} {023018} (\bibinfo {year} {2024})}\BibitemShut {NoStop}%
\bibitem [{\citenamefont {Li}\ \emph {et~al.}(2020)\citenamefont {Li}, \citenamefont {Wen}, \citenamefont {Fang}, \citenamefont {Li},\ and\ \citenamefont {Zhang}}]{HighFrequencyGWs}%
  \BibitemOpen
  \bibfield  {author} {\bibinfo {author} {\bibfnamefont {F.-Y.}\ \bibnamefont {Li}}, \bibinfo {author} {\bibfnamefont {H.}~\bibnamefont {Wen}}, \bibinfo {author} {\bibfnamefont {Z.-Y.}\ \bibnamefont {Fang}}, \bibinfo {author} {\bibfnamefont {D.}~\bibnamefont {Li}},\ and\ \bibinfo {author} {\bibfnamefont {T.-J.}\ \bibnamefont {Zhang}},\ }\bibfield  {title} {\bibinfo {title} {{Electromagnetic response to high-frequency gravitational waves having additional polarization states: distinguishing and probing tensor-mode, vector-mode and scalar-mode gravitons}},\ }\href {https://doi.org/10.1140/epjc/s10052-020-08429-2} {\bibfield  {journal} {\bibinfo  {journal} {Eur. Phys. J. C}\ }\textbf {\bibinfo {volume} {80}},\ \bibinfo {pages} {879} (\bibinfo {year} {2020})},\ \Eprint {https://arxiv.org/abs/1712.00766} {arXiv:1712.00766 [gr-qc]} \BibitemShut {NoStop}%
\bibitem [{\citenamefont {Kramer}\ \emph {et~al.}(2004)\citenamefont {Kramer}, \citenamefont {Backer}, \citenamefont {Cordes}, \citenamefont {Lazio}, \citenamefont {Stappers},\ and\ \citenamefont {Johnston}}]{PTA_SKAStrongFieldTests}%
  \BibitemOpen
  \bibfield  {author} {\bibinfo {author} {\bibfnamefont {M.}~\bibnamefont {Kramer}}, \bibinfo {author} {\bibfnamefont {D.~C.}\ \bibnamefont {Backer}}, \bibinfo {author} {\bibfnamefont {J.~M.}\ \bibnamefont {Cordes}}, \bibinfo {author} {\bibfnamefont {T.~J.~W.}\ \bibnamefont {Lazio}}, \bibinfo {author} {\bibfnamefont {B.~W.}\ \bibnamefont {Stappers}},\ and\ \bibinfo {author} {\bibfnamefont {S.}~\bibnamefont {Johnston}},\ }\bibfield  {title} {\bibinfo {title} {{Strong-field tests of gravity using pulsars and black holes}},\ }\href {https://doi.org/10.1016/j.newar.2004.09.020} {\bibfield  {journal} {\bibinfo  {journal} {New Astron. Rev.}\ }\textbf {\bibinfo {volume} {48}},\ \bibinfo {pages} {993} (\bibinfo {year} {2004})},\ \Eprint {https://arxiv.org/abs/astro-ph/0409379} {arXiv:astro-ph/0409379} \BibitemShut {NoStop}%
\bibitem [{\citenamefont {Alves}\ and\ \citenamefont {Tinto}(2011)}]{PTA1}%
  \BibitemOpen
  \bibfield  {author} {\bibinfo {author} {\bibfnamefont {M.~E. d.~S.}\ \bibnamefont {Alves}}\ and\ \bibinfo {author} {\bibfnamefont {M.}~\bibnamefont {Tinto}},\ }\bibfield  {title} {\bibinfo {title} {Pulsar timing sensitivities to gravitational waves from relativistic metric theories of gravity},\ }\href {https://doi.org/10.1103/PhysRevD.83.123529} {\bibfield  {journal} {\bibinfo  {journal} {Phys. Rev. D}\ }\textbf {\bibinfo {volume} {83}},\ \bibinfo {pages} {123529} (\bibinfo {year} {2011})}\BibitemShut {NoStop}%
\bibitem [{\citenamefont {Qin}\ \emph {et~al.}(2019)\citenamefont {Qin}, \citenamefont {Boddy}, \citenamefont {Kamionkowski},\ and\ \citenamefont {Dai}}]{PTA2}%
  \BibitemOpen
  \bibfield  {author} {\bibinfo {author} {\bibfnamefont {W.}~\bibnamefont {Qin}}, \bibinfo {author} {\bibfnamefont {K.~K.}\ \bibnamefont {Boddy}}, \bibinfo {author} {\bibfnamefont {M.}~\bibnamefont {Kamionkowski}},\ and\ \bibinfo {author} {\bibfnamefont {L.}~\bibnamefont {Dai}},\ }\bibfield  {title} {\bibinfo {title} {Pulsar-timing arrays, astrometry, and gravitational waves},\ }\href {https://doi.org/10.1103/PhysRevD.99.063002} {\bibfield  {journal} {\bibinfo  {journal} {Phys. Rev. D}\ }\textbf {\bibinfo {volume} {99}},\ \bibinfo {pages} {063002} (\bibinfo {year} {2019})}\BibitemShut {NoStop}%
\bibitem [{\citenamefont {Wu}\ \emph {et~al.}(2022)\citenamefont {Wu}, \citenamefont {Chen},\ and\ \citenamefont {Huang}}]{PTADataConstraints}%
  \BibitemOpen
  \bibfield  {author} {\bibinfo {author} {\bibfnamefont {Y.-M.}\ \bibnamefont {Wu}}, \bibinfo {author} {\bibfnamefont {Z.-C.}\ \bibnamefont {Chen}},\ and\ \bibinfo {author} {\bibfnamefont {Q.-G.}\ \bibnamefont {Huang}},\ }\bibfield  {title} {\bibinfo {title} {Constraining the polarization of gravitational waves with the parkes pulsar timing array second data release},\ }\href {https://doi.org/10.3847/1538-4357/ac35cc} {\bibfield  {journal} {\bibinfo  {journal} {The Astrophysical Journal}\ }\textbf {\bibinfo {volume} {925}},\ \bibinfo {pages} {37} (\bibinfo {year} {2022})}\BibitemShut {NoStop}%
\bibitem [{\citenamefont {Gair}\ \emph {et~al.}(2015)\citenamefont {Gair}, \citenamefont {Romano},\ and\ \citenamefont {Taylor}}]{PTAMapping}%
  \BibitemOpen
  \bibfield  {author} {\bibinfo {author} {\bibfnamefont {J.~R.}\ \bibnamefont {Gair}}, \bibinfo {author} {\bibfnamefont {J.~D.}\ \bibnamefont {Romano}},\ and\ \bibinfo {author} {\bibfnamefont {S.~R.}\ \bibnamefont {Taylor}},\ }\bibfield  {title} {\bibinfo {title} {Mapping gravitational-wave backgrounds of arbitrary polarisation using pulsar timing arrays},\ }\href {https://doi.org/10.1103/PhysRevD.92.102003} {\bibfield  {journal} {\bibinfo  {journal} {Phys. Rev. D}\ }\textbf {\bibinfo {volume} {92}},\ \bibinfo {pages} {102003} (\bibinfo {year} {2015})}\BibitemShut {NoStop}%
\bibitem [{\citenamefont {Lee}\ \emph {et~al.}(2008)\citenamefont {Lee}, \citenamefont {Jenet},\ and\ \citenamefont {Price}}]{PTAPulsarAmount}%
  \BibitemOpen
  \bibfield  {author} {\bibinfo {author} {\bibfnamefont {K.~J.}\ \bibnamefont {Lee}}, \bibinfo {author} {\bibfnamefont {F.~A.}\ \bibnamefont {Jenet}},\ and\ \bibinfo {author} {\bibfnamefont {R.~H.}\ \bibnamefont {Price}},\ }\bibfield  {title} {\bibinfo {title} {Pulsar timing as a probe of non-einsteinian polarizations of gravitational waves},\ }\href {https://doi.org/10.1086/591080} {\bibfield  {journal} {\bibinfo  {journal} {The Astrophysical Journal}\ }\textbf {\bibinfo {volume} {685}},\ \bibinfo {pages} {1304} (\bibinfo {year} {2008})}\BibitemShut {NoStop}%
\bibitem [{\citenamefont {Salvatelli}\ \emph {et~al.}(2016)\citenamefont {Salvatelli}, \citenamefont {Piazza},\ and\ \citenamefont {Marinoni}}]{WLMeasurement2}%
  \BibitemOpen
  \bibfield  {author} {\bibinfo {author} {\bibfnamefont {V.}~\bibnamefont {Salvatelli}}, \bibinfo {author} {\bibfnamefont {F.}~\bibnamefont {Piazza}},\ and\ \bibinfo {author} {\bibfnamefont {C.}~\bibnamefont {Marinoni}},\ }\bibfield  {title} {\bibinfo {title} {{Constraints on modified gravity from Planck 2015: when the health of your theory makes the difference}},\ }\href {https://doi.org/10.1088/1475-7516/2016/09/027} {\bibfield  {journal} {\bibinfo  {journal} {JCAP}\ }\textbf {\bibinfo {volume} {09}},\ \bibinfo {pages} {027}},\ \Eprint {https://arxiv.org/abs/1602.08283} {arXiv:1602.08283 [astro-ph.CO]} \BibitemShut {NoStop}%
\bibitem [{\citenamefont {Daniel}\ and\ \citenamefont {Linder}(2010)}]{WeakLensingMeasurement}%
  \BibitemOpen
  \bibfield  {author} {\bibinfo {author} {\bibfnamefont {S.~F.}\ \bibnamefont {Daniel}}\ and\ \bibinfo {author} {\bibfnamefont {E.~V.}\ \bibnamefont {Linder}},\ }\bibfield  {title} {\bibinfo {title} {{Confronting General Relativity with Further Cosmological Data}},\ }\href {https://doi.org/10.1103/PhysRevD.82.103523} {\bibfield  {journal} {\bibinfo  {journal} {Phys. Rev. D}\ }\textbf {\bibinfo {volume} {82}},\ \bibinfo {pages} {103523} (\bibinfo {year} {2010})},\ \Eprint {https://arxiv.org/abs/1008.0397} {arXiv:1008.0397 [astro-ph.CO]} \BibitemShut {NoStop}%
\bibitem [{\citenamefont {Garcia-Quintero}\ \emph {et~al.}(2020)\citenamefont {Garcia-Quintero}, \citenamefont {Ishak},\ and\ \citenamefont {Ning}}]{WeakLensingMeasurement2}%
  \BibitemOpen
  \bibfield  {author} {\bibinfo {author} {\bibfnamefont {C.}~\bibnamefont {Garcia-Quintero}}, \bibinfo {author} {\bibfnamefont {M.}~\bibnamefont {Ishak}},\ and\ \bibinfo {author} {\bibfnamefont {O.}~\bibnamefont {Ning}},\ }\bibfield  {title} {\bibinfo {title} {{Current constraints on deviations from General Relativity using binning in redshift and scale}},\ }\href {https://doi.org/10.1088/1475-7516/2020/12/018} {\bibfield  {journal} {\bibinfo  {journal} {JCAP}\ }\textbf {\bibinfo {volume} {12}},\ \bibinfo {pages} {018}},\ \Eprint {https://arxiv.org/abs/2010.12519} {arXiv:2010.12519 [astro-ph.CO]} \BibitemShut {NoStop}%
\bibitem [{\citenamefont {Mueller}\ \emph {et~al.}(2018)\citenamefont {Mueller}, \citenamefont {Percival}, \citenamefont {Linder}, \citenamefont {Alam}, \citenamefont {Zhao}, \citenamefont {S\'anchez}, \citenamefont {Beutler},\ and\ \citenamefont {Brinkmann}}]{WLMeasurement3}%
  \BibitemOpen
  \bibfield  {author} {\bibinfo {author} {\bibfnamefont {E.-M.}\ \bibnamefont {Mueller}}, \bibinfo {author} {\bibfnamefont {W.}~\bibnamefont {Percival}}, \bibinfo {author} {\bibfnamefont {E.}~\bibnamefont {Linder}}, \bibinfo {author} {\bibfnamefont {S.}~\bibnamefont {Alam}}, \bibinfo {author} {\bibfnamefont {G.-B.}\ \bibnamefont {Zhao}}, \bibinfo {author} {\bibfnamefont {A.~G.}\ \bibnamefont {S\'anchez}}, \bibinfo {author} {\bibfnamefont {F.}~\bibnamefont {Beutler}},\ and\ \bibinfo {author} {\bibfnamefont {J.}~\bibnamefont {Brinkmann}},\ }\bibfield  {title} {\bibinfo {title} {{The clustering of galaxies in the completed SDSS-III Baryon Oscillation Spectroscopic Survey: constraining modified gravity}},\ }\href {https://doi.org/10.1093/mnras/stx3232} {\bibfield  {journal} {\bibinfo  {journal} {Mon. Not. Roy. Astron. Soc.}\ }\textbf {\bibinfo {volume} {475}},\ \bibinfo {pages} {2122} (\bibinfo {year} {2018})},\ \Eprint {https://arxiv.org/abs/1612.00812} {arXiv:1612.00812 [astro-ph.CO]} \BibitemShut {NoStop}%
\bibitem [{\citenamefont {Abbott}\ \emph {et~al.}(2023)\citenamefont {Abbott} \emph {et~al.}}]{DESYear3Lensing}%
  \BibitemOpen
  \bibfield  {author} {\bibinfo {author} {\bibfnamefont {T.~M.~C.}\ \bibnamefont {Abbott}} \emph {et~al.} (\bibinfo {collaboration} {DES}),\ }\bibfield  {title} {\bibinfo {title} {{Dark Energy Survey Year 3 results: Constraints on extensions to \ensuremath{\Lambda}CDM with weak lensing and galaxy clustering}},\ }\href {https://doi.org/10.1103/PhysRevD.107.083504} {\bibfield  {journal} {\bibinfo  {journal} {Phys. Rev. D}\ }\textbf {\bibinfo {volume} {107}},\ \bibinfo {pages} {083504} (\bibinfo {year} {2023})},\ \Eprint {https://arxiv.org/abs/2207.05766} {arXiv:2207.05766 [astro-ph.CO]} \BibitemShut {NoStop}%
\bibitem [{\citenamefont {Aghanim}\ \emph {et~al.}(2020)\citenamefont {Aghanim} \emph {et~al.}}]{WLPlanck2018}%
  \BibitemOpen
  \bibfield  {author} {\bibinfo {author} {\bibfnamefont {N.}~\bibnamefont {Aghanim}} \emph {et~al.} (\bibinfo {collaboration} {Planck}),\ }\bibfield  {title} {\bibinfo {title} {{Planck 2018 results. VI. Cosmological parameters}},\ }\href {https://doi.org/10.1051/0004-6361/201833910} {\bibfield  {journal} {\bibinfo  {journal} {Astron. Astrophys.}\ }\textbf {\bibinfo {volume} {641}},\ \bibinfo {pages} {A6} (\bibinfo {year} {2020})},\ \bibinfo {note} {[Erratum: Astron.Astrophys. 652, C4 (2021)]},\ \Eprint {https://arxiv.org/abs/1807.06209} {arXiv:1807.06209 [astro-ph.CO]} \BibitemShut {NoStop}%
\bibitem [{\citenamefont {Bertolami}\ and\ \citenamefont {Páramos}(2008)}]{NMCScalarTensorAnalogy}%
  \BibitemOpen
  \bibfield  {author} {\bibinfo {author} {\bibfnamefont {O.}~\bibnamefont {Bertolami}}\ and\ \bibinfo {author} {\bibfnamefont {J.}~\bibnamefont {Páramos}},\ }\bibfield  {title} {\bibinfo {title} {{On the non-trivial gravitational coupling to matter}},\ }\href {https://doi.org/10.1088/0264-9381/25/24/245017} {\bibfield  {journal} {\bibinfo  {journal} {Class. Quant. Grav.}\ }\textbf {\bibinfo {volume} {25}},\ \bibinfo {pages} {245017} (\bibinfo {year} {2008})},\ \Eprint {https://arxiv.org/abs/0805.1241} {arXiv:0805.1241 [gr-qc]} \BibitemShut {NoStop}%
\bibitem [{\citenamefont {van~de Weygaert}\ and\ \citenamefont {Platen}(2011)}]{CosmicVoids}%
  \BibitemOpen
  \bibfield  {author} {\bibinfo {author} {\bibfnamefont {R.}~\bibnamefont {van~de Weygaert}}\ and\ \bibinfo {author} {\bibfnamefont {E.}~\bibnamefont {Platen}},\ }\bibfield  {title} {\bibinfo {title} {{Cosmic Voids: structure, dynamics and galaxies}},\ }\href {https://doi.org/10.1142/S2010194511000092} {\bibfield  {journal} {\bibinfo  {journal} {Int. J. Mod. Phys. Conf. Ser.}\ }\textbf {\bibinfo {volume} {01}},\ \bibinfo {pages} {41} (\bibinfo {year} {2011})},\ \Eprint {https://arxiv.org/abs/0912.2997} {arXiv:0912.2997 [astro-ph.CO]} \BibitemShut {NoStop}%
\bibitem [{\citenamefont {Abazajian}\ \emph {et~al.}(2022)\citenamefont {Abazajian} \emph {et~al.}}]{PrimordialGWs}%
  \BibitemOpen
  \bibfield  {author} {\bibinfo {author} {\bibfnamefont {K.}~\bibnamefont {Abazajian}} \emph {et~al.} (\bibinfo {collaboration} {CMB-S4}),\ }\bibfield  {title} {\bibinfo {title} {{CMB-S4: Forecasting Constraints on Primordial Gravitational Waves}},\ }\href {https://doi.org/10.3847/1538-4357/ac1596} {\bibfield  {journal} {\bibinfo  {journal} {Astrophys. J.}\ }\textbf {\bibinfo {volume} {926}},\ \bibinfo {pages} {54} (\bibinfo {year} {2022})},\ \Eprint {https://arxiv.org/abs/2008.12619} {arXiv:2008.12619 [astro-ph.CO]} \BibitemShut {NoStop}%
\end{thebibliography}%

\end{document}